\documentclass[sn-mathphys-num]{sn-jnl}

\usepackage{graphicx}%
\usepackage{multirow}%
\usepackage{amsmath,amssymb,amsfonts}%
\usepackage{amsthm}%
\usepackage{mathrsfs}%
\usepackage[title]{appendix}%
\usepackage{xcolor}%
\usepackage{textcomp}%
\usepackage{manyfoot}%
\usepackage{booktabs}%
\usepackage{algorithm}%
\usepackage{algorithmicx}%
\usepackage{algpseudocode}%
\usepackage{listings}%
\usepackage{physics}
\usepackage{array}
\usepackage{amsmath}
\usepackage{amsthm}
\usepackage{xcolor} 

\usepackage{subfig} 
\usepackage{comment}
\usepackage{array}
\usepackage{tabularx}
\usepackage{multirow}
\usepackage{longtable}
\usepackage{makecell}

\theoremstyle{thmstyleone}%
\newtheorem{theorem}{Theorem}
\newtheorem{proposition}[theorem]{Proposition}%

\theoremstyle{thmstyletwo}%

\theoremstyle{thmstylethree}%
\def\Plus{\texttt{+}} 

\begin{document}

\title[Article Title]{Quantum Machine Learning Algorithms for Anomaly Detection: a Review}


\author*[1,2]{\fnm{Sebastiano} \sur{Corli}}\email{sebastiano.corli@polimi.it}

\author*[3]{\fnm{Lorenzo} \sur{Moro}}\email{lorenzo.moro@polimi.it}

\author[4]{\fnm{Daniele} \sur{Dragoni}}\email{daniele.dragoni@ext.leonardo.it}

\author[5]{\fnm{Massimiliano} \sur{Dispenza}}\email{massimiliano.dispenza@leonardo.it}

\author*[2,6]{\fnm{Enrico} \sur{Prati}}\email{enrico.prati@unimi.it}

\affil*[1]{\orgdiv{Department of Physics}, \orgname{Politecnico di Milano}, \orgaddress{\city{Milano}, \state{Italy}}}

\affil[2]{\orgdiv{Istituto di Fotonica e Nanotecnologie}, \orgname{Consiglio Nazionale delle Ricerche}, \orgaddress{\street{Piazza Leonardo da Vinci, 32}, \city{Milano}, \state{Italy}}}

\affil[3]{\orgdiv{QBrain}, \city{Milano}, \state{Italy}}

\affil[4]{\orgdiv{High Performance Computing Laboratory}, \orgname{Leonardo s.p.a.}, \orgaddress{\street{Via R. Pieragostini, 80}, \city{Genova}, \state{Italy}}}

\affil[5]{\orgdiv{Quantum Technologies Leonardo Lab}, \orgname{Leonardo S.p.A.}, \orgaddress{\street{Via Tiburtina, KM 12, 400}, \city{Roma}, \state{Italy}}}

\affil[6]{\orgdiv{Department of physics}, \orgname{Università degli studi di Milano}, \street{Via G. Celoria, 16} \city{Milano}, \country{Italy}}


\abstract{The advent of quantum computers has justified the development of quantum machine learning algorithms, based on the adaptation of the principles of machine learning to the formalism of qubits. Among such quantum algorithms, anomaly detection represents an important problem crossing several disciplines from cybersecurity, to fraud detection to particle physics.
We summarize the key concepts involved in quantum computing, introducing the formal concept of quantum speed up. The review provides a structured map of anomaly detection based on quantum machine learning.  We have grouped existing algorithms according to the different learning methods, namely quantum supervised, quantum unsupervised and quantum reinforcement learning, respectively.  We provide an estimate of the hardware resources to provide sufficient computational power in the future. The review provides a systematic and compact understanding of the techniques belonging to each category. We eventually provide a discussion on the computational complexity of the learning methods in real application domains.}




\maketitle

\section{Introduction}\label{sec1}

\begin{figure}[h]
    \centering
    \includegraphics[width=0.5\textwidth]{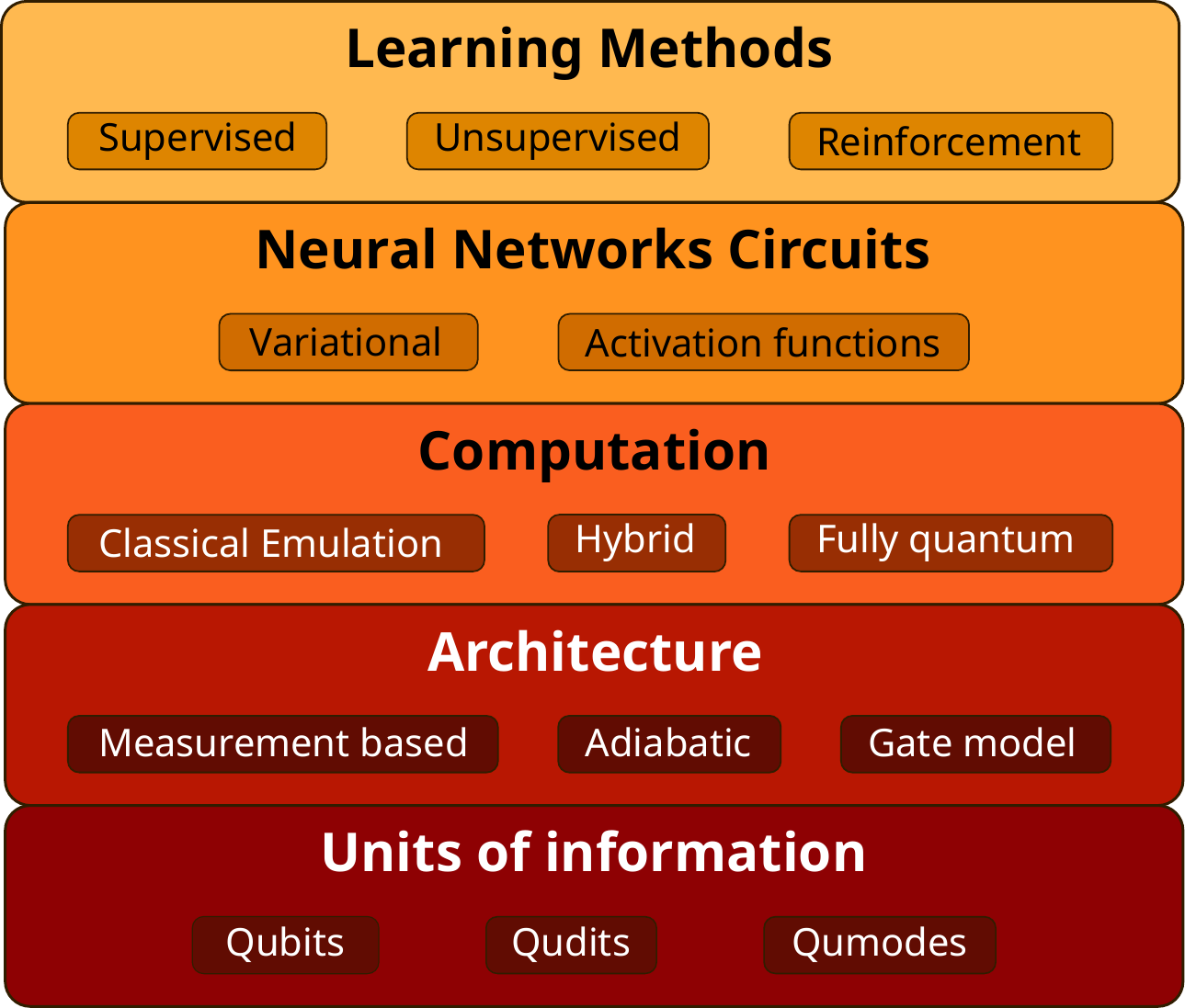}
    \caption{A list of topics to categorize the field of quantum machine learning and its algorithms.}
    \label{fig:taxonomy}
    \centering
    \end{figure}

Anomaly detection takes advantage from a wide range of artificial intelligence algorithms, which -- combined with human supervision -- may raise the degree of protection and of integrity of systems and data.
    On the other hand, the advent of quantum computing has made possible to implement quantum algorithms on real prototypical -- but already commercial -- quantum hardware. Such algorithms include machine learning-related algorithms which may inherit the quantum speed-up typical of quantum algorithms.
    The differentiation among quantum computing architectures (gate based~\cite{barenco1995elementary,schuld2015introduction}, adiabatic~\cite{morita2008mathematical,van2001powerful}, measurement based~\cite{hoban2014measurement,pius2010automatic,corli2022efficient}), encoding (digital~\cite{kopczyk2018quantum} versus continuous variables~\cite{pfister2019continuous,killoran2019continuous,kreis2012classifying,kendon2010quantum}), hardware (several substrates from solid state to trapped ions~\cite{haffner2008quantum,blatt2012quantum} to photons~\cite{o2007optical}), the diversity of the quantum algorithms and the unclear advantage carried by some of them in the field of quantum machine learning, and also the range of potential application domains and the different methods to achieve the same goal, call for a systematic review of what has been done and what is known in the field, in order to address more efficiently the investigation towards meaningful, feasible and relevant applications.
        The quantum machine learning algorithms proposed in literature for anomaly detection purposes are updated to Q1 of 2024, and clustered by applying the criteria of training method. Indeed, the latter represents the criterion which drives the choice among a families of algorithms. Therefore, we classify the quantum algorithms for machine learning according to the same classification of classical algorithms, namely among supervised learning, unsupervised learning and reinforcement learning, respectively.
    Despite its recent birth, the topic of quantum machine learning has been systematically reviewed in the time span 2015-2023~\cite{garcia2022systematic,biamonte2017quantum,zhang2020recent,schuld2015introduction,zeguendry2023quantum,jerbi2023quantum}. We privileged the literature which includes a practical implementation on either an actual quantum computer or the simulator of some existing hardware. The literature reviewed by such sources is integrated by more recent articles not included there.
    In the next sections we describe the summary of aims and AI algorithms used in anomaly detection, the basics of quantum computing and quantum advantage, the key quantum algorithms developed in the field which are relevant for anomaly detection purposes, and we systematically analyze the most recent advancements in the field of quantum machine learning applied to anomaly detection. In the second section, we summarize the application for quantum machine learning in the field of anomaly detection. In the third section, we introduce the concept of quantum advantage, along with a classification for the possible quantum speedups. In the fourth section, the groundings of quantum computing are introduced: from the definition of qubits, qudits and qumodes to the encoding of classical information into these units of computation. In the fifth section, different architectures of quantum computation (adiabatic and circuital models) are introduced, along with the support of classical computers. The sixth section is dedicated to focus on the role of HPC and classical computation to interface with quantum hardware and algorithms. Quantum neural networks and variational circuits, to translate on a quantum device the classical neural networks, are explained in section seven. In the remaining sections, a review on specific quantum algorithms for anomaly detection is provided, classified with respect to the categories of supervised, unsupervised and reinforcement learning. A summary of all the exposed algorithms can be found at Table \ref{tab:summary}.
    \color{black}
    In order to cover all of the classes of quantum algorithms and computational architectures in Figure \ref{fig:taxonomy}, we select and summarize a paramount paper for each of these classes. For instance, Killoran's work~\cite{killoran2019continuous} in 2019 defined how to build continuous variables neural networks for quantum computers, along with their employment for anomaly detection, while Tacchino in 2019~\cite{tacchino2019artificial} proposed a model of fully quantum neural perceptron. The other works we report are from Useche~\cite{useche2022quantum} (2022) for performing classification tasks with qudits, Herr~\cite{herr2021anomaly} (2021) for introducing the QGAN in the field of anomaly detection, Moro~\cite{moro2023anomaly} (2023) to boost the performances on the Restricted Boltzmann Machine via an annealer, the Harrow, Hassidim, Lloyd paper~\cite{harrow2009quantum} (2009), which introduced the namesake HHL algorithm employed for the support vector machines (and beyond), and the paper by Albarràn-Arriagada~\cite{albarran2018measurement} (2018) for the quantum Reinforcement Learning.
    \color{black}

\section{Application domains for quantum anomaly detection}

As this review elaborates on the intersection of quantum machine learning methods with applications in the anomaly detection, we first briefly assess which classes of algorithms are related to such topic, from image recognition and data classification to clustering analysis. Moreover, another topic of interest is provided by the domain of applications: cybersecurity is a major focus for quantum machine learning~\cite{payares2021quantum,killoran2019continuous,wang2022integrating,liu2018quantum,suryotrisongko2022evaluating} (and for all--around machine learning), but a considerable interest arises on disparate topics such as big data for science, i.e. detecting Higgs--boson decay at LHC collider \cite{ngairangbam2022anomaly,wozniak2023quantum}, geophysical analysis \cite{abedi2012support}, detection of new particles at LHC~\cite{wu2021application,schuhmacher2023unravelling} or even audio recognition~\cite{davy2002detection,chai2024quantum}.

Typically, cybersecurity is characterized as a collection of technologies and processes designed to protect computers, networks, programs, and data against malicious activities, attacks, harm, or unauthorized access. In the field of cybersecurity, anomaly detection is of paramount importance. Many datasets exist, including intrusion analysis, malware analysis, and spam analysis, which are used for different purposes~\cite{sarker2021deep}.

All cybersecurity matter, while becoming increasingly crucial to all modern industrial and institutional activities, has also grown in the past decades in terms of complexity of the multiple bodies and structure which have been created to implement cyber defence functionalities. 
SOCs (Security Operation Centers) are for example hugely complex cybersecurity systems, exploited to monitor infrastructures, to supervise networks, to detect threats also able to guarantee early warning and security awareness.
Once security incidents have been detected, they have also to be managed. The so called Computer Emergency Response Teams (CERTs) come into play. 
Such complexity of cyber monitoring and countermeasure systems is strongly related to an equivalent rearrangement of the threat side, where more conventional private cyber crime groups and cyber terrorists or hacker sources were joined by government linked teams. Such groups carry out a so called cyber warfare by systematically implementing cyber attacks to national or institutional IT services~\cite{li2021comprehensive}.
Complexity grows together with continuous and rapid adaptations and modification of threats themselves. Ransomware groups and other malicious players, e.g., are changing their initial access vectors while the digital attack surface and vulnerabilities shift, also exploiting commercial tools to disguise their breaches and deploying new ransomware schemes. 
Anomaly detection~\cite{ravinder2023review} is a fundamental tool for this task and such continuously evolving cyber threat landscape ultimately calls for actions by SOCs and CERTs to be largely become automated only asking for man-in-the-loop in few very critical steps; which on its turn naturally links to the use of machine learning~\cite{wang2022efficient,ayodeji2020new,gomez2023susan,tufan2021anomaly}, as a means of automated response. 

The main purpose of such algorithms is to provide early warning of attack, possibly even before the attack is launched~\cite{wirkuttis2017artificial}.
Cyber intelligence deals with amount of data, their heterogeneity and their high production rate. AI is believed~\cite{truong2020artificial} capable to  enhance cyberspace security more effectively than conventional methods for three reasons, namely:
\begin{itemize}
\item Discovery of new and sophisticated changes in attack flexibility, by better adaptation to detect anomalous, faster and more accurate operations.
\item Naturally handling high volumes of data
\item Learning over time to respond better to threats.
\end{itemize}
Moreover, between different AI methods such as neural networks, fuzzy logic, expert system, machine learning, and deep learning, the two latter bring the most achievements.
In the field of cybersecurity, the applications span mainly on malware detection, intrusion detection (ID), endpoint detection (ED), phishing detection and advanced persistent threat (APT). All of them take advantage of different methods based on a number of possible algorithms (or combinations of them): Naive Bayes method, Support Vector Machines (SVM), Decision Trees and, more recently deep Neural Networks~\cite{yuan2014droid,yuxin2019malware,vinayakumar2019robust, bitter2010application}. One should keep in mind that since spurious transactions are far fewer than the normal ones, the highly imbalanced
data makes fraud detection very challenging and calls for ways to address it beyond the traditional
machine learning approach\cite{wang2022integrating}. Furthermore, the development for instance of new fraud
detection methods is made more difficult due to the severe limitation of the exchange of ideas in fraud detection\cite{kou2004survey}.
More recently, Reinforcement Learning proved to be a robust but flexible method to prevent cyber attacks~\cite{ahmad2021network, sewak2021deep,sjarif2019endpoint}, also thanks to a vast range of available algorithms, such as Deep Deterministic Policy Gradient (DDPG)~\cite{silver2014deterministic}, Trust Region Policy Optimization (TRPO)~\cite{schulman2015trust}, Proximal Policy Optimization (PPO)~\cite{schulman2017proximal}, Generalized Advantage Estimation (GAE)~\cite{schulman2015high}.

In the broader field of anomaly detection, Neural Networks have also been successfully employed for medical and public health domain~\cite{campbell2000linear}, fault detection for mechanical components and structural damage detection~\cite{petsche1995neural,fujimaki2005approach,brotherton2001anomaly}.
As for image and pattern recognition or data text analysis, along with detection of spurious elements in datasets from the corresponding domains, techniques such as Support Vector Machines~\cite{davy2002detection,manevitz2001one}, Neural Networks~\cite{augusteijn2002neural,singh2004approach} and clustering based algorithms~\cite{srivastava2006enabling,srivastava2005discovering} have been deployed. Support Vector Machines have also been addressed to foil phishing attacks, achieving a rationale performance with 99.6\% of True Positive Rate and 0.44\% of False Positive Rate~\cite{gowtham2014comprehensive}.
Finally, also \textit{advanced persistent threat} can be performed by deep learning algorithms, such as dilated convolutional auto-encoders (DCAEs) algorithm~\cite{yu2017network}.

Anomaly detection represents a major method in the field across these different purposes, and it turns out to be promisingly explored also in the field of quantum machine learning. In the next sections we therefore anticipate the key concepts of quantum information and quantum algorithms, so to consistently review anomaly detection from the perspective of the solution of security-related tasks, as outlined above.  

\section{Quantum algorithms and quantum advantage}

    Quantum algorithms belong to computational classes defined by quantum Turing machines~\cite{deutsch1985quantum} instead of the conventional Turing machines.
    In the complexity theory, for both classical and quantum computation, the runtime of an algorithm is measured in terms of number of elementary operations $N$ involved~\cite{montanaro2016quantum}. In the circuit models for quantum computing, such operations match the application of native gates on the hardware for the gate-based architecture.
    Therefore, the same problem can be mapped as NP but not P for classical machines while it can be of class BQP if defined on the Hilbert spaces on which quantum Turing machines rely~\cite{jager2023universal}. Jager and Krems demonstrated that there exists a feature map and a quantum kernel that make variational quantum classifiers and quantum kernel support vector machines efficient solvers for any BQP problem. Therefore, some problems which are classically NP but BQP from a quantum approach, can be solved exponentially faster by using the appropriate quantum algorithm instead of a classical algorithm. Such property is called quantum speed-up. Different degrees of speed-up have been defined by a panel of experts in 2014~\cite{ronnow2014defining}, which can be summarized as follows:
    \begin{description}
        \item[Provable quantum speed-up:] there is a proof that there can be no classical algorithm that performs as well or better than the quantum algorithm. Example: Grover’s algorithm scales quadratically better than classical, provided there exist an oracle to mark the desired state;
        \item[Strong quantum speed-up:] the quantum algorithm performs better than
        the best possible, not necessarily known explicitly, classical algorithm (i.e. lower bound to classical algorithm is not known)
        Example:  Shor’s quantum algorithm to factorize prime numbers (grows polynomially instead of exponentially with the number of digits of the prime number);
        \item[Common quantum speed-up:] the quantum algorithm performs better than the \textit{best available} classical algorithm, as often the best available classical algorithm for strong quantum speed-up is not known;
        \item[Potential quantum speed-up:] if there is no consensus about which is the best available classical algorithm, it refers to the comparison with an arbitrary classical algorithm;
        \item[Limited quantum speed-up] it refers to the benchmark of two corresponding algorithms. Example: classical and quantum annealing.
    \end{description}
    The landscape of quantum algorithms shows a range of possible speed-ups, which is even more difficult to systematize for the domain of quantum machine learning methods, where the (possible) speed-up is sometimes not quantified. The evaluation increases in difficulty as more architectures and more encoding methods are possible (see Section~\ref{SEC:processing_info} and Figure~\ref{fig:taxonomy}).
    \begin{table}
    \begin{tabular}{ c|ccccc}
     \toprule
     \centering 
     \textbf{Methods} & \textbf{Speed-up} \; & \makecell{\textbf{Amplitude} \\ \textbf{Amplification}} & \textbf{HHL} & \textbf{Adiabatic} & \textbf{qRAM} \\
     \midrule
     Bayesian inference & $O(\sqrt N)$ & yes & yes & no & no \\
     \midrule
     Online perceptron & $O(\sqrt N)$  & yes & no & no & optional \\
     \midrule
      Least-squares fitting & $O(\log(N))$ & yes & yes & no & yes \\
     \midrule
     \makecell{Classical \\ Boltzmann machine} & $O(\sqrt N)$  & yes/no & optional/no & no/yes & optional \\
     \midrule
     \makecell{Quantum \\ Boltzmann machine} & $O(\log(N))$ & optional/no & no & no/yes & no \\ 
     \midrule
     Quantum PCA & $O(\log(N))$ & no & yes & no & optional \\
     \midrule
     \makecell{Quantum support \\ vector machine} & $O(\log(N))$ & no & yes & no & yes \\
     \midrule
     \makecell{Quantum \\ reinforcement learning} & $O(\sqrt N)$ &  yes & no & no & no \\
     \bottomrule
    \end{tabular}
    \caption{Speed-up quantification for given quantum machine learning subroutines. The table is taken from~\cite{biamonte2017quantum}. Some of the reported algorithms will be further discussed in detail in the next sections.}
    \label{tab:SpeedUp}
    \end{table}
    
    From the point of view of the implementation of quantum machine learning proposed in literature, three approaches are common. First, by the direct speed-up of machine learning techniques, by using algebra-related algorithms like those in the Table \ref{tab:SpeedUp}, of which the HHL algorithm is paramount~\cite{biamonte2017quantum}. Secondly, by implementing variational quantum circuits and finally multilayer perceptron-based quantum neural networks. Recently, the competitiveness of quantum models based on variational circuits compared to classical models has been raised by the demonstration~\cite{jerbi2023quantum} that explicit models~\cite{schuld2019quantum,havlivcek2019supervised} outperform implicit models, and data re-uploading models exponentially outperforms simple explicit models. Explicit models rely on a parametric definition of the unitary operators $\hat U(\theta)$, $\theta$ collecting the family of parameters. Such models therefore can be easily encoded into any variational quantum circuit. The QAOA algorithm is an example of explicit model.

\section{Encoding data with quantum systems}
    \subsection{From bits to qubits}
        
        The qubit is the fundamental unit of information encoded by a quantum computer. The qubit is a quantum state, defined by a vector $\ket{\psi}$ in a Hilbert space $\mathcal{H}=\mathbb{C}^2$. It is the transposition of the classical bit, but instead of assuming two discrete values, formally $\{0,1\}$, such values are transposed in a vector state~\cite{kopczyk2018quantum}:
        \begin{equation}
            0 \rightarrow \ket{0} = \begin{pmatrix} 1 \\ 0
            \end{pmatrix}, \quad 
            1 \rightarrow \ket{1} = \begin{pmatrix}
            0 \\ 1
            \end{pmatrix} \quad \Rightarrow \quad
            \ket{\psi} = \cos(\frac{\theta}{2}) \ket{0} + e^{i\phi} \sin(\frac{\theta}{2}) \ket{1}
        \end{equation}
        The $\{ \ket{0}, \ket{1} \}$ vectors form a orthonormal basis in the $\mathbb{C}^2$ space, therefore any qubit can be set in a linear combination as in the rightmost expression for $\ket{\psi}$~\cite[pag.94]{schuld2015introduction}. It follows immediately that $\bra{\psi}\ket{\psi}=1$, i.e. the state vector $\ket{\psi}$ is normalized. The $a$ and $b$ coefficients represent the probability for the state to be found either in the $\ket{0}$ or $\ket{1}$ state. All the logic operations on a single qubit are implemented by operators $\hat{U}$ which transform such state from $\mathbb{C}^2$ to $\mathbb{C}^2$: $\hat{U} \ket{\psi} = \ket{\psi'}$. To preserve the normalization of the vector $\ket{\psi}$, the single-qubit operators $\hat{U}$ are given by the unitary group $SU(2)$. The most known single-qubits operators are the NOT gate $\hat X$, the $\hat Z$ gate and the Hadamard gate $\hat H$:
        \begin{equation}
        \label{eq:XYHop}
            \hat X = \begin{bmatrix}
                0 & 1 \\ 1 & 0
            \end{bmatrix}, \quad \hat Z = \begin{bmatrix}
                1 & 0 \\ 0 & -1
            \end{bmatrix}, \quad \hat H = \frac{1}{\sqrt 2} \begin{bmatrix}
                1 & 1 \\ 1 & -1
            \end{bmatrix}
        \end{equation}
        The $\hat X$ gate flips the $\ket{0}$ state into the $\ket{1}$ and vice versa, acting in fact as the classical NOT gate. The Hadamard gate, instead, is an isomorphism on $\mathbb{C}^2$ mapping the computational basis $\{\ket{0}, \ket{1}\}$ into the conjugated one $\{\ket{+}, \ket{-}\}$ one, where $\ket{\pm} = (\ket{0} \pm \ket{1})/\sqrt 2$. For a comparison, see Table \ref{tab:QubitsvsQumodes}.
    
    \subsubsection{Encoding bits into qubits}
    \label{sec:memoryAdvantage}
    
    It is possible to encode bits into qubits in several ways. The most vanilla method consists of a 1-to-1 encoding, making one bit $i$ to be encoded by one quantum state $\ket{i}$. In literature, such encoding is referred to as multi-register encoding~\cite{agliardi2022optimal}, where $N$ is the number of available qubits, and $\ket{i}$ can assume the values of $\ket{0}$ or $\ket{1}$. When all the qubits are initialized in the $\ket{0}$ state, to flip a single qubit it suffices to apply a $\hat X$ NOT gate. Such operation can be performed simultaneously, yielding a circuit depth of $O(1)$ operations to be performed.
    However, an enhancement can be given by the superposition principle: in a register of $N$ qubits, it is possible to encode $n=2^N$ bits, each permutation being given by the superposition of the different states. In fact, the analog encoding makes usage of all the possible permutations of $0$ and $1$. In the following line, the multi-register and the analog encoding are respectively shown~\cite{duan2020survey}:
    \begin{equation}
    \bigotimes_{i=0}^{N-1} \ket{i} \, , \qquad \qquad \sum_{i=0}^{2^N-1} c_i \ket{i}
    \end{equation}
    Therefore, within the analog encoding (rightmost expression) via $N$ qubits it is possible to store $n=2^N$ units of memory, which means $N$ bits require $\log_2(N)$ available qubits. At this point, one should also notice that in the latter case the encoding process requires an exponential number $O(N)$ of steps~\cite{agliardi2022optimal}, calling for methods or approximations to circumvent the issue. In the past, quantum RAM (qRAM) have been proposed to feed directly the register of qubits asked to load the data~\cite{giovannetti2008quantum}, but hitherto no example does exists~\cite{duan2020survey,jaques2023qram,perez2020data}.
    
    One last family of data encoding is given by the Hamiltonian encoding. Within this frame, the data need to be formatted into a Hamiltonian operator, for instance the Ising Hamiltonian. Such a technique turns to be useful mainly in order to perform quantum annealing, see Section \ref{subsec:adiabatic}, and it will be explored in details in Section \ref{subsec:QUBO}.

    \subsection{Qudits}
    
    The qudit is a $d$-dimension generalization of the qubit. Instead of a basis spanned by $2$ vectors, the space of qudits is generated by $D$ vectors. It is possible to choose a computational basis~\cite{konar2021qutrit} $\{\ket{0}, \ket{1}, \dots, \ket{D-1} \}$, therefore a vector in this $\mathbb{C}^d$ space will given by
    \begin{equation}
    \label{eq:qudit}
        \ket{\psi} = \alpha_0 \ket{0} + \alpha_1 \ket{1} + \dots + \alpha_{D-1} \ket{D-1}
    \end{equation}
    The state in Equation~\ref{eq:qudit} must be normalized~\cite{srivastava2016modelling}, so that $\left|\alpha_0\right|^2+\left|\alpha_1\right|^2 +\dots+\left|\alpha_{D-1}\right|^2=1$. Referring to Section~\ref{sec:memoryAdvantage}, given $M$ qudits it is possible to store $D^M$ units of binary memory. A feature comparison between qubits and qudits is provided in Table \ref{tab:QubitsvsQumodes}.
    The qubits operators from Equation \eqref{eq:XYHop} can be adapted in the qudits formalism as well. The phase Hadamard gate should be able to put any $\ket{k}$ element from the computational basis $\{\ket{i} \}_{i=0}^{D-1}$ into a superposition over all the generators of such basis~\cite{wang2020qudits}, along with a generalization for the NOT and the $\hat Z$ operators~\cite{useche2022quantum,bravyi2022hybrid}:
    \begin{equation}
    \label{eq:HadamardQudit}
        \hat H_D = \frac{1}{\sqrt D} \sum_{i,j=0}^{D-1} \exp{i\theta_{i,j}^D} \ket{i} \bra{j}, \qquad
        \hat X^m = \sum_{k=0}^{D-1} \ket{k\oplus m} \bra{k}, \qquad 
        \hat Z = \sum_{k=0}^{D-1} e^{\frac{2\pi i}{k}} \ket{k}\bra{k}
    \end{equation}
    where the angle $\theta_{n,m}^D$ is defined as $\theta_{n,m}^D = \frac{2\pi}{D}nm$. The qudits in their conjugated basis, after a Hadamard transformation, are reported in Table \ref{tab:QubitsvsQumodes}. The $\oplus$ is the sum modulo $D$ (for the qubits, $D=2$). In~\cite{useche2022quantum}, it is introduced the control gates $\hat{CU}$ as a two-qudits operators. Here a target and a control qudits need to be specified: the operation $\hat{U}$ holds on the target qudit only if the control one is set in the $\ket{1}$ state, otherwise it does not. The generalized control gate operator is $\hat{CU}^k$: the $\hat U$ unitary operator is applied on the target qudit when the control one is set in the $\ket{k}$ state.

    \subsection{Qumodes (Continuous Variables)}
    
    In the frame of Continuous Variables (CV) for quantum computing, quantum information is encoded in continuous degrees of freedom such as the amplitudes of the electromagnetic field~\cite{killoran2019continuous}. Specifically, the unit of information we described before as a qubit is substituted by the so-called qumode. In the phase space representation, the state of a single qumode is described by two real and conjugated variables such as $(x,p) \in \mathbb{R}^2$, whether $N$ qumodes are depicted by $(\textbf{x}, \textbf{p}) \in \mathbb{R}^{2N}$. Qumode states also have a representation as vectors or density matrices in the countably infinite Hilbert space spanned by the Fock states $\{ \ket{n} \}_{n=0}^{+\infty}$, which are the eigenstates of the photon number operator $\hat{n} = (\hat{x}^2 + \hat{p}^2 -1 )/2$, where $\hat{x}$ and $\hat{p}$ are the position and momentum operators. A comparison~\cite{pfister2019continuous} between qubit, qudits and qumode architectures is provided in Table~\ref{tab:QubitsvsQumodes}.
    \begin{table}
    \centering
    \begin{tabular}{|c|c|c|}
     \toprule
     \multicolumn{3}{c}{\textbf{Units of information}} \\
     \midrule
     \centering Qubits & \centering Qudits & Qumodes \\
     \midrule
     \multicolumn{3}{c}{\textbf{Computational basis}}\\
     \midrule
     $\{ \ket{0}, \ket{1} \}$ & $\{ \ket{i} \}_{i=0}^{D-1}$ & $\{ \ket{q} \}_{q \in \mathbb{R}}$   \\
     \midrule
     \multicolumn{3}{c}{\textbf{Scalar product}}\\
     \midrule
     $\bra{k}\ket{l}=\delta_{k,l}$, $k,l \in \{0,1\}$ & $\bra{k}\ket{l}=\delta_{k,l}$, $k,l \in \{0,\dots,D-1\}$  & $\bra{q}\ket{q'} = \delta(q-q')$, $q, q' \in \mathbb{R}$  \\
     \midrule
     \multicolumn{3}{c}{\textbf{Superposition}}\\
     \midrule
     $\ket{\psi}=a\ket{0}+b\ket{1}$ & $\ket{\psi} = \sum_{i=0}^{D-1} \alpha_i \ket{i}$ & $\ket{\psi} = \int \dd q \, \psi(q) \ket{q}$ \\
     \midrule
     \multicolumn{3}{c}{\textbf{Conjugated basis}}\\
     \midrule
     $\ket{\pm}=\frac{1}{\sqrt{ 2}}\left( \ket{0} \pm \ket{1} \right)$ & $\ket{\theta_k} = \frac{1}{\sqrt D} \sum_{i=0}^{D-1} \theta_{i,k}^D \ket{i}$  & $\ket{p} = \int \dd q \, e^{i pq} \ket{q}$ \\
     \bottomrule
    \end{tabular}
    \caption{Comparing features from qubits, qudits and qumodes, see Figure~\ref{fig:taxonomy}.}
    \label{tab:QubitsvsQumodes}
    \end{table}

    \subsubsection{Operations in the frame of Continuous Variables}
    \label{sec:CVop}
    
    Killoran et al.~\cite{killoran2019continuous} report some possible operations to implement in the CV frame. First, the position and momentum operators are introduced:
    \begin{equation}
        \hat X = \int_{\mathbb{R}} x \ket{x} \bra{x} \dd x, \quad \hat P = \int_{\mathbb{R}}  p \ket{p} \bra{p} \dd p
    \end{equation}
    Being $\hat X$ and $\hat P$ defined on the entire real line, the orthonormality relations hold:
    \begin{equation}
        \bra{p}\ket{p'}=\delta(p-p'), \quad \bra{x}\ket{x'}=\delta(x-x')
    \end{equation}
    The so-called Gaussian operators implement the linear transformations. On a set of a single $(x,p)$ qumode, the rotation operator $\hat R$ acts between positions and momenta, while the displacement operator $\hat D$ performs the translations over the qumodes:
    \begin{equation}
        \hat R(\phi) : \begin{bmatrix}
        x \\ p
        \end{bmatrix} \to \begin{bmatrix}
        \cos(\phi) & \sin(\phi) \\
        -\sin(\phi) & \cos(\phi)
        \end{bmatrix}
        \begin{bmatrix}
        x \\ p
        \end{bmatrix}, \quad
        \hat D(\alpha) : \begin{bmatrix}
        x \\ p
        \end{bmatrix} \to
        \begin{bmatrix}
        x + \sqrt 2 \Re(\alpha) \\ p + \sqrt 2 \Im(\alpha)
        \end{bmatrix}
    \end{equation}
    Together $\hat R$ and $\hat D$ are able to implement affine transformations on a single qumode. Another Gaussian transformation is given by the beamsplitter $\hat{BS}$, which is a 2-qumodes operator:
    \begin{equation}
    \label{eq:beamsplitter}
        \hat{BS}(\theta) : \begin{bmatrix}
        x_1 \\ x_2 \\ p_1 \\ p_2
        \end{bmatrix} \to \begin{bmatrix}
        \cos(\phi) & \sin(\phi) & 0 & 0 \\
        -\sin(\phi) & \cos(\phi) & 0 & 0 \\
        0 & 0 & \cos(\phi) & \sin(\phi) \\
        0 & 0 & -\sin(\phi) & \cos(\phi)
        \end{bmatrix}
        \begin{bmatrix}
        x_1 \\ x_2 \\ p_1 \\ p_2
        \end{bmatrix}
    \end{equation}
    The last of the Gaussian operator is given by the squeezing one, $\hat S$:
    \begin{equation}
        \hat S(\tau) : \begin{bmatrix}
        x \\ p
        \end{bmatrix} \to \begin{bmatrix}
        e^{-\tau} & 0 \\
        0 & e^{\tau}
        \end{bmatrix}
        \begin{bmatrix}
        x \\ p
        \end{bmatrix} 
    \end{equation}
    Defining $\hat{\textbf{r}}=(\hat{\textbf{X}}, \hat{\textbf{P}})$, it is straightforward to state the uncertainty relation in the CV frame:
    \begin{equation}
        [\hat r_i; \hat r_j] = i\Omega_{ij}, \quad 
        \Omega = \begin{pmatrix}
        \mathbb{O} & \mathbb{I} \\
        \mathbb{I} & \mathbb{O}
        \end{pmatrix}
    \end{equation}
    
    \subsection{Embedding into QUBO problems}
    \label{subsec:QUBO}

        Adiabatic quantum computers can find the optimal solution to a specific class of optimization problems: Quadratic Unconstrained Binary Optimization (QUBO) problems. A QUBO problem is mathematically described as:
        \begin{equation}
        \label{EQ:QUBO}
            \hat H_P\equiv -\sum_{i=1}^Nh_i\hat \sigma_i^z-\sum_{i<j}J_{ij}\sigma_i^z\hat \sigma_j^z\;,
        \end{equation}
        where $\hat \sigma_i^z$ are the Pauli matrices that acting along the $z$-direction, and $J_{ij}$ and $h_i$ represent the parameters to the problem to be solved. We call $J_{ij}$ couplings or weights and $h_i$ biases. Since the eigenvalues of the Hamiltonian $H_P$ represent the possible solution to the problem, the goal is to set the couplings and the biases so that the ground state of $H_P$ represents the optimal solution to the optimization problem.
        
        The QUBO problems can be represented as graphs, where nodes are associated with biases and edges with couplings. Ideally, we want to map the optimization problem graph directly into the quantum annealer QPU. However, in a quantum annealing system, the hardware graph topology, which represents the pattern of physical connections for qubits and the couplers between them, is fixed. Since we cannot modify the qubit connectivity of a specific quantum annealer, we must map the model parameters into the hardware topology by a suitable embedding to solve an optimization problem. The basic idea of embedding is to identify groups of qubits (chains) so that they form the topology of the QUBO problem under investigation by behaving as individual units. The connectivity of each group can be enhanced by creating strong ferromagnetic couplings between the qubits, which forces coupled qubits to stay in the same state. 

\section{Processing quantum information with quantum computers}
\label{SEC:processing_info}

    Three available architectures are provided in the field of quantum computing, see Figure \ref{fig:taxonomy}. Two out of three, the MBQC and the gate models, reproduce on a quantum device the classical Von Neumann-Zuse paradigm, i.e. processing the inputs into outputs via a sequence of commands. Such architectures can be labeled as circuital model, as both of them aim to install a circuit of logical, controlled and sequential operations on the qubits. On the contrary, the adiabatic computation provides a scheme of computation which is unedited for any classical device.

    \subsection{Circuital model}
    
\begin{figure}[ht]
        \includegraphics[width=\textwidth]{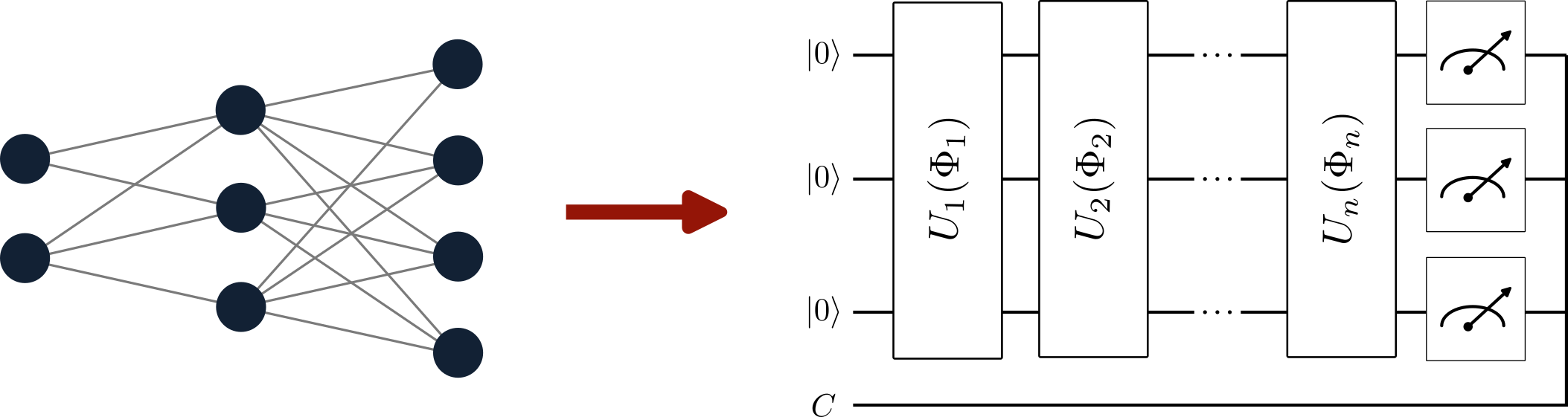}
        \caption{\label{fig:NNtoCirc}In classical deep learning, neural networks are employed for data processing. In quantum computing, circuits replace the typical neural architecture. The blocks of unitary operators $\hat U_i(\Phi_i)$ are the counterparts for the classical layers, the $\Phi_i$ variables are the hyperparameters, which have to be tuned in order to minimize a certain loss function.}
        \end{figure}

    In the gate model, the logic gates are given by certain physical operations on the qubits. Such gates, apart from being described by unitary matrices in an algebraic fashion, can just be taught as the classical logic gates (OR, AND, XOR and so on) transposed in the quantum frame. Such logic transformations can involve just one single qubit, or rather two as well as $n$, see Figure~\ref{fig:NNtoCirc}. Any logic gate can be built by composing a set of universal gates, generally given by two single-qubit and one two-qubit operators~\cite{barenco1995elementary}.
    
    In the measurement-based model instead, such logic gates are built relying on quantum phenomena such as entanglement and measurement\cite{corli2022efficient}. Nevertheless, the gate model can be mapped into the MBQC one~\cite{danos2007measurement, pius2010automatic}, proving that the two models are able to yield the same output.

    \subsection{Adiabatic quantum model}
    \label{subsec:adiabatic}
    
        Quantum annealers are quantum computers capable of finding the optimal solution to a QUBO problem by measuring the ground state of the QPU, i.e., the qubit configuration corresponding to the minimum energy of the system. The basic idea of quantum annealing is to prepare the qubits in the ground state, an easy-to-build configuration described by a Hamiltonian $H_T$, and then let the system evolve until it becomes equal to $H_P$, as in Equation~\ref{EQ:QUBO}. If the evolution is sufficiently slow, the adiabatic theorem~\cite{morita2008mathematical,morita2007convergence} guarantees that the systems stay in the ground state. Therefore, it is possible to find the solution to the optimization problem by simply measuring the system.
        
        Quantum annealers realize quantum annealing by introducing a time-dependant transverse field resulting in a total Hamiltonian:
        \begin{equation}
        \label{eq:adiabaticEvo}
            H(t)=-F(t) \underbrace{\left(\sum_{i<j}J_{ij}\sigma^z_i\sigma^z_j+\sum_ih_i\sigma^z_i\right)}_{H_P} -G(t) \underbrace{\sum_i\sigma^x_i}_{H_T}
        \end{equation}
        where $t$ represents the time, and the functions $F(t)$ and $G(t)$ control the annealing evolution and are referred to as the annealing schedule. At the beginning of the annealing, the system is prepared in the ground state of $H_T$ ($G(t) \gg F(t)$). At the end of the annealing, the system should be in the ground state of $H_P$ ($G(t) \ll F(t)$).

\section{Emulation of quantum computing resources by High-Performance Computing}

Quantum computing is a potentially disruptive computational paradigm that will enable efficient solutions to problems that are inherently difficult for classical digital devices. Although large-scale, error-corrected quantum computers are not yet available, hardware technology is evolving at a rapid pace, and demonstrations of quantum supremacy have already been achieved on current noisy intermediate-scale quantum (NISQ) devices for selected problems of academic interest \cite{arute2019quantum, zhong2021phase, wu2021strong,madsen2022quantum}. From the practical applications standpoint, however, NISQ devices still need to operate alongside classical digital hardware. Real-world applications are in fact characterized by complex computational workflows and large problem instances in which most of the computational burden is necessarily carried by traditional resources. For these reasons, NISQ computers are currently utilized as accelerators or co-processors embedded in hybrid quantum-classical computation.

The orchestration of hybrid hardware resources is currently implemented by means of a loose-integration paradigm, in which the classical and quantum processing is typically performed via local machines with consumer capacities and via remote quantum devices respectively. From an end-user perspective, this paradigm is considered the most effective as it allows the evaluation of alternative vendor solutions while limiting the risks associated with the experimentation of highly prototypical technologies which might suffer from rapid obsolescence. The main drawbacks of such an approach are instead related to the latency associated with the continuous data transfer, along with the additional concerns regarding the exchange with third-parties servers of sensitive or restricted data. 

To mitigate these issues, the scientific community is also starting to investigate on-premises scenarios with the co-location of quantum and digital classical hardware. This is commonly considered as a first step that, in the long term, should yield to a tight-integration paradigm in which quantum and classical processors are both co-located and interconnected via dedicated high-speed, high-capacity links~\cite{humble2021quantum}. The first experimentations in this direction are being conducted in various high-performance computing (HPC) centers, see at \href{https://eurohpc-ju.europa.eu/selection-six-sites-host-first-european-quantum-computers-2022-10-04_en}{euroHPCJU} and \href{https://www.hpcqs.eu/}{HPCQS}.

Beyond providing the means for an effective exploration of hybrid quantum-classical integration paradigms, 
HPC resources enable the emulation of quantum computers with up to the equivalent of 40 to 50 qubits, which is more than what most NISQ devices deliver today. Given that the current quantum hardware is still difficult and expensive to access, HPC emulators provide unique opportunities for conducting impactful R\&D that would not be possible otherwise. Typical activities enabled by HPC emulators are test/development of new algorithms for real-world applications, evaluation of the solution quality and time-to-solution behavior when scaling up the size of the problem, and the investigation of co-design of quantum algorithms and hardware.

A variety of emulators are currently available that can be used to implement a range of quantum algorithms, including those that are presented in this review. Most of them target a qubit architecture that implements the gate-based computational setting, either with an exact quantum state representation (state-vector, density matrix) or with approximate/compressed state representation (tensor networks). Such emulating libraries implement standard linear algebra operations that emulate the behavior of physical gates operated on the qubits register.  Some of the most known software development kits (SDKs) that provide emulator backends are \href{https://qiskit.org/}{Qiskit} (IBM), \href{https://quantumai.google/cirq}{Cirq} (Google), and Pennylane~\cite{bergholm2018pennylane} (Xanadu). 
The majority of such libraries, which have been initially written to run on local machines, are now capable to exploit distributed memory protocols and GPU acceleration, which enable the simulation of intermediate-size quantum states (30+ qubits) and the execution of deeper circuits with acceptable running times. 

In addition, as most of the hybrid quantum machine learning algorithms currently under investigation rely on variational circuits (VQE~\cite{peruzzo2014variational}, QAOA~\cite{farhi2014quantum}, Quantum NN), which are parameterized circuits trained by a classical computer through the optimization of a differentiable loss function, many of these SDKs have also been designed or integrated with state-of-the-art machine learning software packages such as PyTorch~\cite{paszke2017automatic}, TensorFlow~\cite{abadi2016tensorflow}, \href{https://www.paddlepaddle.org.cn/en}{Paddle-Paddle} to leverage automatic differentiation techniques such as backpropagation. These can take advantage of GPU acceleration to reduce the overall execution time but incur additional memory overhead due to the need to store partial derivatives of the forward pass. The utilization of premium, large-memory GPUs that are typically available in HPC centers can boost performances. 

Within the qubit architecture, a few emulators have been developed to deal with the measurement-based computational setting. Examples are Parceval (Quandela)~\cite{heurtel2023perceval} and \href{https://qml.baidu.com/}{Paddle-Quantum} (Baidu), the latter taking also advantage of backpropagation methods.
As for architectures based on qudits, \href{https://quantum-jet.readthedocs.io/en/latest/code/jet.html}{Jet} (Xanadu)~\cite{vincent2022jet} and \href{https://quantumai.google/cirq/build/qudits}{Cirq} (Google) libraries are available. Emulators based on continuous variable architectures are also available. 

\section{Generalization of neural networks in quantum circuits}

    
    The classes of Machine Learning and Deep Learning are often juxtaposed in literature and applications, but indeed the first includes the second as a special case based on neural networks. Deep learning exploits a deep hierarchy of layers of artificial rate neurons, resulting in a non-Von Neumann-Zuse architecture that is virtualized on standard digital CMOS hardware. A software tunes a set of hyperparameters of the NNs, called synaptic weights, to re-elaborate the inputs to outputs.
    
    A ML approach based on traditional and classical computing is straightforward to be translated on a quantum hardware, as it suffices to encode the inputs and the prompts into a quantum circuit. The main bottleneck is the constraint on the current size of the hardware, but from a theoretical perspective the problem can be treated as any classical-to-quantum algorithm. In such a way, it is quite standard to benchmark the performances between any classical algorithm and its quantum counterpart. A comparison for some of the most known classical machine learning methods and routines is given in Table~\ref{tab:SpeedUp}. Instead, neural network may require a paradigm shift towards new architectures. In the following two subsections, we present the two main approaches to neural networks, namely the variational and the quantum perceptron approaches, respectively.

    \subsection{Variational approach}
    \label{sec:VariationalApproach}
    
        As said, several algorithms rely on an Artificial Neural Network (ANN, or more simply NN). NNs allow to transform input data to outputs (labels, actions) encoding the inputs through various layers of artificial synapses. According to the Hornik's theorem~\cite{hornik1991approximation}, a sufficiently complex NN can always approximate the label output given an input.
        
        In quantum computing, many models have been proposed to replace the classical architecture of multiperceptron-based neural networks. The main feature consists of introducing a specific circuit model, able to process the input states of the system through a series of iterations. Of course, shallow circuits belong to this architecture, as well as one-layer classical neural networks.
        In order to process the information, instead of layers of neurons, quantum circuits display a block of unitary operations to be performed. The angles, implemented by unitary operators such as rotations, substitute the synaptic weights. In this frame, a quantum neural network (QNN) can be implemented via a variational quantum circuit (VQC)~\cite{kubler2021inductive, jerbi2024shadows}.
        However, a broad range of quantum neural networks models have been proposed~\cite{suzuki2022natural,huang2020quantum,chalumuri2021hybrid,li2019orthogonal,li2020quantum}: hybrid quantum circuit-classical neural networks approaches~\cite{srikumar2021clustering}, quantum neuron models~\cite{chen2020quantum} as an alternative for the classical activation functions and so on. A classical activation function $f$ is defined as
        \begin{equation}
        \label{eq:classicalActFunc}
            f(\textbf{x}) = \sigma\left( \sum_j W_{ij} x_j + b_i \right)
        \end{equation}
        where $W$ is called the weight function, and $\textbf{b}$ the bias vector. There are several classes of $\sigma$ activation functions, among them the perceptron, the sigmoid, the ReLu and others. The perceptron, in its classical description, given a set of $N$ inputs acts as a step activation function~\cite{maronese2021continuous}\cite[pag.49]{schuld2015introduction}, i.e. $\sigma$ is substituted by the Heaviside step function $\theta$, or rather by a sign function. In Section~\ref{sec:perceptron} a model for a quantum perceptron is presented.
        
        It must be noted that variational circuits are known for suffering the barren plateau effect~\cite{mcclean2018barren}. As random circuits are often proposed as initial guesses for exploring the space of quantum states during optimization of the parameters, one discovers that the exponential dimension of Hilbert space and the gradient estimation complexity make such choice unsuitable on more than a few qubits. In general the probability that the gradient along reasonable directions is non-zero to some fixed precision is exponentially small as a function of the number of qubits. Mitigations to this issue have been proposed thanks to smart inizialization~\cite{grant2019initialization}, also avoiding it thanks to quantum convolutional neural networks~\cite{pesah2021absence}.

    \subsection{Neurons into qubits}

        There are more options to encode neurons into a qubit. Nevertheless, given the definition of a qubit, few encoding options can be defined, bounded by the maximum encoding capability of a register of N qubits. The first option consists of the $1$-\textit{to}-$1$ encoding, where each and every input neuron of the network corresponds to one qubit
       ~\cite{cao2017quantum,hu2018towards,da2016weightless,matsui2009qubit,da2016quantum}.  The information is provided as a string of bits assigned to classical base states of the quantum state space.
        Similarly, a 1-{\it to}-1 method consists of storing a
        superposition of binary data as a series of bit strings in a multi-qubit state.  Such
        quantum neural networks refer to the concept of the quantum associative memory
       ~\cite{ventura2000quantum,da2017neural}. A different $1$-\textit{to}-$1$ option is
        given by the quron (quantum neuron)~\cite{schuld2014quest}.  A quron is a qubit whose $0$
        and $1$ states stand for the resting and active neural firing state,
        respectively\cite{schuld2014quest}.  
        
        Alternatively, a radically different encoding option consists of
        storing the information as coefficients of a superposition of quantum states
       ~\cite{shao2018quantum,tacchino2019artificial,kamruzzaman2019quantum,tacchino2020quantum,maronese2021continuous,molteni2023optimization}.  The
        encoding efficiency becomes exponential as a $n$-qubit state is an element of a
        $2^n$-dimensional vector space, but one has to remember that also operations required to store the state increase exponentially. From one hand,  loading a real image classification problem of few megabits in a quantum neural network makes the $1$-\textit{to}-$1$
        option currently not viable~\cite{pritt2017satellite}, while the choice $n$-\textit{to}-$2^n$ allows to encode a
        megabit image in a state by using $\sim20$ qubits only.  In the latter case one should anyway deal with the difficulty of preparing such a state, unless it is for instance generated by a circuit which approximates some aimed  distribution, or alternatively it comes directly from the physical conversion of flying qubits containing quantum data acquired by some quantum sensing system.


\section{Quantum supervised learning}

    Supervised Learning is the branch of Machine Learning which has been more transposed in a quantum formulation. Here we present the most significant quantum algorithms relevant for cybersecurity tasks, along with a description of their classical counterparts: activation functions for binary decisions~\cite{tacchino2019artificial,tacchino2020quantum,tacchino2020variational}, Support Vector Machine (SVM)~\cite{rebentrost2014quantum,delilbasic2021quantum,cervantes2020comprehensive,abedi2012support,zhang2004wavelet,ding2021quantum,li2015experimental} and kernel methods in general~\cite{schuld2019quantum}.
    One should notice that while current cybersecurity data are fundamentally classical in nature, in the future incoming quantum data from either quantum sensors or quantum communication networks may carry quantum (entangled) data, which in turn can be classified for instance by quantum tensor networks, as demonstrated by one of the authors~\cite{lazzarin2022multi}.
    \color{black}
    In the broader field of classical anomaly detection, a paramount role is played by image classification, e.g. to spot medical diseases~\cite{campbell2000linear, jeon2024anomaly,wei2018anomaly} or mechanical defects in industrial processes~\cite{fujimaki2005approach,liu2024deep, petsche1995neural}, therefore a specific Section of our analysis is dedicated to their quantum counterpart.
    \color{black}
    Classification by quantum tensor network on reduced MNIST with 4 categories has shown to return the same performances as best supervised learning algorithms, but more interestingly, it was able to discriminate quantum ground states carrying entanglement.
    \color{black}
    In the following Section, we introduce a range of techniques which vary from the implementation of natively quantum perceptron to the employment of adiabatic computation to improve performances of the Restricted Boltzmann Machine. Moreover, we show how to encode quantum neural networks on the continuous variables and classification tasks on qudits. All of the aforementioned techniques performs well when dealing with sampling from probabilistic distributions. Eventually, we show how to achieve quantum advantage on the Support Vector Machines via the HHL algorithm.
    \color{black}

    \subsection{Quantum feed-forward Neural Network for binary decisions}
    \label{sec:perceptron}
        
        Recently~\cite{tacchino2020quantum,tacchino2020variational}, a model of perceptron implemented by a quantum circuit has been proposed. This model features an input vector $\vec i = (i_0, \dots, i_{m-1})$ and a weight vector $\vec w=(w_0, \dots, w_{m-1})$, such that the activation response depends on their $\vec i \cdot \vec w$ scalar product. In such scenario, the components of the vectors $i_k, w_k=\pm 1 \, \forall k$. The vectors can be encoded into a quantum register by the following states:
        \begin{equation}
            \ket{\psi_i} = \frac{1}{\sqrt m} \sum_{j=0}^{m-1} i_j \ket{j}, \qquad \ket{\psi_w} = \frac{1}{\sqrt m} \sum_{j=0}^{m-1} w_j \ket{j}
        \end{equation}
        where the state $\ket{j}$ belongs to $\{\ket{0\dots00}, \ket{0\dots01}, \dots, \ket{1\dots11} \}$. Being $i_k$, $w_k$ all $\pm 1$, and $m$ being the dimension of the input and weight vectors, $\ket{\psi_i}$ and $\ket{\psi_w}$ are real equally-weighted (REW) superpositions of all the computational basis states $\ket{j}$. The states $\ket{j}$ live in a $\mathcal{H}^{\otimes N}$ Hilbert space, where $N=\log_2(m)$.\\
        The inner product between $\ket{\psi_i}$ and $\ket{\psi_w}$ returns $\vec i \cdot \vec w$ times $m$. To prepare the input state, the following transformation can be implemented as
        \begin{equation}
            \hat{U}_i \ket{0}^{\otimes N} = \ket{\psi_i}
        \end{equation}
        where $\hat{U}_i$ can be composed by any $m \times m$ matrix with $\vec i$ on the first column~\cite{tacchino2019artificial}. To perform the $\vec i \cdot \vec w$ inner product, it is possible to define a unitary operator $\hat{U}_w$ such that
        $
            \hat{U}_w \ket{\psi_w} = \ket{1}^{\otimes N} = \ket{m-1}
        $.
        To do so, choose any unitary $m \times m$ with $\vec w$ on the last row. The inner product between $\ket{\psi_i}$ and $\ket{\psi_w}$ can thus be performed by
        \begin{equation}
            \bra{\psi_w} \ket{\psi_i} = \bra{\psi_w} \hat{U}_w^\dagger \hat{U}_w \ket{\psi_i} = \bra{m-1} \ket{\psi_{i,w}} = c_{m-1}
        \end{equation}
        where $\ket{\psi_{i,w}}=1/\sqrt{m}\sum_{j=0}^{m-1}c_j\ket{j}$ is simply $\hat{U}_w\ket{\psi_i}$. Therefore, $c_{m-1}$ yields the scalar product $(\vec i \cdot \vec w)m$. The coefficient $c_{m-1}$ can be obtained, in a circuital computation, by entangling $\ket{\psi_{i,w}}$ with an ancilla $\ket{0}$, through a multi CNOT gate (i.e. a CNOT whose control qubits are given by the $\ket{j}$ state). The sole state on which such multi-CNOT operator $\hat{C}_{\ket{j},m}$ acts on is the $\ket{1}^{\otimes (m-1)}$ state, i.e. the last one ($\ket{m-1}$):
        \begin{equation}
            \hat{CX}_{\ket{m-1}, \ket{m}}\ket{\psi_{i,w}} \ket{0} = \frac{1}{\sqrt{m}} \left[\sum_{j=0}^{m-2}c_j\ket{j} \ket{0} + c_{m-1}\ket{m-1}\ket{1} \right]
        \end{equation}
        where $\ket{m-1}$ is the multi-qubits control state and $\ket{m}$ the target one. Measuring the ancilla qubit on the $\ket{1}$ basis, it is possible to activate the perceptron with probability $\left| c_{m-1} \right|^2$. Such achievement reproduces the perceptron in a quantum circuit. One should notice that such activation function ends the circuit with the measurement process, so the quantum information cannot travel further to other nodes. The issue has been addressed by one of the Authors~\cite{maronese2022quantum} by replacing the measurement process with a quantum circuit performing the Taylor series of the aimed activation function. Such method enables to program a multilayered perceptron.

    \subsection{Quantum restricted Boltzmann Machine}
    \label{sec:QRBM}
\begin{figure}
        \centering
        \includegraphics[width=0.4\textwidth]{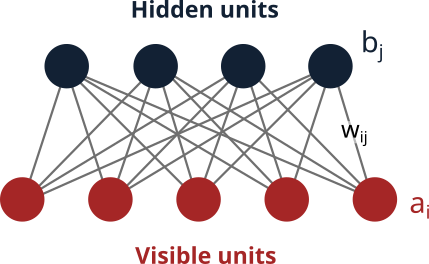}
        \caption{A graphical representation of a Restricted Boltzmann Machine. Each undirected edge represents a weighted dependency between two nodes, while each node is associated with a bias. The network has four hidden units (blue) and six visible units (red).}
        \label{FIG:RBM}
    \end{figure}

    Restricted Boltzmann Machines (RBMs) are neural network generative models first introduced by Hinton et al. in 1983 to improve upon the Hebbian learning method used in Hopfield networks. These models are designed to learn the underlying probability distributions of a dataset by using the Boltzmann distribution in their sampling function. A RBM consists of two layers: a layer of visible binary units (representing the input/output) and a layer of hidden binary units (which help the model mimic the dataset's structure). The units are connected by real-weighted connections, as illustrated in Figure~\ref{FIG:RBM}. RBMs do not allow connections between units within the same layer, resulting in a bipartite system.

    RBMs are flexible neural network models that can be used for various tasks, such as generating samples, making recommendations, or extracting features. They can also be used as classifiers by using different techniques, such as using them as feature extractors and appending a separate classifier or training them supervised with the label appended to the input data. These supervised RBMs are called discriminative restricted Boltzmann machines (DRBMs), which combine descriptive power with classification ability. The idea is to train the DRBM with a dataset where the label is appended to the input, then remove it from unseen inputs and reconstruct it using the RBM.

    The RBM is an energy-based model where every specific configuration of visible and hidden units is associated with an energy $E(\mathbf{v}, \mathbf{h}) = -\sum_i a_i v_i -\sum_j b_j h_j -\sum_{i,j} v_i W_{i j} h_j$, where $\mathbf{a}$, $\mathbf{b}$ are biases and $\mathbf{W}$ are the weights that represent the connection strength between units. Specifically, the joint probability of a configuration is given by the Boltzmann distribution
    \begin{equation}
        \label{EQ:jointProbRBM}
        P(\mathbf{v}, \mathbf{h}) = \frac{\exp(-E(\mathbf{v}, \mathbf{h}))}{\sum_{\mathbf{v}, \mathbf{h}} \exp(-E(\mathbf{v}, \mathbf{h})) }.
    \end{equation}

    The objective of training an RBM is to adjust the model's weights to increase the energy of states in the training dataset and lower the energy of all other configurations, allowing the model to learn how to generate and reconstruct the critical information encoded in the dataset. However, training an RBM can be challenging due to the large number of states that increases exponentially with the number of visible and hidden units, making it impractical to compute the partition function. Although an exact computation is not possible, several classical methods can be used to train the model, such as Contrastive Divergence~\cite{hinton2002training} (CD), Persistent Contrastive Divergence~\cite{tieleman2008training} (PCD), and Lean Contrastive Divergence~\cite{ning2018lcd} (LCD).

    Although these methods are effective in practice, RBMs can be more difficult and costly to train than other models that rely on backpropagation techniques, such as neural networks. Their training is often unstable and requires significant computational resources, and the approximations made during training can affect overall performance. Quantum computers provide an alternative approach to training RBMs, allowing for faster computation and better gradient estimates by querying the quantum processing unit. D-Wave quantum annealers, which are commonly used to sample the ground state of a QUBO problem, can also be used to train RBMs, resulting in faster computation and a better gradient estimate. These RBMs trained on a quantum annealer are called Quantum Restricted Boltzmann Machines (QRBMs).
    
    The basic idea to train a RBM on a quantum computer \cite{maronese2022quco,rocutto2021quantum,rocutto2021complete} is to extract a batch of samples from the quantum machine, which are dispersed according to the Boltzmann distribution associated to the RBM. If the computational cost of initializing the quantum computer is neglected, the quantum algorithm computational complexity to obtain a single sample scales as $O(1)$. The advantage of employing the D-Wave adiabatic quantum machine to exploit RBMs could emerge as an increase of performance metrics, such as the accuracy and the likelihood, or as a reduction in the computational complexity or computational times depending on the specific problem under consideration. The quantum RBM has been used to address anomaly detection of IP traffic data, performing 64x faster than classic hardware in the inference \cite{moro2023anomaly}.
    More general machines called Boltzmann machines, based on a complete (not bipartite) graph, have also been addressed on an adiabatic quantum computer \cite{rocutto2023fast}.

    \subsection{Neural networks in Continuous Variables}
    \label{sec:CVNN}

    An architecture to set up a Neural Network (NN) by continuous variables (CV) has been provided by Killoran et al.~\cite{killoran2019continuous}. It is shown that through the gates of CV encoding it is possible to reproduce the classical layer for a NN:
    \begin{equation}
    \label{eq:classicalNN}
        \mathcal{L}(\textbf{x}) = \varphi(W \textbf{x} + \textbf{b})
    \end{equation}
    where $\varphi$ is the activation function, $W$ is the weight matrix and $\textbf{b}$ is the bias vector. Such layer can be embedded in the CV formalism via the following sequence of operators/logic gates:
    \begin{equation}
    \label{eq:CVlayer}
        \hat{\mathcal{L}} = \hat{\Phi} \circ \hat{\mathcal{D}} \circ \hat{\mathcal{U}}_2 \circ \hat{\mathcal{S}} \circ \hat{\mathcal{U}}_1
    \end{equation}
    Here $\hat{\mathcal{D}}=\bigotimes_{i=1}^N \hat{D}_i(\alpha_i)$, $\hat{\mathcal{S}}=\bigotimes_{i=1}^N \hat{S}_i(\tau_i)$, where $\hat{D}$ and $\hat{S}$ are the Gaussian operators in Section~\ref{sec:CVop}, while the $\hat{\mathcal{U}}_i$ operators are given by a composition of beamsplitter $\hat{BS}(\theta)$. Instead, $\hat{\Phi}$ is a new non-Gaussian operation we are going to define in this Section.
    To perform Machine Learning tasks in CV, is therefore possible to implement a variational circuit built by a set of layers such as in Equation~\ref{eq:CVlayer}. In the following, it is shown how a quantum neural network has been built in ref.~\cite{killoran2019continuous}. The first three operations can be decomposed into a direct sum of two blocks:
    \begin{equation}
        \hat{\mathcal{U}}_2 \circ \hat{\mathcal{S}} \circ \hat{\mathcal{U}}_1 = 
        \begin{bmatrix}
        M_2 & \mathbb{O} \\
        \mathbb{O} & M_2
        \end{bmatrix}
        \begin{bmatrix}
        \Sigma & \mathbb{O} \\
        \mathbb{O} & \Sigma^{-1}
        \end{bmatrix}
        \begin{bmatrix}
        M_1 & \mathbb{O} \\
        \mathbb{O} & M_1
        \end{bmatrix} =
        \begin{bmatrix}
        M_2 \Sigma M_1 & \mathbb{O} \\
        \mathbb{O} & M_2 \Sigma^{-1} M_1
        \end{bmatrix}
    \end{equation}
    where the first block on the diagonal acts over the $\hat{\textbf{x}}$ variables, the second one over $\hat{\textbf{p}}$, in a similar fashion as the beamsplitter operator in Equation~\ref{eq:beamsplitter}. Afterwards, it is possible to apply the shifting by the displacement operator, so that the initial state $\ket{\textbf{x}}$, up to this point, is morphed into
    \begin{equation}
        \ket{\textbf{x}} \to \hat{\mathcal{L}}\ket{\textbf{x}} = \ket{M_2 \Sigma M_1 \textbf{x} + \sqrt{2} \boldsymbol{\alpha}_r}
    \end{equation}
    where $\boldsymbol\alpha_r=\Re(\boldsymbol\alpha)$. The next step is to implement a non-linear transformation, which is given by the non-Gaussian operations $\hat \Phi$. To build up such single-qumode gate, define a non-linear transformation $\phi(x)$, which can be written in a Taylor expansion of $\hat X$ for a certain degree of approximation, thereafter implement the operation in the form
    \begin{equation}
        \exp\left( -\frac{i}{2} \phi(\hat X_1) \otimes \hat P_2 \right) \ket{x}_1 \ket{0}_2 = 
        \exp\left( -\frac{i}{2} \phi(x) \hat P_2 \right) \ket{x}_1 \ket{0}_2
    \end{equation}
    Now the unitary operation is nothing but a translation over the second qubit:
    \begin{equation}
        \exp\left( -\frac{i}{2} \phi(x) \hat P_2 \right) \ket{x}_1 \ket{0}_2 = \ket{x}_1 \ket{\phi(x)}_2
    \end{equation}
    Eventually, a similar operation as from Equations~\ref{eq:classicalActFunc} and \ref{eq:classicalNN} has been implemented in the framework of Continuous Variables, as stated in Equation \eqref{eq:classicalNN}:
    \begin{equation}
    \label{eq:CVfulllayers}
        \hat{\mathcal{L}}(\textbf{r}) = \phi(M\textbf{r} + \sqrt 2 \boldsymbol\alpha)
    \end{equation}
    where $M=M_2\Sigma M_1\oplus M_2 \Sigma^{-1}M_1$.
    \begin{figure}[h]
    \centering
    \includegraphics[width=.8\textwidth]{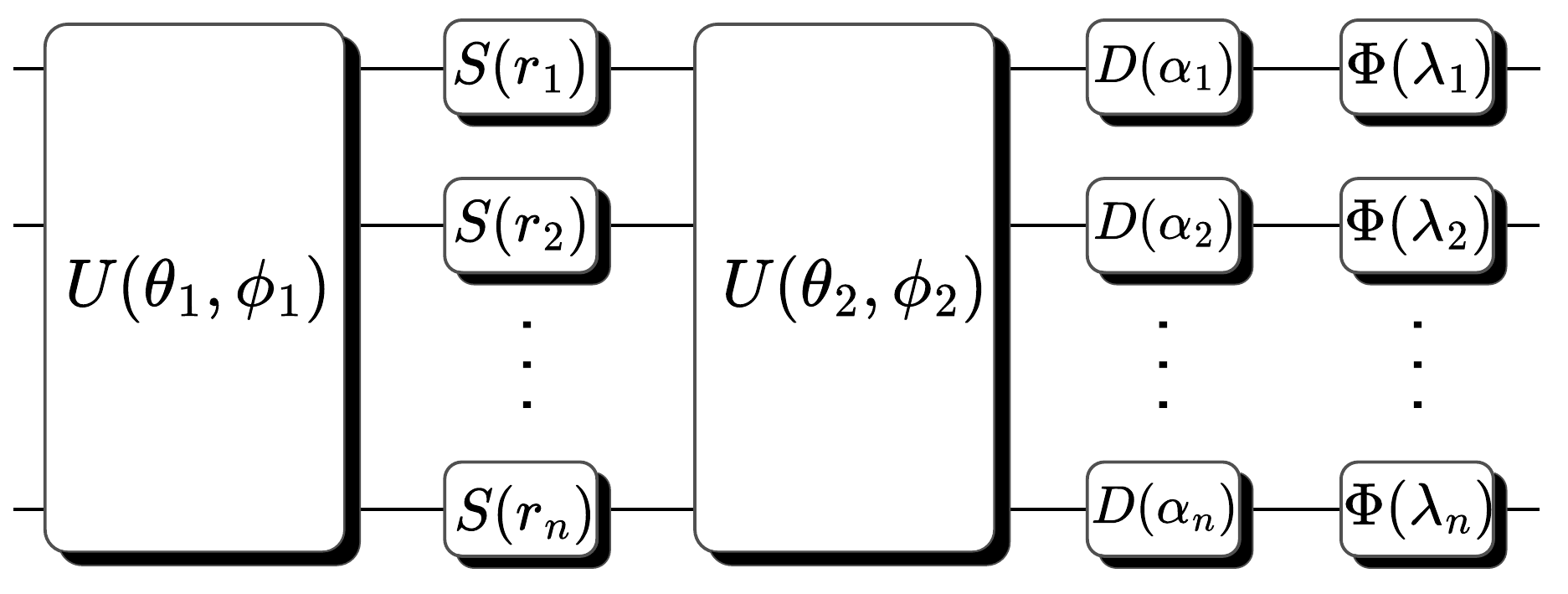}
    \caption{The picture represents how to build up a layer $\mathcal{L}$ in the frame of Continuous Variables, following the scheme from Equation~\ref{eq:CVlayer}.}
    \end{figure}

    \subsubsection{CV neural networks for fraud detection}
    
    From the formalism in Section~\ref{sec:CVNN}, it is possible to develop a hybrid algorithm, where it is possible to alternate classical Neural Network with quantum circuits for Supervised Learning tasks. As both the hyperparameters from the classical NNs and the variational circuit need to be tuned, a backpropagation can be performed by training the data on a classical device. In Ref.~\cite{killoran2019continuous}, a mean square error (MSE) was introduced as loss function:
    \begin{equation}
        L = \frac{1}{N} \sum_{i=1}^N [f(x_i) - \bra{\psi(x_i)} \hat X \ket{\psi(x_i)}]^2
    \end{equation}
    The model, when correctly trained, should return $\bra{\psi(x)} \hat X\ket{\psi(x)}=f(x) \; \forall x$. In~\cite{killoran2019continuous}, such model of hybrid Supervised Learning was employed to detect fraudulent transactions. The performance of the training outputted a ROC curve with area $0.945$, whereas the ideal curve should return a unitary area.

    \subsubsection{Potential advantages of CV neural networks}
    
    In the implementation of Killoran et al..~\cite{killoran2019continuous}, three potential advantages are proposed for employing the CV quantum neural networks. In the first place, CV neural networks can be performed on any photonic device, as the operation to implement them are universal in the photonic technology platform.
    
    Secondly, Hornik's theorem, as explained in Section~\ref{sec:VariationalApproach}, guarantees that any Lebesgue measurable function can be reproduced by a neural network. Quantum neural networks embed this property along with effects such as superposition and entanglement, which are intrinsic of the quantum realm. Furthermore, dealing with qumodes $(\textbf{x},\textbf{p})$ it is possible to rely on both the positions and momenta representations, $\textbf{x}$ and $\textbf{p}$ being the Fourier transform of each other.
    
    As the last point, quantum neural network in the CV framework can be employed for nonlinear transformations over distributions of probability. For instance, given a single-mode state $\ket{\psi}$, it is possible to encode its amplitude and transform it via a unitary transformation $\mathcal{W}$, due to the transformations acting inside the layers in Equation~\eqref{eq:CVfulllayers}:
    \begin{equation}
        \ket{\psi} = \int \psi(x) \dd x \Rightarrow \tilde \psi(x) = \int \mathcal{W}(x,x') \psi(x') \dd x'
    \end{equation}

    \subsection{Classification with qudits}
    \label{subsec:DMKDC}

        Given a finite set of vectors $x$ in $\mathbb R^m$, partitioned between $D$ classes, the Density Matrix Kernel Density Classification method (DMKDC) is an algorithm developed by Useche et al.~\cite{useche2022quantum} which aims to reproduce the probability functions $P_j$ for an element $x$ to belong to a certain class $j \in \{0,...,D-1 \}$.
        
        Given a system of qudits in a $\mathbb{C}^d$ space, suppose to have a number of classes $D\leq d$. Thereafter, a collection of training data $\{x_i \}_{i=1}^N \subseteq \mathcal{X}$ is provided, along with a feature map $\psi : x_i \to \ket{\psi(x_i)}$. Such a map can be set by a softmax encoding, see Ref.~\cite{gonzalez2021classification}, rather than via random Fourier features (RFF), see Ref.~\cite{gonzalez2021classification,sutherland2015error}. Both of these encodings would provide a normalized vector such that $\bra{\psi(x)} \ket{\psi(x)}=1$, $\ket{\psi(x)}$ being encoded in the fashion of a qudit vector, as in Equation \eqref{eq:qudit}, whose coefficients are computed by the chosen encoding methods. In second place, the density matrix $\hat \rho$, associated to such states, is constructed as a maximally mixed state over all the samples, see the below Equation. At the same time, it is possible to define a specific density matrix $\rho_j$ corresponding to each $j$-th class:
        \begin{equation}
            \hat \rho = \frac{1}{N} \sum_{i=1}^N \ket{\psi(x_i)}\bra{\psi(x_i)}, \qquad
            \hat \rho_j = \frac{1}{N_j} \sum_{i=1}^{N_j} \ket{\psi(x_i)}\bra{\psi(x_i)}
        \end{equation}
        $N$ being the cardinality of the entire $\{x_i\}_{i=1}^N$ dataset. The frequency $\pi_j=N_j/N$ accounts how many times a data $x_j$ belongs to the $j$-th class, $N_j$ counting the number of samples into the $j$-th class.
        The posterior probability for a generic $x$ sample to belong to the $j$-th class reads as~\cite{gonzalez2021classification}
        \begin{equation}
        \label{eq:probQudits}
            P_j(x) = \frac{\pi_j \bra{\psi(x)}\hat\rho_j\ket{\psi(x)}}{\sum_{k=0}^{D-1} \pi_k\bra{\psi(x)}\hat\rho_k \ket{\psi(x)}}
        \end{equation}
        The aim of the algorithm is to sample $\bra{\psi(x)}\hat\rho_k\ket{\psi(x)} \; \forall k$ in order to get the probability $P_j$.
        As the $\ket{\psi(x_k)}$ are known from the data, and $\hat \rho_j$ being a Hermitian operator, it is possible to diagonalize it via a $\hat U_j$ transformation:
        \begin{equation}
            \hat \rho_j = \hat U_j \hat \Lambda_j \hat U_j^\dagger = \hat U_j \left( \sum_{i=0}^{D-1} \lambda_{ji} \ket{i} \bra{i} \right) \hat U_j^\dagger
        \end{equation}
        The $\lambda_{ji}$ are the $i$-th eigenvalues for the $\hat \rho_j$ operators, $i$ and $j$ ranking from $0$ to $D-1$. For each operator $\hat \rho_j$, there exists a specific $\hat U_j$ unitary transformation capable to diagonalize the density matrix $\hat \rho$ into the computational basis $\ket{i}$, where $i=0,\dots,D-1$. The expectation $\bra{\psi(x)} \hat \rho_j \ket{\psi(x)}$ can therefore be written as
        \begin{equation}
        \label{eq:expDensity}
            \bra{\psi(x)} \hat \rho_j \ket{\psi(x)} = \bra{\psi(x)} \hat U_j \left( \sum_{i=0}^{D-1} \lambda_{ji} \ket{i} \bra{i} \right) \hat U_j^\dagger \ket{\psi(x)} = \sum_{i=0}^{D-1} \lambda_{ji} \left| \bra{i} \hat U_j^\dagger \ket{\psi(x)} \right|^2
        \end{equation}
        Afterwards, it is possible to introduce the $\hat{U}_{\lambda_j}$ operator:
        \begin{equation}
            \hat{U}_{\lambda_j} \ket{0} = \sum_{i=0}^{D-1} \sqrt{\lambda_{ij}} \ket{i}
        \end{equation}
        Before to show the qudit implementation of the DMKDC circuit, we sum up the pipeline for the training of the training process based on the density matrix estimation:
        \begin{enumerate}
            \item map the data $x_i \in \mathcal{X}$ into a qudit vector, thanks to a RFF or a softmax encoding;
            \item sample the $\pi_j$ frequencies, thus estimating the $\hat \rho_j$ probability densities;
            \item introduce the $\hat U_j$ operator able to diagonalize the $\hat \rho_j$ observables.
        \end{enumerate}
        It is worthy to notice that such training procedure does not involve iterative operations. The training samples are solely employed to prepare the $\hat \rho$ matrices, the time complexity of the algorithm scaling linearly on the size of training dataset. In fact, as remarked by Gonzalez~\cite{gonzalez2021classification}, the complexity of the algorithm is $O(ND^2 )$ for the estimation of the $\hat \rho_j$ elements, $N$ being again the cardinality of the dataset, and $O(D^3)$ for the diagonalization of the same probability densities.
        \begin{figure}[h]
        \centering
        \includegraphics[width=\textwidth]{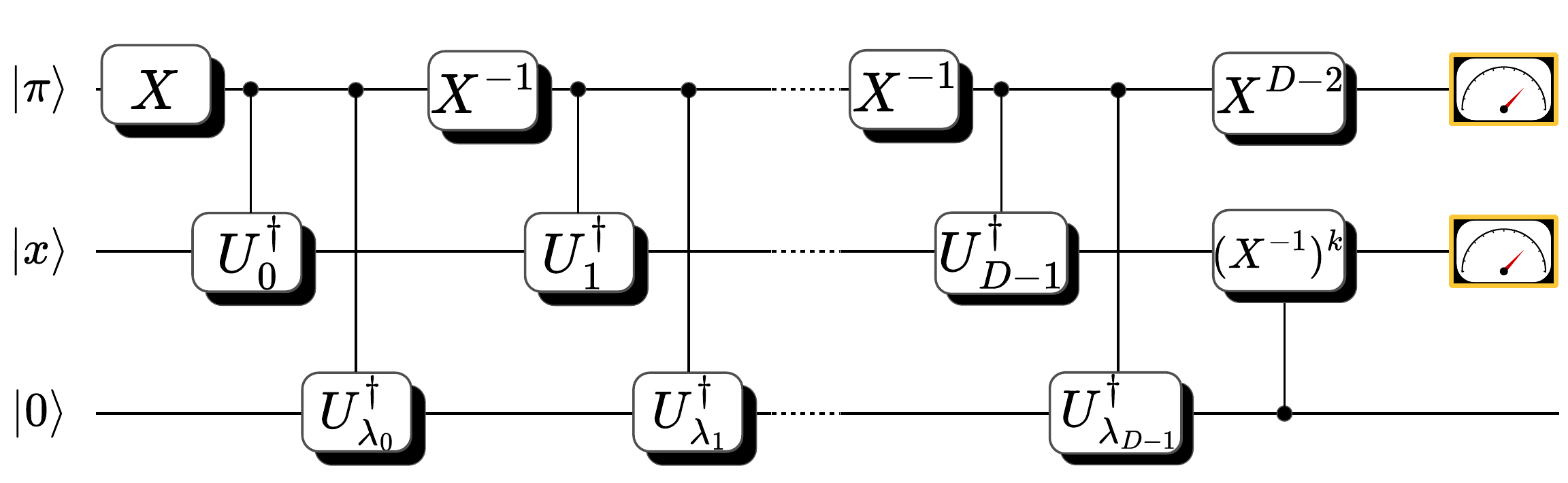}
        \caption{A scheme for the DMKDC circuit.}
        \end{figure}
    
    \subsubsection{The DMKDC circuit}
    
    To prepare the DMKDC register, in the first place a qudit with all the frequencies $\pi_j$ is prepared as follows:
    \begin{equation}
        \ket{\pi} = \sum_{i=0}^{D-1} \sqrt{\pi_i} \ket{i}
    \end{equation}
    In the second place, a qudit $\ket{\psi(x)}$ encodes the classical data to be classified, at last a $\ket{0}$ ancilla. The overall state, before to compute, results in
    \begin{equation}
        \ket{\pi} \otimes \ket{\psi(x)} \otimes \ket{0} = \left( \sum_{i=0}^{D-1} \sqrt{\pi_i} \ket{i} \right) \otimes \ket{\psi(x)} \otimes \ket{0}
    \end{equation}
    As a first step, apply a $\hat X$ gate on the $\ket{\pi}$ qudit, which consists of a sum modulo $D$ over the $\ket{i}$ generators, so that the new state reads
    \begin{equation}
        \left( \sum_{i=0}^{D-1} \sqrt{\pi_i} \ket{i\oplus1} \right) \otimes \ket{\psi(x)} \otimes \ket{0}
    \end{equation}
    where $\oplus$ is the sum modulo $D$. Afterwards, apply a $\hat{CU}_0^\dagger$ and a $\hat{CU}_{\lambda_0}$ gates, with the first qudit as control and the second and third respectively as targets:
    \begin{equation}
        \sqrt{\pi_0} \ket{1} \otimes \hat{U}_0^\dagger \ket{\psi(x)} \otimes \ket{\lambda_0} + \left( \sum_{i=1}^{D-1} \sqrt{\pi_i} \ket{i\oplus1} \right) \otimes \ket{\psi(x}) \otimes \ket{0}
    \end{equation}
    where the first controlled gate apply a $\hat U_0$ on the second qudit, while the second c-gate morphs the $\ket{0}$ qudit into $\ket{\lambda_0}$. Applying back a $\hat{X}^{-1}$ gate, the state turns to be
    \begin{equation}
        \sqrt{\pi_0} \ket{0} \otimes \hat{U}_0^\dagger \ket{\psi(x)} \otimes \ket{\lambda_0} + \left( \sum_{i=1}^{D-1} \sqrt{\pi_i} \ket{i} \right) \otimes \ket{\psi(x)} \otimes \ket{0}
    \end{equation}
    As a second step, from the above state apply the $\hat{CU}_1^\dagger$ and the $\hat{CU}_{\lambda_1}$ gates, which outputs
    \begin{equation}
        \sqrt{\pi_0} \ket{0} \otimes \hat{U}_0^\dagger \ket{\psi(x)} \otimes \ket{\lambda_0} + \sqrt{\pi_1} \ket{1} \otimes \hat{U}_1^\dagger \ket{\psi(x)} \otimes \ket{\lambda_1} + \left( \sum_{i=2}^{D-1} \sqrt{\pi_i} \ket{i} \right) \otimes \ket{\psi(x)} \otimes \ket{0}
    \end{equation}
    To iterate the process, apply the $\hat{X}^{m}$ gate on the first qudit, thereafter the $\hat{CU}_m^\dagger$ and $\hat{CU}_{\lambda_m}$ gates, for $m=2,\dots,D-1$. Eventually the circuit returns the following state:
    \begin{equation}
        \sum_{i=0}^{D-1} \left( \sqrt{\pi_i} \ket{i}  \otimes \hat{U}^\dagger_i \ket{\psi(x)} \otimes \ket{\lambda_i} \right)
    \end{equation}
    The second and the third qudits can be rewritten as
    \begin{equation}
        \hat{U}^\dagger_i \ket{\psi(x)} \otimes \ket{\lambda_i} = \sum_{l=0}^{D-1} \ket{l} \bra{l} \hat{U}_i^\dagger \ket{\psi(x)} \otimes \sum_{m=0}^{D-1} \sqrt{\lambda_{mi}} \ket{m} = 
        \sum_{l=0}^{D-1} a_{li} \ket{l} \otimes \sum_{m=0}^{D-1} \sqrt{\lambda_{mi}} \ket{m}
    \end{equation}
    where the $a_{li}$ coefficients are $\bra{l} \hat{U}_i^\dagger \ket{\psi(x)}$. It is possible to recombine the tensor product coupling the diagonal terms and the off-diagonal apart:
    \begin{equation}
        \sum_{l=0}^{D-1} a_{li} \sqrt{\lambda_{li}} \ket{ll} + \sum_{\substack{m,l=0 \\ m \neq l }}^{D-1} a_{li} \sqrt{\lambda_{mi}} \ket{lm}
    \end{equation}
    Applying the $\hat{C(X^{-1})}^k$ gate using the second qudit as target and the third as control, it leads to
    \begin{equation}
        \sum_{i=0}^{D-1} \sqrt{\pi_i} \ket{i} \otimes \left( \sum_{l=0}^{D-1} a_{li} \sqrt{\lambda_{li}} \ket{0l} + \sum_{\substack{m,l=0 \\ m \neq l }}^{D-1} a_{li} \sqrt{\lambda_{mi}} \ket{(l-m)m} \right)
    \end{equation}
    The probability $P_j$ in Equation~\ref{eq:probQudits} can be achieved by measuring the first qudit in the $j$-th element and the second one in $\ket{0}$:
    \begin{equation}
        P_{j0} = \pi_j \sum_{l=0}^{D-1} \left| a_{lj}\right|^2 \lambda_{lj} = \pi_j \bra{\psi(x)} \hat{\rho}_j \ket{\psi(x)}
    \end{equation}
    The second passage has made usage of Equation~\ref{eq:expDensity}. At the end of the process, in the phase of testing, it is possible to point out which class the data $x$ belongs to by maximizing the probability: 
    \begin{equation}
        \max_j [\pi_j \bra{\psi(x)} \hat{\rho}_j \ket{\psi(x)}]
    \end{equation}

    \subsection{Classical and Quantum Support Vector Machines}
    \label{subsec:SVM}

    In the field of quantum machine learning, support vector machines (SVM) have been deployed for instance to distinguish anomalies from normal activities. More specifically, such algorithms has been employed to spot fraudulent credit card transactions or spurious bank loan~\cite{wang2022integrating}, to address malware detection~\cite{barrue2023quantum} or rather to prevent cyber attacks, such as DDoS attacks~\cite{payares2021quantum}.
    Support Vector Machines are a classical supervised learning algorithm which aims to learn from the training samples $\textbf{x}_i$ in order to classify a new data sample into positive or negative class~\cite{kariya2021investigation}. The data samples are given in the form $\{ (\textbf{x}_1, y_1), (\textbf{x}_2,y_2), \dots, (\textbf{x}_N, y_N) \}$, with say two possible classes $A$ and $B$ and a relation to satisfy given by~\cite{boser1992training}
    \begin{equation}
    \label{eq:yclassification}
        \begin{cases}
        y_k = +1 \quad \text{if} \; \, x_k \in A\\
        y_k = -1 \quad \text{if} \; \, x_k \in B
        \end{cases}
    \end{equation}
    The space where the data $\textbf{x}_i$ are set is $\mathbb{R}^d$. In such a formulation, it is possible to define a new set of coordinates $\textbf{z}$ such that $\phi(\textbf{x})=\textbf{z}$. The $\phi$ are called feature maps, mapping the $\textbf{x}_i$ to the space of the $\textbf{z}$, i.e. $\phi: \mathbb{R}^d \to \mathbb{R}^M$, with $M \geq d$. 
    The purpose is to set a hyper-plane, given by the $\textbf{w} \cdot \textbf{z} + b = 0$ equation, where $\textbf{z}$ are the generic coordinates in a $\mathbb{R}^M$ space and $\{\textbf{w},b\}$ define the parameters for the hyperplane. The vector $\textbf{w}$ of parameters is defined as
    \begin{equation}
    \label{eq:wweights}
        \textbf{w} = \sum_{i=1}^d \alpha_i \phi( \textbf{x}_i )
    \end{equation}
    where $\textbf{x}_i$ are the data for the training, and $\alpha_i$ their corresponding weights.\\
    For the classification to succeed, at the end of the training $\textbf{w}$ and $\textbf{b}$ should be set such that $\textbf{w} \cdot \textbf{z} + b  \geq 1$ for a training point $\textbf{x}_i$ in the positive class, and $\textbf{w} \cdot \textbf{z} + b  \leq - 1$ for a training point $\textbf{x}_i$ in the negative class. Via the formulation in Equation~\ref{eq:wweights}, the hyperplane $D(\textbf{x})=0$ can be defined as~\cite{zhang2004wavelet}
    \begin{equation}
    \label{eq:hyperplaneEq}
        D(\textbf{x}) = \sum_{i=1}^d \alpha_i \phi(\textbf{x}_i) \cdot \textbf{z} + b = \sum_{i=1}^d \alpha_i \phi(\textbf{x}_i) \cdot \phi(\textbf{x}) + b = \sum_{i=1}^d \alpha_i k(\textbf{x}_i,\textbf{x}) + b
    \end{equation}
    where $K_{ij}=k(\textbf{x}_i,\textbf{x}_j)$ is called kernel function.

    \subsubsection{Kernel models}
    
    In Equation~\ref{eq:hyperplaneEq}, the kernel function is defined as the inner product between feature maps, but many other definitions may arise. Linear kernels are defined as~\cite{li2015experimental,suykens1999least,wu2021application}
    \begin{equation}
        k(\textbf{x}_i,\textbf{x}_j) = \phi(\textbf{x}_i) \cdot \phi(\textbf{x}_j)
    \end{equation}
    In such case, $d$ and $M$, the dimensions of the data and the feature space, are equal.
    Nevertheless, many models of different kernels may arise. Depending on the nature of the problem to be tackled, different kernels may induce different metrics for the classification tasks.
    The polynomial kernel is defined as~\cite{ding2021quantum,abedi2012support,wu2021application}
    $
        k_s(\textbf{x}_i,\textbf{x}_j) = (\lambda \phi(\textbf{x}_i) \cdot \phi(\textbf{x}_j) + r)^s
    $, 
    where $s$ is the polynomial degree and $\lambda$, $r$ are constants to be tuned. Another class is given by the gaussian kernel~\cite{ding2021quantum,suykens1999least,abedi2012support},
    $
        k_{g}(\textbf{x}_i,\textbf{x}_j) = \exp\left( -\frac{\| \phi(\textbf{x}_i) - \phi(\textbf{x}_j) \|^2}{2 \sigma^2} \right)
    $, 
    where $\sigma$ is again a constant to be tuned. The last kernel model frequently cited in literature is the the Radial Basis Function kernel (RBF)~\cite{wu2021application},
    $
        k_{RBF}(\textbf{x}_i, \textbf{x}_j) = \exp\left( - \gamma \| \phi(\textbf{x}_i) - \phi(\textbf{x}_j) \|^2 \right)
    $.
    Again, $\gamma$ is a parameter to tune for best fitting the real model.
    
    In any of these formulations, $K_{ij}$ still remains a symmetric matrix.
    
    \subsubsection{From hard to soft margins}
    
\begin{figure}
    \centering
    \subfloat[\label{subfig:SVM1}]{{\includegraphics[width=0.3\textwidth]{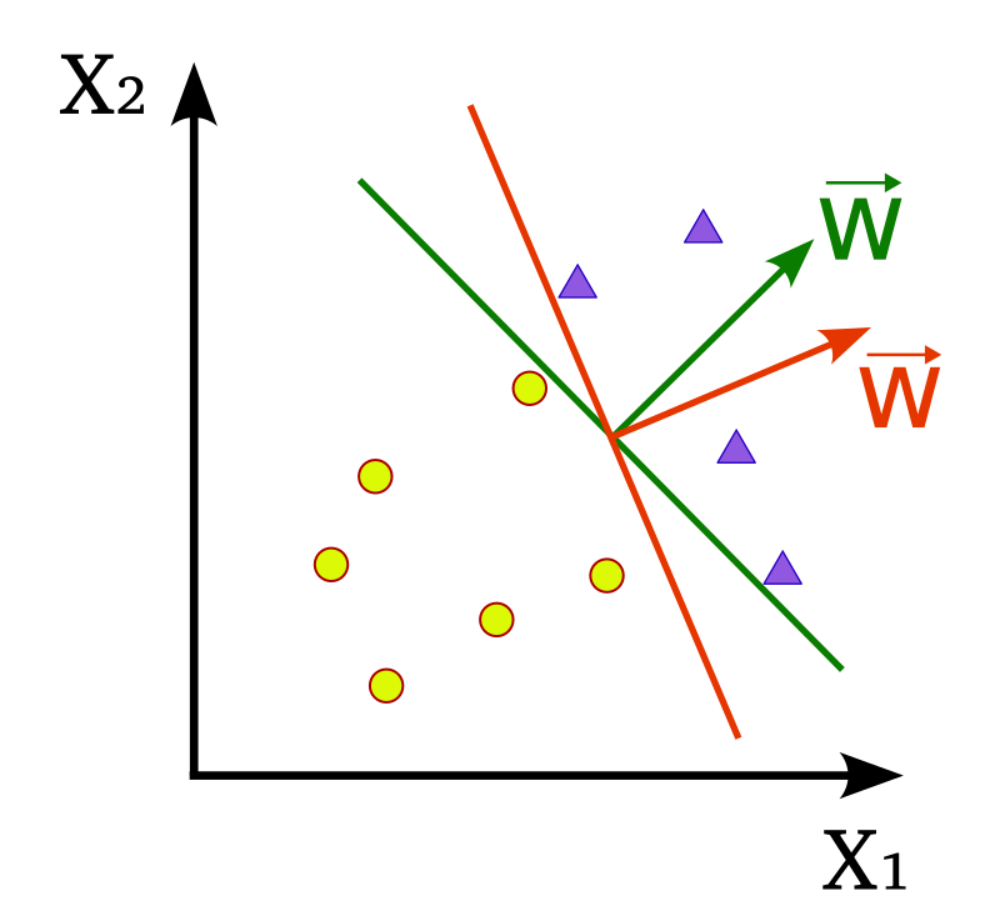} }}
    \quad
    \subfloat[\label{subfig:SVM2}]{{\includegraphics[width=0.3\textwidth]{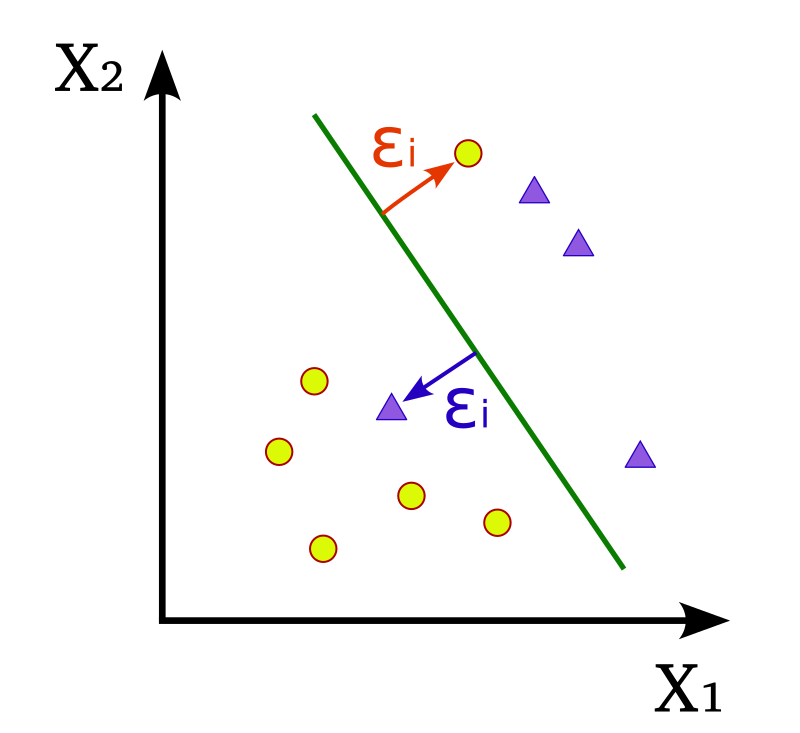} }}
    \quad 
    \subfloat[\label{subfig:SVM3}]{{\includegraphics[width=0.3\textwidth]{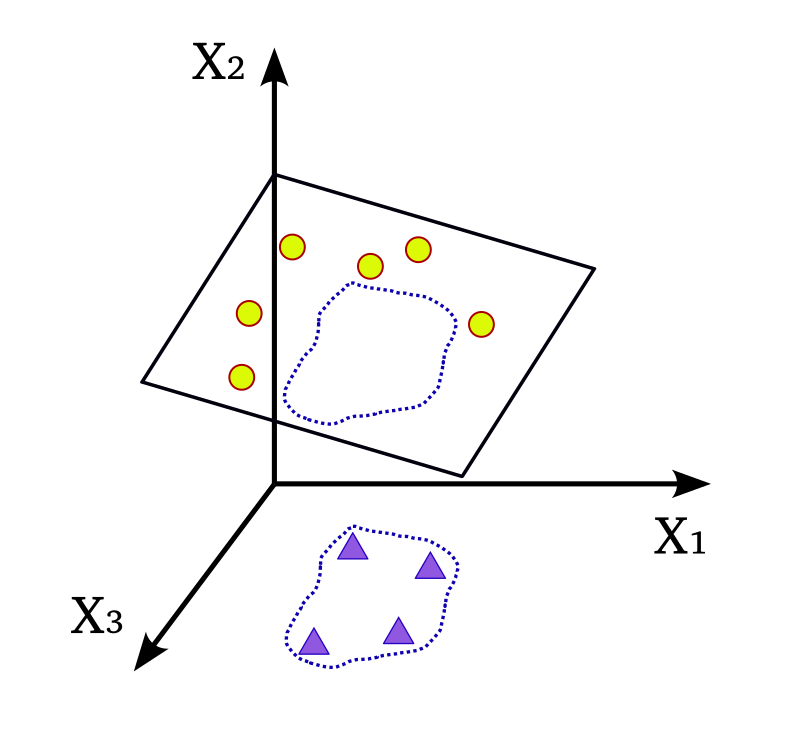} }}
    \caption{The first two images represent a $2D$ region, split by two axes $X_1$ and $X_2$ (a) Tuning $\textbf{w}$ changes the direction of the hyperplane, and therefore the distance between the data $\textbf{x}_i$ and the hyperplane itself. The $\textbf{w}$ vector is orthogonal to the hyperplane itself (b) Some data may follow into a disputed area, thereafter an error $\varepsilon_i$ (soft variables) is introduced to correct these margins (c) Introducing feature maps, a linear classification is made feasible, where otherwise, in a lower dimension space, it would not be possible.}
    \label{fig:hyperplanes}
    \end{figure}

    Regardless of the choice for the kernel, the final goal of the algorithm should be to reproduce the function in Equation~\ref{eq:yclassification} and ref.~\cite{rebentrost2014quantum}:
    \begin{equation}
        y(\textbf{x}) = \text{sgn}\left( D(\textbf{x}) \right) = \text{sgn}\left( \sum_{i=1}^m \alpha_i k(\textbf{x}_i,\textbf{x}) + b \right)
    \end{equation}
    and therefore, the hyperparameters which need to be trained are now $\{b, \vec \alpha \}$. A way to express the affiliation of a $\textbf{x}_i$ data to one of the two classes, for $\phi(\textbf{x})=\textbf{x}$ ($\phi$ thus being the identity map) is the following~\cite{zhang2020recent}:
    \begin{equation}
    \label{eq:constraintHyper}
        (\textbf{w}\cdot \textbf{x}_i + b ) y_i \geq 1
    \end{equation}
    With such a formulation, the closest data $\textbf{x}_i$ to the hyperplane yield an equation $\textbf{w}\cdot \textbf{x}_i + b = \pm 1$. In Appendix \ref{sec:Distance}, we prove $2/\|\textbf{w}\|$ to be the distance between such points, thus calling $\|\textbf{w}\|$ to be minimized for the classification to succeed at best.
    However, some data $\textbf{x}_i$ may fall into a so-called grey region, with distance $\varepsilon_i$ with respect to the corresponding hyperplane. The $\varepsilon_i$ can be thought as errors, or soft-variables (because the margins of the hyperplanes are now ``soft''). The condition in Equation~\ref{eq:constraintHyper} can be translated into an equality:
    \begin{equation}
    \label{eq:contstraintHyper2}
        (\textbf{w} \cdot \textbf{x}_i +b) y_i = 1 - \varepsilon_i \; \Rightarrow  \;
        (\textbf{w} \cdot \textbf{x}_i +b) = y_i (1 - \varepsilon_i)
    \end{equation}
    $y_i$ being $\pm 1$. We will refer to this condition in the next steps. Being $\textbf{w}$ the normal vector to the hyperplane with coordinates $\textbf{x}$, the distance between the two separation hyperplanes for the two classes is given by $\textbf{w}^T \cdot \textbf{w}$, i.e. by the norm of $\textbf{w}$. Choosing the direction of $\textbf{w}$, it is possible to minimize the distance between the two regions in the phase space which define the two classes, reducing therefore the probability to find an error $\varepsilon_i$. The purpose is now to minimize such distance, so that any object falling in the between of the two planes can be classified with no ambiguity. Such geometrical deduction can be pursued in Figure~\ref{subfig:SVM1}. Nevertheless, when some outliers inevitably occur, as in Figure~\ref{subfig:SVM2}, we are even interested in reducing the $\varepsilon_i$ total distance. Summing these conditions with the constraint in Equation~\ref{eq:contstraintHyper2}, the following system is provided:
    \begin{equation}
        \begin{cases}
        \min_\textbf{w} \langle \textbf{w}, \textbf{w} \rangle = \|\textbf{w}\|^2 \\
        \min_{\varepsilon_i} \frac{\gamma}{2} \sum_i \varepsilon^2_i \\
        (\textbf{w} \cdot \textbf{x}_i +b) = y_i (1 - \varepsilon_i)
        \end{cases}
    \end{equation}
    where $\gamma$ is the sensitivity to the total amount of errors $\varepsilon_i$. This optimization problem can be formulated via the Lagrangian multipliers $L(x) = f(x) - \lambda g(x)$, where the condition $f(x)$ is given by the inner product of $\textbf{w}$ and the sum over the $\varepsilon_i$, while the constraint $g(x)$ by the last equation of the system~\cite{cervantes2020comprehensive}:
    \begin{equation}
        L(\textbf{w}, b, \vec \alpha) = \frac{\|\textbf{w}\|^2}{2} + \frac{\gamma}{2} \sum_i \varepsilon_i^2 - \sum_i \alpha_i [\langle \textbf{w}, \textbf{x}_i \rangle + b - y_i (1-\varepsilon_i) ]
    \end{equation}
    The $\alpha_i$ coefficients play the role for the Lagrangian multipliers. To get the best parameters, we derive the Lagrangian $L(\textbf{w}, b, \vec \alpha)$ with respect to $\textbf{w}$, $b$, $\varepsilon_i$ and $\alpha_i$:
    \begin{equation}
    \label{eq:systemLagrangian}
        \begin{cases}
        \vec \nabla_\textbf{w} L = \textbf{w} - \sum_i \alpha_i \textbf{x}_i \overset{!}{=} 0 \Rightarrow \textbf{w} = \sum_i \alpha_i \textbf{x}_i \\
        \frac{\partial L}{\partial b} = - \sum_i \alpha_i \overset{!}{=} 0 \Rightarrow \sum_i \alpha_i = 0 \\
        \frac{\partial L}{\partial \varepsilon_j} = \gamma \varepsilon_j - \alpha_j y_j \overset{!}{=} 0 \Rightarrow \varepsilon_j = \frac{\alpha_j y_j}{\gamma} \\
        \frac{\partial L}{\partial \alpha_j} \overset{!}{=} 0 \Rightarrow \textbf{w} \cdot \textbf{x}_j +b = y_j - \varepsilon_j y_j
        \end{cases}
    \end{equation}
    Afterwards, substitute the first and the third equation in the fourth one, which yields
    \begin{equation}
        \sum_i \alpha_i \textbf{x}_i \cdot \textbf{x}_j +b = y_j - \frac{\alpha_j}{\gamma}
    \end{equation}
    as $y_i^2=1$. The same expression can be rewritten in terms of the kernel $K_{ij}$ and with all the parameters $\{b, \vec \alpha \}$ on the left member:
    \begin{equation}
    \label{eq:KernelSVMindexForm}
        \sum_i \alpha_i K_{ij} + \frac{\alpha_j}{\gamma} + b = y_j
    \end{equation}

    \subsubsection{Quantum advantage due to HHL algorithm}

    Taking into account the constraint $\sum_i \alpha_i=0$ from Equation~\ref{eq:systemLagrangian}, the same expression in Equation~\ref{eq:KernelSVMindexForm} can be reformulated into a matrix fashion as
    \begin{equation}
    \label{eq:SVMlinear}
        \begin{pmatrix}
        0 & \vec 1^T \\
        \vec 1 & K + \gamma^{-1} \hat{I}
        \end{pmatrix}
        \begin{pmatrix}
        b \\ \vec \alpha
        \end{pmatrix} = F
        \begin{pmatrix}
        b \\ \vec \alpha
        \end{pmatrix} = 
        \begin{pmatrix}
        0 \\ \vec y
        \end{pmatrix}
    \end{equation}
    The first row holds because of the third equation in the system from Equation~\ref{eq:systemLagrangian}. The dimension of the kernel matrix $K$ and the identity $\mathbb{I}$ which multiplies $\gamma^{-1}$ is $M$, i.e. the number of data from the training. Therefore, it is possible to obtain the parameters $\{b, \vec \alpha \}$ by just inverting the $F$ matrix:
    \begin{equation}
        \begin{pmatrix}
        b \\ \vec \alpha 
        \end{pmatrix} = F^{-1}
        \begin{pmatrix}
        0 \\ \vec y
        \end{pmatrix}
    \end{equation}
    The $F$ matrix can be expressed as $J+K_\gamma$, where
    \begin{equation}
    \label{eq:sparseMatrices}
        J = \begin{pmatrix}
        0 & \vec 1^T \\
        \vec 1 & \mathbb{O}
        \end{pmatrix}, \qquad
        K_\gamma = \begin{pmatrix}
        0 & \vec 0^T \\
        \vec 0 & K + \gamma^{-1} \mathbb{I}
        \end{pmatrix}
    \end{equation}
    Back to Equation~\ref{eq:SVMlinear}, it is possible to encode the training parameters $\{b, \vec \alpha \}$ as
    \begin{equation}
        \ket{b} \ket{\vec \alpha} = \frac{1}{\sqrt C} \left( b\ket{0} + \sum_{i=1}^M \alpha_k \ket{k} \right)
    \end{equation}
    where the normalization constant is set to be $C=b^2 + \sum_{i=1}^M \alpha_k^2$. As in the classical case, there are many available definitions of kernels, the first one of which can be
    \begin{equation}
        K = \sum_{i,j} \bra{\phi_i} \ket{\phi_j} \ket{i}\bra{j}
    \end{equation}
    Just for instance, in order to reproduce a polynomial kernel, it is possible to embed a state vector $\ket{x_j}$ in a higher dimensional space~\cite{rebentrost2014quantum},
    $
        \ket{\phi_j} = \ket{x_j}^{\otimes s}
    $,
    and therefore the inner product $\bra{\phi_j}\ket{\phi_i} = \bra{x_j}\ket{x_i}^s$. To invert the matrix in Equation~\ref{eq:SVMlinear}, it is possible to apply the HHL algorithm, achieving an exponential speed-up. An overall explanation for the HHL algorithm is provided in Appendix \ref{sec:HHL}. In the first place, the linear system in Equation~\ref{eq:SVMlinear} can be embedded into a quantum transformation as
        $\hat F \ket{b, \vec \alpha} = \ket{\vec y}$,
    where the matrix $F$ (and consequently the $\hat F$ operator) is defined as $(J + K + \gamma^{-1} \mathbb{I})/\Tr{J + K + \gamma^{-1}\mathbb{I}}$. The $F$ matrix has a $(M +1) \times (M +1)$ dimension, along with a norm $\|F\|=\max_{j} \| F \textbf{x}_j \|/\| \textbf{x}_j \| \leq 1$ ($\textbf{x}_j$ being its eigenvectors) because of the trace normalization. Thanks to the Lie-Trotter formula, it is possible to decompose the exponentiation of $\hat{F}$ as
    \begin{equation}
        e^{i \hat F \Delta t} = e^{-i \hat{\mathbb{I}} \gamma^{-1} \Delta t/C} e^{-i \hat J \Delta t/C} e^{i \hat K \Delta t/C} + O(\Delta t^2)
    \end{equation}
    where $C=\Tr{J + K + \gamma^{-1}\mathbb{I}}$ is the trace normalization. In order to apply the HHL algorithm, the state $1\ket{\vec y}$ must be endowed with an ancillary qubit, in order to store the eigenvalues of $e^{i \hat F \Delta t}$, and decomposed into a basis for $\hat F$:
    \begin{equation}
        \ket{\vec y} \to \sum_{j=1}^{M+1} \bra{u_j}\ket{\vec y} \ket{u_j} \ket{0}
    \end{equation}
    Thus, applying the HHL algorithm, the $\ket{\vec y}\ket{0}$ state transforms into
    \begin{equation}
        \sum_{j=1}^{M+1} \bra{u_j}\ket{\vec y} \ket{u_j}/\lambda_j
    \end{equation}

    \subsection{Natively quantum kernel methods}
    \label{Subsec:NativeKernel}
    
        The kernel method consists of embedding a set of data into a higher-dimensional space (even infinite-dimensional) called feature space~\cite{schuld2019quantum}. Given the space of data $\mathcal{X}$, the kernel $k$ is defined as a map
        $
            k : \mathcal{X} \times \mathcal{X} \to \mathbb{R}
        $.
        Such a map can be considered as a metric in the $\mathcal{X}$ space of the data. In the previous section, the kernel function $k$ has been alternatively defined through the feature maps $\Phi$, where
        $
            k(x,x') = \bra{\Phi(x)} \ket{\Phi(x')}
        $,
        where $x,x' \in \mathcal{X}$. The feature map $\Phi$ is a map between the space of data $\mathcal{X}$ and the feature space $\mathcal{H}$,
        $
            \Phi : \mathcal{X} \to \mathcal{H}
        $.
        A visual representation for this encoding can be seen in Figure~\ref{subfig:SVM3}. Such method turns to be useful in quantum machine learning, where classical data need to be encoded into a Hilbert space to perform some computation. It follows that the kernel function, encoding the data into the quantum circuit, induces a norm on the Hilbert space $\mathcal{H}$, and therefore a distance in the $\mathcal{X}$ space. Such feature accomplishes the purpose for a classification algorithm: given a target $y$, it is possible to compute the distance $k(x,y)$ thanks to the embedding in the feature space $\mathcal{H}$. By this approach, the data can be directly analyzed into a Hilbert space of features, where it is possible to deploy linear classifiers, relying on the inner products between quantum states. Increasing the size of the Hilbert space, such kernels turn to be classically intractable.
        
        A way to encode the information into qubits (i.e. in the Hilbert space) could be formulated by introducing an operator $\hat{U}_\Phi$, acting as $\hat{U}_\Phi(x) \ket{0} = \ket{x}$~\cite{wu2021application}. In such formulation, the $\hat{U}_\Phi$ operator acts as a creator over the vacuum state $\ket{0}$. Thus the kernel can be estimated by confronting the $\hat{U}$ operators:
        \begin{equation}
        \label{eq:kernelestimation}
            k(x,x') = \left| ^{N \otimes}\bra{0} \hat{U}_\Phi(x)^\dagger \hat{U}_\Phi(x)\ket{0}^{\otimes N} \right|^2
        \end{equation}
        By such formulation, the kernel entry can be evaluated on a quantum computer by measuring the $\hat{U}_\Phi(x)^\dagger \hat{U}_\Phi(x)\ket{0}^{\otimes N}$ state in the computational basis with repeated measurement shots and recording the probability of collapsing the output
        into the $\ket{0}^{\otimes N}$ state.

\section{Quantum unsupervised learning}

    \color{black}
    Recently, unsupervised learning gained wide success due to generative techniques, which allow to produce genuine new data mimicking the original dataset. Such techniques rely on sampling data from an unknown distribution, according to which the original ones are distributed. Quantum devices allow to generate samples from any distribution very efficiently, due to the intrinsic probabilistic nature of quantum mechanics. For instance, in~\cite{moro2023anomaly} it has been proved that a Boltzmann Machine on an adiabatic quantum computer performs 64 times faster than its classical counterpart, involving tasks and data concerning the field of cybersecurity. In the next Section, we detail how classical generative-adversary techniques are designed and suited for quantum computers and anomaly detection purposes.
    \color{black}

    \subsection{QGAN for anomaly detection}
    \label{subsec:QGAN}
    
        Differently from the approach based on SVM mentioned in the previous section, another approach has been proposed by Herr et al.~\cite{herr2021anomaly} relying on hybrid quantum GANs for the anomaly detection task. 
        
        GAN (Generative Adversary Network) is an unsupervised algorithm of machine learning. This algorithm is based on two agents, the discriminative and generative one. Given a distribution of data $\textbf{x} \in p_r(\textbf{x})$ and a set of labels to pair $p_r(\textbf{x})$ with, the former model tries to fit the best conditional probability $p(y|\textbf{x})$, the latter how to generate the joint probability $p(y,\textbf{x})$ for the data distribution. It is possible to introduce some latent variables, $\textbf{z}$, to mimic the distribution for the $\textbf{x}$ data. Given a distribution $p_g(\textbf{z})$ for the latent variables $\textbf{z}$~\cite{cao2020connecting}, the purpose for the generative model is to get the best parameters $\theta$ to build a function
        $
            G_\theta : \mathcal{Z} \to \mathcal{X}
        $.
        The data generated from $G_\theta$ will be distributed according to a probability distribution $\mathbb{P}_\theta$, while the true (unknown) distribution of data in the $\mathcal{X}$ space will be given by $\mathbb{P}_t$ ($t$ standing for true). The purpose of the discriminative model, now that the hidden variables $\textbf{z}$ are embedded in the $\mathcal{X}$ space, is to distinguish which variables are distributed according to $\mathbb{P}_t$ rather than $\mathbb{P}_\theta$:
        $
            D_\omega : \mathcal{X} \to [0;1]
        $.
        The problem can be reformulated as a \textit{minmax} one. In the WGAN (Wasserstrein GAN) formulation~\cite{herr2021anomaly}, given a set of data $\textbf{x}$ distributed accordingly to $\mathbb{P}_t$ (written as $\textbf{x} \sim \mathbb{P}_t$ ), and a set of $\textbf{z}$ with a $\mathbb{P}_z$ distribution, the purpose is to maximize the expectation function over $D_\omega$, neutralizing at the same time the ``fraudulent'' action of $G_\theta$:
        \begin{equation}
        \label{eq:minmaxW}
            \min_G \max_{D \in \mathcal{D}} \; \, \mathbb{E}_{\textbf{x} \sim \mathbb{P}_t}[D(\textbf{x})] - \mathbb{E}_{\textbf{z} \sim \mathbb{P}_z}[D(G(\textbf{z}))]
        \end{equation}
        where $\mathcal{D}$ is the class of all the $1$-Lipschitz functions. A $\alpha$-Lipschitz function $f$ is defined such that
        \begin{equation}
        \label{eq:Lipschitz}
            \| f(x) - f(x_0) \|_{\mathbb{C}} \leq \alpha \| x - x_0 \|_{\mathbb{D}}
        \end{equation}
        where $0<\alpha\leq1$ (for $\alpha=1$, $f$ is defined 1-Lipschitz), ${\mathbb{D}}$ and ${\mathbb{C}}$ are respectively the domain and codomain of the function $f$ and are metric spaces endowed with a distance and a norm (expressed by $\| . \|$ in Equation~\ref{eq:Lipschitz}). Thereafter, it is possible to state the proposition by Gulrajani et al.~\cite{gulrajani2017improved}:
        \begin{proposition}
            Let $\mathbb{P}_t$ and $\mathbb{P}_g$ two distributions in a $\mathcal{X}$ compact metric space. Then, there is a $1$-Lipschitz function $f^*$ which is the optimal solution of $\max_{\| f \|\leq L} \mathbb{E}_{y\sim \mathbb{P}_t}[f(y)] - \mathbb{E}_{x\sim \mathbb{P}_g}[f(x)]$.
        \end{proposition}
        As a corollary to the proposition, it is stated that $f^*$ has gradient norm $1$ almost everywhere under $\mathbb{P}_t$ and $\mathbb{P}_g$. Without entering the mathematical details and proof for such proposition, which we recommend to Gulrajani's paper~\cite{gulrajani2017improved}, it is possible to set the minmax problem in Equation~\ref{eq:minmaxW} with Lagrangian multipliers:
        \begin{equation}
        \label{eq:WGANloss}
            \mathcal{L}_c(\omega,\theta) = \mathbb{E}_{\textbf{z} \sim \mathbb{P}_z}[D_\omega(G_\theta(\textbf{z}))] - \mathbb{E}_{\textbf{x} \sim \mathbb{P}_t}[D_\omega(\textbf{x})] + \lambda \mathbb{E}_{\mathbb{P}_{\tilde{\textbf{x}}}(\tilde{\textbf{x}})}[(\| \nabla_{\tilde{\textbf{x}}} D_\omega(\tilde{\textbf{x}}) \| - 1)^2]
        \end{equation}
        where $\tilde{\textbf{x}}=\textbf{x}+\epsilon( G(\textbf{z}) - \textbf{x} )$, $\epsilon$ being uniformly distributed in the range $(0,1)$, i.e. $\epsilon \sim U(0,1)$, while $\textbf{x}\sim \mathbb{P}_t$ and $\textbf{z}\sim \mathbb{P}_g$ still. The multiplier term is called gradient penalty term to the critic loss function $\mathcal{L}_c(\omega, \theta)$.
    
    \subsubsection{Training the NNs as competitors}
    
    The training for a WGAN consists of two steps: in the first one, given a set of generated $G_\theta(\textbf{z})$ and true $\textbf{x}$ inputs, the purpose is to find the global minimum for the Lagrangian loss function defined in Equation~\ref{eq:WGANloss}, w.r.t. $\omega$ hyperparameters for the $D_\omega$ model via a stochastic gradient descent technique.
    
    In the second step, the purpose is instead to enhance the pursuit of the generator. The way to achieve such goal is to maximize the first Lagrangian term, w.r.t. both the $\theta$ and $\omega$ hyperparameters:
    \begin{equation}
    \label{eq:GenerativeLayers}
        \mathcal{L}_g(\theta) = -\mathbb{E}_{\textbf{z} \sim \mathbb{P}_z}[D_\omega(G_\theta(\textbf{z}))]
    \end{equation}
    In its classical fashion, the generative model $G_\theta$ is built up by a series of $L$ layers. In the first place, the latent variables $\textbf{z}\sim U(0,1)$ are reshaped through a series of maps $g_i: \mathbb{R}^N \to \mathbb{R}^N$, so that the overall NN model results in
    \begin{equation}
        g(\theta, \textbf{z}) = (g_{\theta_L} \circ g_{\theta_{L-1}} \dots \circ g_{\theta_1})(\textbf{z})
    \end{equation}
    where $\theta = (\theta_L, \theta_{L-1}, \dots, \theta_1)$. A set of activation functions can be applied for each $g_{\theta_i}$ layer: in~\cite{herr2021anomaly} leaky ReLU were deployed. Secondly, the final form of the generator will be given by the $W : \mathbb{R}^N \to \mathbb{R}^M$ function:
    \begin{equation}
        G_\theta(\textbf{z}, \phi) = W[g(\theta, \textbf{z}), \phi]    
    \end{equation}
    $W$ in Ref.~\cite{herr2021anomaly} is set to be a sigmoid function on $w_{ij} z_j + b_i$, with $\phi$ collecting the corresponding hyperparameters of the weight matrix $w$ and the bias vector $\textbf{b}$. To update the hyperparameters, it is possible to adopt any gradient descent method. The updating proceeds by the usual chain rule in the derivation process:
    \begin{equation}
    \label{eq:backpropagation}
        \frac{\partial \mathcal{L}_g}{\partial \theta_m} = - \mathbb{E}_{\mathbb{P}(\textbf{z})}\left[\frac{\partial D}{\partial G_\theta}\frac{\partial G_\theta}{\partial g_m}\frac{\partial g_m}{\partial \theta_m}\right] 
    \end{equation}
    The discriminative model $D_\omega : \mathbb{R}^M \to \mathbb{R}$ is a NN endoewd with several hidden layers, mapping the data in a $\mathbb{R}^M$ space to the label space in $\mathbb{R}$.
    
    \subsubsection{Quantum variational circuit for generative models}
    
    %
    Quantum circuits support the generative procedure of the model. In fact, quantum computers are expected to sample efficiently from distributions which are hard in a classical way~\cite{aaronson2011computational,bremner2016average,arute2019quantum,sweke2021quantum,agliardi2022optimal}. On the contrary, the critic model needs lots of classical data, which requires too much time to be loaded and makes such transposition unfeasible in the NISQ era~\cite{aaronson2015read}.
    
    When implementing the generative model on a quantum device, the $\textbf{z}$ latent variables are given into a uniform distribution $\textbf{z} \sim U(-\pi, \pi)$, whereas the encoded state $\ket{\textbf{z}}$ is given by the preparation layer $\hat S(\textbf{z})\ket{0}^{\otimes N}$. The operator $\hat S$, which implements such preparation layer, is composed as $\hat S(\textbf{z}) = \bigotimes_{i=1}^N R_i^X(z_i)$, $R^X_i$ being the X-rotation over the $z_i$ angle. After the state has been encoded, it follows a layer of rotations $\hat U_\nu(\vec \theta )$ in all the $\{ \hat X, \hat Y, \hat Z \}$ basis, alternated to CNOT gates. Therefore, two hyperparameters are given: $\nu$ stores the basis on which to perform rotations, $\vec \theta$ the angles. While $\nu$ encodes the architecture of the circuit, the $\vec \theta$ angles are the variational  parameters to optimize on. The multilayer generative function $g$ in Equation~\ref{eq:GenerativeLayers} is transposed in an expectation value over $\hat{\textbf{Z}}=\bigotimes_{i=1}^N\hat Z_i$:
    \begin{equation}
        g(\theta, \textbf{z}) = \bra{\textbf{z}} \hat U_\nu(\vec \theta) \hat{\textbf{Z}} \hat U_\nu(\vec \theta) \ket{\textbf{z}}    
    \end{equation}
    where instead of composing $\dots \circ g_{i+1} \circ g_{i} \circ \dots$ layers of activation functions, there is a sequence of $\hat{U}_\nu(\vec \theta)$ operators. At the end, a classical activation function $W$ is applied to compose the last layer for the generative model. Beside this difference, the gradient descent method is applied in the same manner as from Equation~\ref{eq:backpropagation}, but instead the derivative of $g_m$ is given by
    \begin{equation}
        \frac{\partial g}{\partial \theta_m} = \frac{\partial \langle \hat{\textbf{Z}} \rangle_{\hat{\textbf{z}}, \vec \theta, \nu}}{\partial \theta_m}
    \end{equation}

    \subsection{Other approaches to quantum anomaly detection by unsupervised learning methods}

        Generally speaking, quantum anomaly detection has been intensively explored in the past few years~\cite{liu2018quantum}. Anomaly detection for cybersecurity can take advantage of its development in other fields.
        For instance, a field of application of anomaly detection is particle physics. There, a number of algorithms have been proposed. 
        For instance, Alve et al.~\cite{alvi2022quantum} have applied QAD to an analysis characterized by a low statistics dataset. They have explored anomaly detection task in the four-lepton final state at the Large Hadron Collider that is limited by a small dataset, by a semi-supervised mode, without finding any evidence of speed-up. On the contrary, other examples sharing the unsupervised approach provided quantum speed-up, as follows. 

        \subsubsection{Quantum auto-encoders}
            Quantum auto-encoders have been assessed for unsupervised machine learning models based on artificial neural
            networks. The aim consists of learning background distributions by quantum auto-encoders based on
            variational quantum circuits, as problem of anomaly detection at the LHC collider. For representative signals, it turns out that a simple quantum auto-encoder outperforms classical auto-encoders~\cite{ngairangbam2022anomaly}. There, a quantum auto-encoder has been developed, consisting of a circuit divided into three blocks, namely the state preparation that encodes classical inputs into quantum states, the unitary evolution circuit that evolves the input states, and the  measurement and postprocessing part that measures the evolved state and processing the obtained observables.
        
        \subsubsection{Amplitude estimation-based method}
            An anomaly detection algorithm based on density estimation (ADDE) has been proposed by Liang et al.~\cite{liang2019quantum} to potentially express exponential speed-up, but it was later found not executing. Then, another group demonstrated such an exponential speed-up based on a modified version~\cite{guo2022quantum}. Such a new quantum ADDE algorithm is based on amplitude estimation. It is shown that such algorithm can achieve exponential speed-up on the number M of training data points compared with the classical counterpart. 
        
        
        \subsubsection{Natively quantum kernels and hardware benchmarking}
            In 2022, anomaly detection for credit card fraud detection have been demonstrated by quantum
            kernels on 20 qubits by authors including HSBC Bank affiliation~\cite{kyriienko2022unsupervised}. The benchmarks consist of kernel-based approaches, in particular unsupervised modeling 
            on one-class support vector machines (OC-SVM). Quantum
            kernels are applied to different type of anomaly detection, leading to observe that quantum fraud detection challenges 
            the equivalent classical protocols at increasing number of features, which are equal to the number of qubits
            for data embedding. The better precision has been achieved by combining quantum
            kernels with re-uploading, with the advantage increasing with the size of the
            system. The Authors claim that with 20 qubits the quantum-classical separation of average precision
            is equal to 15\%. The Authors estimate the computational cost to estimate the Gram matrix representing the kernel is $O(N_s^2)$ where $N_s$ is the number of samples, while the continuous retraining to update  on-the-fly the kernel is $O(N_s)$. Instead, the time needed for inference (to assign a label fraud or not) for detecting $N_d$ new-coming samples is of $O(N_{d} N_{s})$ kernel evaluations. The report is of particular interest as an evaluation is made for what concerns such inference time for three different hardware platforms: 1) superconducting circuits, 2) trapped ions and 3) optical systems.
            One can evaluate the training time for a dataset of 500 elements.
            In superconducting qubits, operation happen at MHz speed. A reproducible kernel measurements may require at least $N_{shots} = 10^{3-6}$ measurement shots. With $10^5$ kernel evaluations, the training time is 100 s - 28 h training time at optimistic MHz
            rate same cost of 28 hours. The inference time is significantly smaller down to 0.5 s in the case of reduced dataset.
            Instead, for large datasets (100000 samples), it may raise to 16 weeks, which could only be reduced with partial inference of Gram matrix. 
            In trapped ions, for 10 kHz per shots, with $10^5$ kernel evaluations the training time and $N_{shots} = 10^{3-6}$ is 3 h - 17 weeks. 
            With photons on deterministic gates, which is currently still an open field of research, the expected time ranges between 10 ms and 10 s.

\section{Quantum Approximate Optimization Algorithm}
    \label{sec:QAOA}

    Data clustering is the process of identifying natural groupings or clusters within multidimensional data based on some similarity measure. Clustering is a fundamental process in many different disciplines~\cite{omran2007overview}, for instance, it can be employed to divide the data set into a specified number of clusters, trying to minimize certain criteria (e.g. a square error function) falling into the class of optimization problems. Moreover, clustering algorithms are employed to perform network traffic identification~\cite{li2022robust} and for graph-based network security~\cite{lagraa2023review}.
    \color{black}
    In literature, in fact, a common approach consists of representing the servers as nodes of a graph, and the flow of data between them as the edges of the graph itself~\cite{lagraa2017botgm}. By monitoring the topology of the graph, relying on a technique called graph similarity, any anomaly can be straightforwardly detected. In 2022, Li et al. proposed an algorithm of clustering in order to monitor the traffic flow on the web~\cite{li2022robust}. Nevertheless, despite the effectiveness of such approach, the graph encoding for anomaly detections turns the problem to an NP-hard one by scaling with the number of nodes -- see~\cite{lagraa2017botgm, lagraa2023review}. Instead, quantum computation is able to tackle graph-based problems in a polynomial time. In the next Section, we provide a paramount example about how an NP-hard graph problem could be leveraged by a quantum machine learning approach.
    \color{black}

    \subsection{The MaxCut problem}
    
    The MaxCut algorithm is a NP-hard combinatorial problem~\cite{festa2002randomized,burer2002rank} which can be set as follows. Given a $G=(V,E)$ graph, $V=\{1, \dots ,n\}$ being the vertices of the graph and $E$ their connections, the weights of the $(i,j)\in E$ connections are given by the weight matrix $w_{ij}$. The purpose in the MaxCut problem is to find the best subset $S \subset V$ of vertices and its complement $\bar S$ to maximize the sum over the weights connecting the two subsets~\cite{festa2002randomized}:
    $
        w(S,\bar S) = \frac{1}{2} \sum_{i \in S, j \in \bar S} w_{ij}
    $.
    The MaxCut problem can be formulated as the following integer quadratic program~\cite{festa2002randomized,burer2002rank} and therefore mapped into a Hamiltonian formulation, inspired by the Ising model:
    \begin{equation}
    \label{eq:Hmaxcut}
        \max \;  \sum_{1\leq i < j \leq n} \frac{w_{ij}}{2} (1-z_i z_j) \to
        \hat H_C = \sum_{ ij } \frac{w_{ij}}{2} (\hat{I} - \hat{Z}_i \hat{Z}_j)
    \end{equation}
    The constraint $z_i=\pm 1$ holds, allowing to replace such classical variables with the third Pauli matrix from $\mathfrak{su}(2)$ algebra, i.e. the $\hat{Z}$ operator. Nonetheless, it holds that $z_i=\langle \hat Z_i\rangle$. The cut is defined by the condition $S=\{i\in V| z_i=+1\}$, and it can be set by maximizing the $E_C=\langle \hat H_C \rangle$ observable~\cite{zhou2020quantum}.
    When two vertices belong to the same subset $S$ or $\bar S$, it follows that $z_i=z_j$ and the contribution to $E_C$ is null. Thus, the set of $S$ and $\bar S$ can be thought as a partition of the system in up and down spins.
    
     The MaxCut problem can be employed in the field of data mining and machine learning~\cite{ding2001min}, with special regards to unsupervised learning~\cite{beaulieu2021max}: it is possible to recreate unsupervised learning clustering of data by mapping the problem to a graph optimization problem and finding the minimum energy for a MaxCut problem formulation.

    \subsection{QAOA formulation}
    
    The quantum approximate optimization algorithm (QAOA) has been many times applied to tackle the MaxCut problem~\cite{farhi2014quantum,proietti2022native,beaulieu2021max,corli2023max}. Such algorithm consists of preparing a register of $n$-qubits in the eigenstate of a Hamiltonian $\hat{H}_B$:
    \begin{equation}
        \ket{\Psi(t=0)} = \ket{\Plus{}}^{\otimes N}, \quad \hat{H}_B = \bigotimes_{i=1}^N \hat{X}_i  \Rightarrow \hat{H}_B \ket{\Psi(0)} = \ket{\Psi(0)}
    \end{equation}
    More specifically, the state $\ket{\Plus{}}^{\otimes N}$ is the maximum for the Hamiltonian $\hat H_B$. The purpose is now making $\ket{\Psi(t=0)}$ evolve to the maximum eigenstate for the $-\hat H_C$ Hamiltonian from Equation~\ref{eq:Hmaxcut} via the adiabatic theorem, i.e. to the minimum eigenstate for $\hat H_C$. The next step is thus to encode a time-dependent Hamiltonian $\hat H(t)$ to make the state $\ket{\Psi}$ evolve~\cite{van2001powerful}:
    \begin{equation}
    \label{eq:Hadiabatic}
        \hat H(t) = \left[ 1 - \frac{t}{T} \right] \hat H_B  -  \frac{t}{T} \hat H_C, \qquad 0 \leq t \leq T
    \end{equation}
    The annealing schedules are set by the $1 - t/T$ and $t/T$ terms, the $\hat H_P$ Hamiltonian is shaped on the form of the $\hat H_C$ Hamiltonian in Equation~\ref{eq:Hmaxcut}, while the transverse field Hamiltonians ($\hat H_T$ and $\hat H_B$) still keep the same form.
    Via the adiabatic theorem, it is possible to make $\ket{\Psi(t)}$ evolve from the higher energy state of $\hat H_B$ to the higher energy state of $-\hat H_C$ (and therefore to the ground state of $\hat H_C$). As the Hamiltonian in Equation \eqref{eq:Hadiabatic} depends on time, the corresponding time evolution is given by
    \begin{equation}
        \ket{\Psi(T)} = \exp\left(-\frac{i}{\hbar} \int_0^T \dd t \; \hat H(t)\right) \ket{\Psi(0)} = \exp\left(-\frac{iT}{\hbar} \int_0^1 \dd s \; \hat H(s)\right) \ket{\Psi(0)}
    \end{equation}
    where $s = t/T$. It is possible to approximate the above evolution by splitting the continuous trajectory along $t$ (or $s$) in a patchwork of $p$ small, discrete steps of duration $\varepsilon=1/p$~\cite{van2001powerful,an2022quantum}. By applying the Trotter formula~\cite{sun2018adiabatic}, the operator in the above equation can be approximated as 
    \begin{equation}
    \label{eq:middlepointApprox}
        \exp\left(-\frac{iT}{\hbar} \int_0^1 \dd s \; \hat H(s)\right) \approx
        \prod_{k=0}^{p-1} 
        \exp\left(-\frac{iT}{p\hbar} \hat H_k\right)
    \end{equation}
    where $\hat H_k = \hat H(k/p)$. In the limit for $p\to \infty$, it is possible to involve again the Lie-Trotter formula~\cite{farhi2014quantum,streif2019comparison} to split each $\hat H_k$ Hamiltonian into the $\hat H_C$ and $\hat H_B$ terms:
    \begin{equation}
    \label{eq:LieTrotter}
        \prod_{k=1}^p \exp\left(-\frac{iT}{p\hbar} \left[ 1 - \frac{k}{p} \right]\hat H_C \right) \exp\left(-\frac{iT}{p\hbar} \frac{k}{p} \hat H_B\right) \ket{\Psi(0)}
    \end{equation}
    The bigger is $p$, the better both the approximations in Equations~\ref{eq:middlepointApprox} and~\ref{eq:LieTrotter} work. It is possible to treat the terms in the round brackets as a set of angles $\gamma_j$, $\beta_j$, to map the overall evolution into
    \begin{equation}
        \ket{\Psi(\boldsymbol \gamma, \boldsymbol \beta) } = e^{-i \beta_p \hat{H}_B} e^{i \gamma_p \hat{H}_C} \dots e^{-i \beta_1 \hat{H}_B} e^{i \gamma_1 \hat{H}_C} \ket{+}^{\otimes n}
    \end{equation}
    In such formulation, the time evolution via a parameter $t$ has been substituted by $2p$ unitary transformations parameterized by a set of $(\boldsymbol\gamma, \boldsymbol\beta)$ angles. When implementing such operator on a circuit, the number of repetitions over the $e^{-i \beta_j \hat{H}_B} e^{i \gamma_j \hat{H}_C}$ layers stands for the depth $p$ of the circuit itself. The state $\ket{x}$ is therefore evolved naturally to the solution under the action of the following unitary operator:
    \begin{equation}
    \label{eq:evolutionQAOA}
        \ket{\Psi(\boldsymbol\gamma, \boldsymbol\beta)} := \left( \prod_{i=1}^p \hat{U}_B(\beta_i) \hat{U}_C(\gamma_i) \right) \ket{\Plus{}}^{\otimes n}
    \end{equation}
    Recall the cost function $E_C$ in terms of the number $p$ of layers which are inserted in the evolution from Equation~\ref{eq:evolutionQAOA}, e.g. $E_p$:
    $
        E_p(\boldsymbol \gamma, \boldsymbol \beta) = \bra{\Psi(\boldsymbol \gamma, \boldsymbol \beta)} \hat{H}_C \ket{\Psi(\boldsymbol \gamma, \boldsymbol \beta) }
    $.
    More layers are inserted (i.e. the higher is $p$), the more the solution is supposed to be exact. Afterwards, it is possible to define the maximum value over the expectation of $E_p$:
    $
        M_p = \max_{\boldsymbol \beta, \boldsymbol \gamma} E_p
    $.
    Therefore, by the adiabatic theorem, it is possible to state that
    $
        M_{p+1} \geq M_p
    $.
    Eventually, it is possible to map the adiabatic process into an optimization for the $2p$ parameters $\boldsymbol\gamma$, $\boldsymbol\beta$, which can be achieved by a hybrid algorithm combining gradient descent methods on CPUs/GPUs with backpropagation on the quantum circuits. A useful metric, to assess how far the state  is from the solution (i.e. the ground state of $\hat H_C$), is given by the approximation ratio parameter $r$, formulated as~\cite{zhou2020quantum, weggemans2022solving, choi2019tutorial, lotshaw2021empirical, lee2021parameters, wurtz2021fixed, pan2022automatic,wang2018quantum, sack2021quantum, wu2021application}
    $
        r=E/E_{max}
    $.
    Here $E=\bra{\Psi} \hat H_C \ket{\Psi}$, with $\ket{\Psi}$ being the actual state of the system and $E_{max}$ the maximum eigenvalue of the Hamiltonian operator, whose corresponding eigenstate is the goal of the problem under consideration. When $r$ tunes to $1$, the exact solution is provided. The approximation ratio, by such definition, returns the cost function (in our case, the Hamiltonian spectrum) normalized in the $[0;1]$ codomain.    
    
    Said this, some critical considerations have to be done. Once the QAOA was proposed for finding approximate solutions to combinatorial optimization problems~\cite{farhi2014quantum}, it was subsequently shown that QAOA solves the combinatorial problem Max E3LIN2 with better approximation ratio with respect to any polynomial-time classical algorithm known, at the time, but soon a better classical algorithm with better approximation ratio was found~\cite{barak2015beating}. The assignment of a quantum algorithm to a class of speed-up may suggest the priority around the aspects which can be investigated.


\section{Quantum reinforcement learning}
\label{sec:QRL}

Reinforcement learning (RL) has been poorly explored in the field of quantum information, and just in the last years some interest has been raising towards this branch of Machine Learning~\cite{cardenas2018multiqubit,mishra2021quantum,martin2022quantum}.
For instance, Chen et al.~\cite{chen2022variational} proposed a variational quantum reinforcement learning algorithm via evolutionary optimization with no evidence of quantum speed-up. Another variational implementation is due to Acuto et al.~\cite{acuto2022variational}. Dalla Pozza et al.~\cite{dalla2022quantum} developed a quantum RL framework to solve a quantum maze with speed-up,  and Cherrat et al.~\cite{cherrat2022quantum} show a quadratic speed-up under certain conditions for their quantum RL based on policy iteration. 
The field looks currently less developed with respect to quantum supervised and unsupervised paradigms and more development should be expected before prospecting an evident impact on security related tasks. Nevertheless, for sake of completeness, this Section outlines some key aspects of quantum RL, which may inspire future research around its intersection with cybersecurity.

According to one of the first proposals, when transposing classical algorithms of reinforcement learning in the quantum domain, the actions and the states of the system can be described as elements spanning two different Hilbert spaces $\mathcal{A}=\{\ket{a_i}\}$ and $\mathcal{S}=\{\ket{s_i}\}$, or even $\mathcal{E}$ (where $\mathcal{E}$ stands for environment)~\cite{dunjko2016quantum, dong2008quantum}. Apart from the qubits belonging to these two $\mathcal{A}$ and $\mathcal{S}$ systems, an auxiliary system $\mathcal{R}$, called register, can be added~\cite{albarran2018measurement,cardenas2018multiqubit, yu2019reconstruction}. In such case, when initializing the overall system, the state will be presented as
\begin{equation}
    \ket{\Psi} = \ket{\mathcal{A}} \ket{\mathcal{E}} \ket{\mathcal{R}}
\end{equation}
As from the classical RL algorithms, three functions are required: a policy function, a reward function (RF) and a value function (VF)~\cite{albarran2018measurement,albarran2020reinforcement}. The reward function is the criterion to evaluate the goodness of an action taken by the agent, with respect to the fixed task. The value function evaluates the general convergence of the algorithm to the goal it has to be achieved. The policy function defines which action to take with respect to the fixed purpose. However, due to the nature of quantum mechanics, even extracting information from the environment to the space of actions needs a decision problem, which task is relied to the policy function. The process of extracting information from the environment to the actions can be though as an interaction with the two systems.

The simplest case deals with one qubit for the environment $\mathcal{E}$ and one qubit for the action space $\mathcal{A}$. Depending on the RL protocol to implement, the register space can be endowed with one or two qubits. Generally, such states are initialized to $\ket{0}$, i.e.
$
    \ket{\Psi} = \ket{0}_\mathcal{A} \ket{0}_\mathcal{E} \ket{0}_\mathcal{R} \ket{0}_\mathcal{R}
$.
In the first step, the data need to be uploaded into the $\mathcal{E}$ space:
\begin{equation}
\label{eq:QRLencoding}
    \ket{\Psi} = \ket{0}_\mathcal{A} \biggr [\cos(\theta/2) \ket{0}_\mathcal{E} + e^{i \phi} \sin(\theta/2) \ket{1}_\mathcal{E} \biggr ] \ket{0}_\mathcal{R} \ket{0}_\mathcal{R}
\end{equation}
In the second place, apply a set of CNOT gates with $\ket{a}_\mathcal{E}$ as control and the $\ket{0}_\mathcal{R}$ as targets:
\begin{equation}
    \ket{\Psi} = \ket{0}_\mathcal{A} \biggr [\cos(\theta/2) \ket{0}_\mathcal{E} \ket{0}_\mathcal{R} \ket{0}_\mathcal{R} + e^{i \phi} \sin(\theta/2) \ket{1}_\mathcal{E} \ket{1}_\mathcal{R} \ket{1}_\mathcal{R} \biggr ] 
\end{equation}

\subsection{Quantum adaptation algorithm}

The quantum adaptation algorithm, proposed by F. Albarrán-Arriagada et al.~\cite{albarran2018measurement} and applied by Shang Yu et al. (including Albarrán-Arriagada himself) in a semiquantum way~\cite{yu2019reconstruction} has been tested to rebuild a quantum state, in order to describe a quantum system. It consists of the following steps: start from a system where all of the $\mathcal{A}$, $\mathcal{E}$ and $\mathcal{R}$ subsystems take into account a single qubit. In the first place, encode the overall system in a similar fashion as in Equation~\ref{eq:QRLencoding}:
\begin{equation}
    \ket{\Psi} = \ket{0}_\mathcal{A} \biggr [\cos(\theta/2) \ket{0}_\mathcal{E} + e^{i \phi} \sin(\theta/2) \ket{1}_\mathcal{E} \biggr ] \ket{0}_\mathcal{R}
\end{equation}
Therefore, apply a $\hat{CX}_{\mathcal{E}\mathcal{R}}$ operator (CNOT on $\mathcal{E}$ as control, $\mathcal{R}$ as target):
\begin{equation}
    \ket{\Psi} = \ket{0}_\mathcal{A} \biggr [\cos(\theta/2) \ket{0}_\mathcal{E} \ket{0}_\mathcal{R} + e^{i \phi} \sin(\theta/2) \ket{1}_\mathcal{E} \ket{1}_\mathcal{R} \biggr ] 
\end{equation}
Afterwards, perform a measurement over $\mathcal{R}$ in the computational basis $\{\ket{0}, \ket{1}\}$, so that the $\mathcal{E}$ state is going to collapse in $\ket{0}_{\mathcal{E}}$ or $\ket{1}_{\mathcal{E}}$ with probabilities $\cos^2(\theta/2)$ or $\sin^2(\theta/2)$, respectively. If the superposition collapses to $\ket{0}$, the environment $\mathcal{E}$ and the action $\mathcal{A}$ share the same state, otherwise the latter needs to be updated. To update $\ket{0}_\mathcal{A}$, introduce the following operator:
\begin{equation}
    \hat{U}^{(k)}_\mathcal{A}(\alpha^{(k)}, \beta^{(k)}) = e^{-i \hat{Z}_\mathcal{A} \alpha^{(k)}} e^{-i \hat{X}_\mathcal{A} \beta^{(k)}}
\end{equation}
Here $\hat{Z}$ and $\hat{X}$ stand for the elements from Pauli algebra $\mathfrak{su}(2)$, and the suffix $\mathcal{A}$ shows that they are acting over the action space. The index $k$ stands for the iteration over the process. The angles of rotation are defined in the following range:
$
    \alpha^{(k)}, \beta^{(k)} \in [ -\Delta^{(k)}/2; \Delta^{(k)}/2 ]
$,
where $\Delta^{(k)}$ is the parameter to update per iteration $k$. The operator $\hat{U}_\mathcal{A}$ acts on the $\ket{\mathcal{A}}$ state depending on the outcome from the measurement:
\begin{equation}
    \hat{\mathcal{U}}^{(k)}_\mathcal{A} = [ m_k \hat{U}^{(k)}_\mathcal{A}(\alpha^{(k)}, \beta^{(k)}) + (1-m_k) \hat{I}_\mathcal{A} ]
\end{equation}
where $m_k$ is the outcome from the $k$-th measurement, $1$ if $\ket{1}$, otherwise $0$. To update the $\Delta$ parameter, the following rule has been proposed:
\begin{equation}
    \Delta^{(k+1)} = [ (1-m_k) \mathcal{R} + m_k \mathcal{P} ] \Delta^{(k)}
\end{equation}
$\mathcal{R}$ and $\mathcal{P}$ are called the reward and punishment ratios, respectively $\mathcal{R}=\epsilon<1$ and $\mathcal{P}=1/\epsilon>1$, so that every time the outcome is $\ket{0}$ the value of $\Delta$ is reduced, when $\ket{1}$ it is increased. $\epsilon$ is a hyperparameter to tune for every set of simulations, the better the $\epsilon$ parameter, the higher the fidelity between the simulated qubit and the initial state.

At the $k$-th iteration, the system will be set in the state
$
    \ket{\Psi} = \mathbb{U}^{(k)}_\mathcal{A} \ket{0}_\mathcal{A} \ket{\psi}_\mathcal{E} \ket{0}_\mathcal{R}
$,
where the $\mathbb{U}^{(k)}_\mathcal{A}$ operator accounts into memory all the previous actions over $\mathcal{A}$:
\begin{equation}
    \mathbb{U}^{(k)}_\mathcal{A} = \mathcal{U}^{(k)}_\mathcal{A}(\alpha^{(k)}, \beta^{(k)}) \mathbb{U}^{(k-1)}_\mathcal{A}
\end{equation}

\begin{table}
  \caption{Taxonomy for Quantum feed-forward Neural Networks for all the reviewed algorithms. AF acronym stands for activation functions. All the fields in the table refer to Figure~\ref{fig:taxonomy}.}
  \label{tab:summary}
  \begin{tabular}{c|ccccc}
    \toprule
    \textbf{Algorithm} & \textbf{Units} & \textbf{Learning} & \textbf{Architecture} & \textbf{NNs} & \textbf{Computation} \\
    \midrule
    \hyperref[sec:perceptron]{QFFNN} & Qubits & Supervised & Circuital & AF & Fully quantum\\
    \midrule
    \hyperref[sec:QRBM]{QRBM} & Qubits & Supervised & Adiabatic & Variational & Fully quantum \\
    \midrule
    \hyperref[sec:CVNN]{CVNN} & Qumodes & Supervised & Circuital & AF & Fully quantum\\
    \midrule
    \hyperref[subsec:DMKDC]{DMKDC} & Qudits & Supervised & Circuital & Variational & Classical emulation\\
    \midrule
    \hyperref[subsec:SVM]{QSVM} & Qubits & Supervised & Circuital & Variational & Hybrid\\
    \midrule
    \hyperref[subsec:QGAN]{QGAN} & Qubits & Unsupervised & Circuital & Variational & Hybrid\\
    \midrule
    \hyperref[sec:QAOA]{QAOA} & Qubits & Unsupervised & Circuital & Variational & Hybrid\\
    \midrule
    \hyperref[sec:QRL]{QRL} & Qubits & Reinforcement & Circuital & Variational & Hybrid\\
  \bottomrule
\end{tabular}
\end{table}

\begin{table}
  \caption{List of generated or available datasets exploited by the authors of the reviewed papers.}
  \label{tab:datasets}
  \begin{tabular}{c|c|c}
    \toprule
    \textbf{Algorithm} & \textbf{Dataset} & \textbf{Task} \\
    \midrule
    \hyperref[sec:perceptron]{QFFNN}~\cite{tacchino2019artificial,tacchino2020quantum,tacchino2020variational} & Binary images (from 2 $\times$ 2 to 4 $\times$ 4) & Image classification \\
    \midrule
    \hyperref[sec:QRBM]{QRBM}~\cite{moro2023anomaly} & KDD CUP 99~\cite{tavallaee2009detailed}, IDS2018~\cite{sharafaldin2018toward} & Web monitoring \\
    \midrule
    \hyperref[sec:CVNN]{CVNN}~\cite{killoran2019continuous} & Credit card transaction dataset~\cite{dal2015calibrating} & Fraud detection \\
    \midrule
    \hyperref[subsec:DMKDC]{DMKDC}~\cite{useche2022quantum} & \href{https://scikit-learn.org/stable/auto_examples/classification/plot_classifier_comparison.html#sphx-glr-auto-examples-classification-plot-classifier-comparison-py}{Moons and circles} dataset, & Image classification \\
     & approximation of probability density functions & \\
    \midrule
    \hyperref[subsec:QGAN]{QGAN}~\cite{herr2021anomaly} & \href{https://www.kaggle.com/datasets/mlg-ulb/creditcardfraud}{Kaggle} credit card fraud detection & Fraud detection \\
    \midrule
    \hyperref[sec:QRL]{QRL}~\cite{albarran2018measurement,albarran2020reinforcement} & Gridworld (20 $\times$ 20 size) & Strategy planning \\
  \bottomrule
\end{tabular}
\end{table}

\section{Concluding remarks}

Anomaly detection performed on quantum computers by quantum machine learning algorithms is at its infancy, but reveals high potential. At the same time, one may expect a transition for what concerns the kind of data and the applications to be managed. Indeed, quantum machine learning suffers of the bottleneck of the data loading issue. Given that no qRAM does still exist, the $1-to-2^n$ parallelized encoding of data in qubits is currently not viable because of its exponential data loading cost. Therefore, three options can be considered: (i) a robust but qubit-expensive  $1-to-1$ encoding of classical data to be loaded by the register used as input of the quantum algorithm, or, alternatively -- in some special cases -- (ii) to generate the data  by a pre-trained quantum circuit returning an approximate probability distribution (derived from another probability distribution easier to generate) which can introduce entangled states as input, or (iii)  to feed the quantum algorithms by quantum data -- another option which potentially inputs an entangled state.
While it is still under investigation to which extent the quantum machine learning can be more precise and faster than classical machine learning methods on classical data, it is likely that the major advantage appears when quantum data are considered. Indeed, a quantum circuit naturally manages quantum states, while instead this is not straightforward in machine learning.  
In several cases, there is no knowledge whether strong quantum advantage does hold or not. Algorithms with \textit{common} quantum advantage should be better explored by looking at demonstrating some stronger quantum speed-up degree while empirically evaluating the trend of its performances when scaling for instance the number of qubits. One should be aware that the empirical search of asymptotic behavior may change the estimate of the trend as soon as larger number of qubits are achieved.
Cybersecurity inherits the algorithms from quantum machine learning, but carries the specificities of dealing with large datasets. Here, the mutually exclusive choice between  kernel-based and variational learning shows the tradeoff: kernel-based guarantees optimal kernel can always be found, but it scales with $O(N^2)$, while for 
variational learning it is possible in time $O(N)$. 
Any decision concerning application of machine learning to real anomaly detection datasets should begin with a real problem based on the dataset, on which a direct benchmarking comparison with classical methods could be evaluated.

\begin{appendices}

\section{The HHL algorithm}
    \label{sec:HHL}

    \begin{figure}[h]
    \centering
    \includegraphics[width=\textwidth]{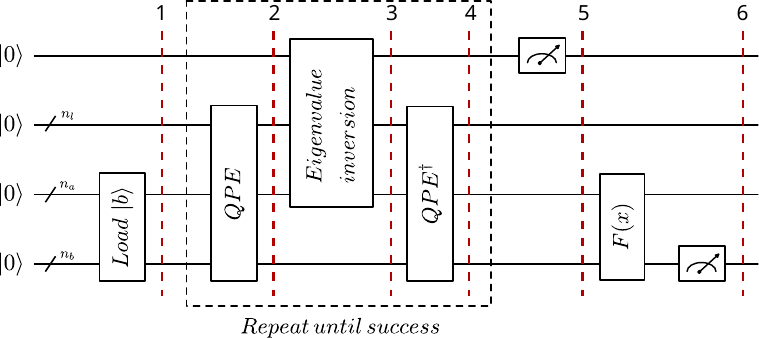}
    \caption{A scheme for the HHL circuit, as reported from \href{https://qiskit.org/textbook/ch-applications/hhl_tutorial.html}{qiskit documentation}.}
    \label{fig:HHL}
    \centering
    \end{figure}
    The Harrow, Hassidim and Lloyd (HHL) algorithm aims to solve a linear system $A \textbf{x} = \textbf{b}$ using a quantum computer~\cite{biamonte2017quantum}. Such algorithm was proposed in~\cite{harrow2009quantum}. The classical method known as the best scales roughly $O(Ns\sqrt(k)\log(1/\epsilon))$ operations, versus a $O(\log(N)k^2 s^2/\epsilon)$ steps on a quantum computer, yielding an exponential speed-up in terms of number of operations $N$~\cite{harrow2009quantum}. The other parameters, in the time-scaling complexity, are the sparsity $s$ of the matrix, i.e. the most number of non-null entries from the rows of the $A$ matrix~\cite{duan2020survey}, the condition number $k$, i.e. the ratio between the largest and smallest eigenvalues of $A$~\cite{harrow2009quantum} and eventually the error $\epsilon$. Thus, the HHL algorithm, to be better performing than the best classical algorithms, requires some caveats, as the $A$ matrix to be sparse and to read out an expectation value over $\textbf{x}$, such as $\bra{x}\hat M \ket{x}$ ($\hat M$ being an observable), rather than outputting the exact $\ket{x}$ state. Nevertheless, such routines are quite common in quantum computation, and may pave the road to future applications in quantum machine learning.
    
    In fact, the matrix inversion is a common routine for many computational processes. The HHL algorithm is a frequent subroutine for many machine learning methods. As in Fig. (\ref{fig:HHL}), the HHL algorithm consists of three main blocks:
    \begin{enumerate}
    \item encoding the $\textbf{b}$ vector into a quantum state $\ket{b}$ (or assume $\ket{b}$ to be already prepared);
    \item perform a quantum phase estimation (QPE), apply a conditioned rotation on an auxiliary qubit by the achieved result and transform back the state by an inverse QPE;
    \item measure the ancilla qubit.
    \end{enumerate}
    Until the qRAM or other techniques of encoding will be leveraged, the first step could turn to be the main overhead~\cite{biamonte2017quantum}, in terms of number $N$ of operations, as the classical information, encoded in $2^n$ bits, needs to be compressed into $n$ qubits.
    Once such step has been accounted, the QPE algorithm takes as input a state $\ket{\psi}_m \in \mathbb{C}^{2^m}$, an ancilla $\ket{0}_n \in \mathbb{C}^{2^n}$ and a unitary operation $\hat U \in \mathbb{C}^{2^m\times 2^m}$ to perform on $\ket{\psi}_m$. The $\ket{\psi}_m$ vector must be eigenvector for $\hat{U}$, under which hypothesis the QPE works in the following manner:
    \begin{equation}
    \hat{QPE} \left[ \hat{U} \ket{0}_n \ket{\psi}_m \right] = \ket{\theta}_n \ket{\psi}_m
    \end{equation}
    with $\hat U$ acting over $\ket{\psi}_m$. From $\ket{\theta}_n$, it is possible to get the binary encoding of $2^n \theta/(2\pi)$. Therefore, to get $\theta$ it is mandatory to divide the result by $2^{n-1}$ and multiply by $\pi$. Just for instance, suppose to apply a $\hat T$ gate on the $\ket{1}$ qubit:
    \begin{equation}
    \hat T \ket{1} = \begin{bmatrix}
    1 & 0 \\
    0 & e^{i\frac{\pi}{4}}
    \end{bmatrix} \begin{pmatrix}
    0 \\ 1
    \end{pmatrix}
    \end{equation}
    For $n=3$, the QPE algorithm, applied on $\hat T \ket{1}$, returns $001$, which is the binary encoding for $1$. Thus, in order to get the correct phase, multiply by $\pi$ and divide by $2^2$, obtaining $\theta=\pi/4$.

    To apply the QPE for the HHL algorithm, the unitary operator $\hat{U}$ can be decomposed as the complex exponentiation of a Hermitian generator $\hat A$:
    \begin{equation}
    \hat U = e^{i \hat A t} = \sum_{j=0}^{2^m -1} e^{i \lambda_j t} \ket{u_j} \bra{u_j}
    \end{equation}
    where $\lambda_j$ are the eigenvalues for $\hat A$ and $\ket{u_j}$ its eigenvectors. Secondly, as any $\mathbb{C}^{2^m\times 2^m}$ Hermitian operator can generate a basis in $\mathbb{C}^{2^m}$ by its eigenstates, the $\ket{\psi}_m$ vector can be decomposed into its $\ket{u_j}$ generators:
    \begin{equation}
    \ket{\psi}_m = \sum_{j=0}^{2^m-1} b_j \ket{u_j}
    \end{equation}
    $b_j$ being the coefficients for $\ket{\psi}_m$ in the $\{\ket{u_j}\}_{j=0}^{2^m-1}$ basis. Afterwards, it is possible to apply the QPE transformation:
    \begin{equation}
    \hat{QPE} \left[ \ket{0}_n e^{i \hat A t} \ket{\psi}_m \right] =
    \hat{QPE} \left[\ket{0}_n \sum_{j=0}^{2^m-1} e^{i\lambda_j t} b_j \ket{u_j}\right] = \sum_{j=0}^{2^m-1} b_j \ket{\lambda_j} \ket{u_j} 
    \end{equation}
    Up to this point, the $\textbf{b}$ vector has been encoded into a $\ket{\psi}_m = \sum_{j=0}^{2^m-1} b_j \ket{u_j}$ state, on which a QPE routine has been acted. The next step is to introduce another ancilla qubit in the $\ket{0}$ state on which to perform a conditioned rotation, using the $\ket{\lambda_j}$ as control qubits:
    \begin{equation}
    \sum_{j=0}^{2^m-1} b_j \ket{\lambda_j} \ket{u_j} \ket{0} \longrightarrow \sum_{j=0}^{2^m-1} b_j \ket{\lambda_j} \ket{u_j} \left[ \sqrt{1-\frac{\Lambda^2}{\lambda_j^2}} \ket{0} + \frac{\Lambda}{\lambda_j} \ket{1} \right]
    \end{equation}
    with $\left|\Lambda \right| < \min_j \lambda_{j}$. Applying the inverse for the QPE yields
    \begin{equation}
    \sum_{j=0}^{2^m-1} b_j \ket{0} \ket{u_j} \left[ \sqrt{1-\frac{\Lambda^2}{\lambda_j^2}} \ket{0} + \frac{\Lambda}{\lambda_j} \ket{1} \right]
    \end{equation}
    Measuring in $\ket{1}$ the ancillary qubit (otherwise the algorithm needs to be run again), the global output turns to be
    \begin{equation}
    \Lambda \sum_{j=0}^{2^m-1} \frac{b_j}{\lambda_j} \ket{0} \ket{u_j}
    \end{equation}
    Apart from a normalization factor, the final output is the state encoding $A^{-1}\textbf{b}$ ($\lambda_j^{-1}$ being the eigenvalues for $A^{-1}$). In case $ A$ not being Hermitian, it is always possible to fix it by building up the following matrix:
    \begin{equation}
    \tilde A = \begin{pmatrix}
    \mathbb{O} & A \\
    A^\dagger & \mathbb{O}
    \end{pmatrix}
    \end{equation}
    with the linear system turning to be
    \begin{equation}
    \tilde A \begin{pmatrix}
    0 \\ \textbf{x}
    \end{pmatrix} = 
    \begin{pmatrix}
    \textbf{b} \\ 0
    \end{pmatrix}
    \end{equation}

    \section{Distance between two hyperplanes}
    \label{sec:Distance}

\begin{figure}[h]
    \centering
    \includegraphics[width=\textwidth]{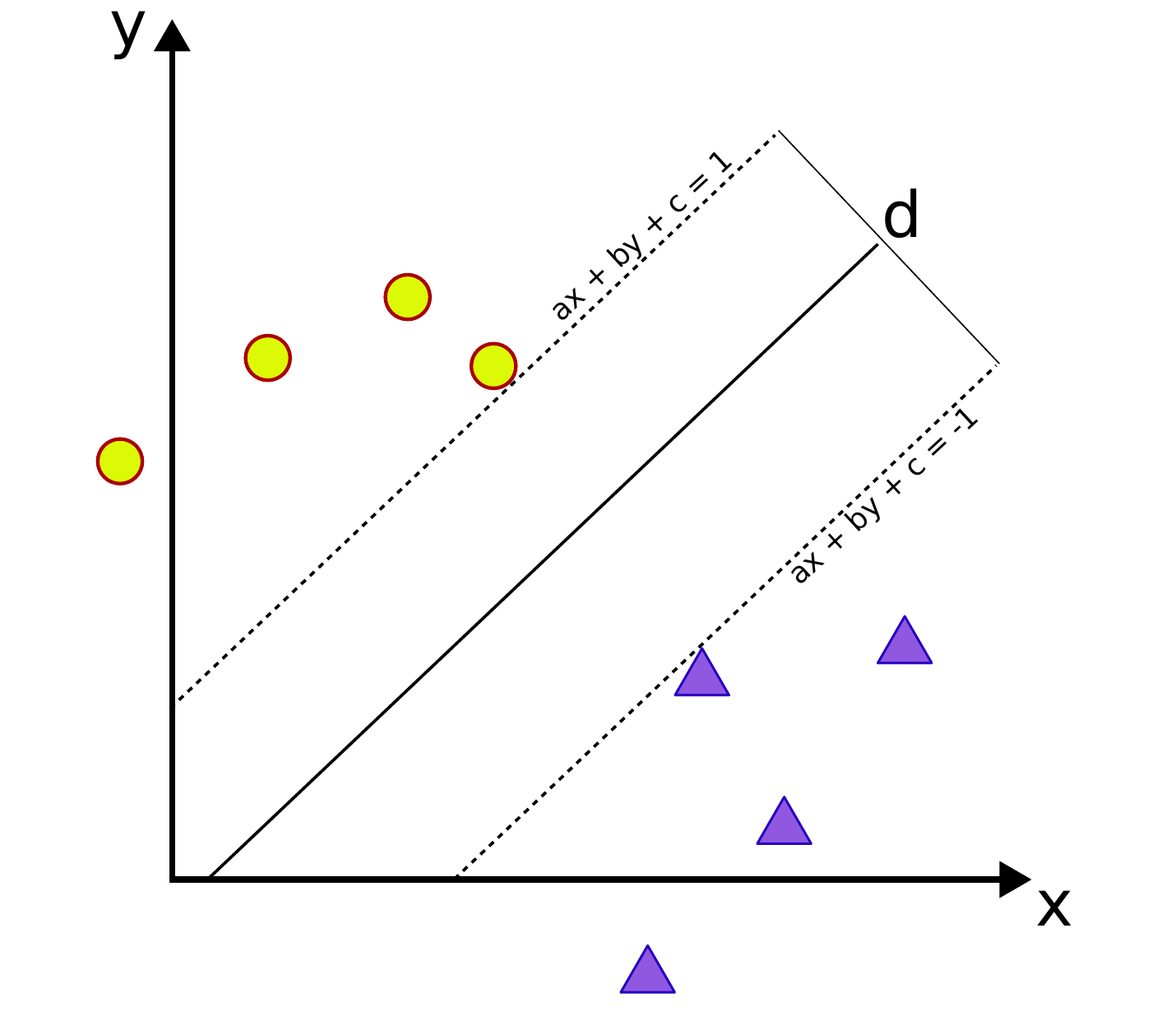}
    \caption{The distance $d=2/\|\textbf{w}\|$} for the $2D$ case, as stated in Equation \eqref{eq:SVMdistance}. In such case, the hyperplane consists of a line.
    \label{fig:SVMdistance}
    \centering
    \end{figure}

    In the following, we prove the distance between the $\textbf{x}_i$ data closest to the hyperplane $\textbf{w} \cdot x + \textbf{b} = 0$ to be $2/\|\textbf{w}\|$. For such points, it holds that $\textbf{w} \cdot \textbf{x} + b = \pm 1$. In the first place, we choose two points $\textbf{x}_1$, $\textbf{x}_2$ such that $\textbf{w} \cdot \textbf{x}_1 + b = 1$, $\textbf{w} \cdot \textbf{x}_2 + b = -1$ and their midpoint $\textbf{x}_M$ to lie on the hyperplane, i.e. $\textbf{w} \cdot \textbf{x}_M + b = 0$, as pictured in Figure \ref{fig:SVMdistance} for the 2D case. The distance $d$ between the two points is given by the sum of the distances between the points and the hyperplane, which we call $d/2$:
    \begin{equation}
        \frac{d}{2} = \frac{\left| \textbf{w} \cdot \textbf{x}_i + c \right|}{\| \textbf{w} \|} = \frac{1}{\|\textbf{w}\|}
    \end{equation}
    Therefore, the overall distance $d$ is set to be 
    \begin{equation}
    \label{eq:SVMdistance}
        d = \frac{d}{2} + \frac{d}{2} = \frac{2}{\|\textbf{w}\|}
    \end{equation}

\end{appendices}



\begin{thebibliography}{9}
\bibitem{barenco1995elementary}
  Barenco, Adriano, et al. \href{https://journals.aps.org/pra/pdf/10.1103/PhysRevA.52.3457}{Elementary gates for quantum computation}. Physical review A 52.5 (1995): 3457.

\bibitem{schuld2015introduction}
Schuld, Maria, Ilya Sinayskiy, and Francesco Petruccione. \href{https://arxiv.org/pdf/1409.3097.pdf}{An introduction to quantum machine learning}. Contemporary Physics 56.2 (2015): 172-185.


\bibitem{morita2008mathematical}
Morita, Satoshi, and Hidetoshi Nishimori. \href{https://pubs.aip.org/aip/jmp/article/49/12/125210/231148}{Mathematical foundation of quantum annealing}. Journal of Mathematical Physics 49.12 (2008).


\bibitem{van2001powerful},
Van Dam, Wim, Michele Mosca, and Umesh Vazirani. \href{https://ieeexplore.ieee.org/stamp/stamp.jsp?arnumber=959902&tag=1}{How powerful is adiabatic quantum computation?}. Proceedings 42nd IEEE symposium on foundations of computer science. IEEE, 2001.



\bibitem{hoban2014measurement}
Hoban, Matty J., et al. \href{https://journals.aps.org/prl/pdf/10.1103/PhysRevLett.112.140505}{Measurement-based classical computation}. Physical review letters 112.14 (2014): 140505.


\bibitem{pius2010automatic}
Pius, Einar. \href{https://www.researchgate.net/profile/Einar-Pius/publication/265352974_Automatic_Parallelisation_of_Quantum_Circuits_Using_the_Measurement_Based_Quantum_Computing_Model/links/555bb89908ae91e75e76683f/Automatic-Parallelisation-of-Quantum-Circuits-Using-the-Measurement-Based-Quantum-Computing-Model.pdf}{Automatic parallelisation of quantum circuits using the measurement based quantum computing model}. High Performance Computing. 2010.


\bibitem{corli2022efficient}
Corli, S., Prati, E.: \href{https://ieeexplore.ieee.org/document/10062779}{An efficient algebraic representation for graph states for measurement-based quantum computing}. In: 2022 IEEE International Conference on Rebooting Computing (ICRC), pp. 1–6 (2022).


\bibitem{kopczyk2018quantum},
Kopczyk, D.: \href{https://arxiv.org/pdf/1804.10068}{Quantum machine learning for data scientists}. arXiv preprint
arXiv:1804.10068 (2018)


\bibitem{pfister2019continuous}
Pfister, O.: \href{https://iopscience.iop.org/article/10.1088/1361-6455/ab526f/pdf}{Continuous-variable quantum computing in the quantum optical
frequency comb}. Journal of Physics B: Atomic, Molecular and Optical Physics 53(1), 012001 (2019)


\bibitem{killoran2019continuous}
Killoran, N., Bromley, T.R., Arrazola, J.M., Schuld, M., Quesada, N., Lloyd, S.: \href{https://journals.aps.org/prresearch/pdf/10.1103/PhysRevResearch.1.033063}{Continuous-variable quantum neural networks}. Physical Review Research 1(3), 033063 (2019)


\bibitem{kreis2012classifying}
Kreis, K., Loock, P.: \href{https://journals.aps.org/pra/pdf/10.1103/PhysRevA.85.032307}{Classifying, quantifying, and witnessing qudit-qumode hybrid entanglement}. Physical Review A 85(3), 032307 (2012)


\bibitem{kendon2010quantum}
Kendon, V.M., Nemoto, K., Munro, W.J.: \href{https://www.jstor.org/stable/pdf/25699189.pdf}{Quantum analogue computing}. Philosophical Transactions of the Royal Society A: Mathematical, Physical and Engineering Sciences 368(1924), 3609–3620 (2010)


\bibitem{haffner2008quantum}
H{\"a}ffner, H., Roos, C.F., Blatt, R.: \href{https://www.sciencedirect.com/science/article/pii/S0370157308003463}{Quantum computing with trapped ions}. Physics reports 469(4), 155–203 (2008)


\bibitem{blatt2012quantum}
Blatt, R., Roos, C.F.: \href{https://www.nature.com/articles/nphys2252}{Quantum simulations with trapped ions}. Nature Physics 8(4), 277–284 (2012)


\bibitem{o2007optical}
O’brien, J.L.: \href{https://www.science.org/doi/full/10.1126/science.1142892}{Optical quantum computing}. Science 318(5856), 1567–1570
(2007)


\bibitem{garcia2022systematic}
Peral-García, David, Juan Cruz-Benito, and Francisco José García-Peñalvo. \href{https://www.sciencedirect.com/science/article/pii/S1574013724000030}{Systematic literature review: Quantum machine learning and its applications}. Computer Science Review 51 (2024): 100619.


\bibitem{biamonte2017quantum}
 Biamonte, J., Wittek, P., Pancotti, N., Rebentrost, P., Wiebe, N., Lloyd, S.: \href{https://www.nature.com/articles/nature23474}{Quantum machine learning}. Nature 549(7671), 195–202 (2017)


\bibitem{zhang2020recent}
Zhang, Y., Ni, Q.: \href{https://onlinelibrary.wiley.com/doi/pdf/10.1002/que2.34}{Recent advances in quantum machine learning}. Quantum
Engineering 2(1), 34 (2020)


\bibitem{zeguendry2023quantum}
Zeguendry, A., Jarir, Z., Quafafou, M.: \href{https://www.mdpi.com/1099-4300/25/2/287}{Quantum machine learning: A review and case studies}. Entropy 25(2), 287 (2023)


\bibitem{jerbi2023quantum}
Jerbi, S., Fiderer, L.J., Poulsen Nautrup, H., K{\"u}bler, J.M., Briegel, H.J., Dunjko, V.: \href{https://www.nature.com/articles/s41467-023-36159-y}{Quantum machine learning beyond kernel methods}. Nature Communications 14(1), 1–8 (2023)


\bibitem{tacchino2019artificial}
Tacchino, F., Macchiavello, C., Gerace, D., Bajoni, D.: \href{https://www.nature.com/articles/s41534-019-0140-4}{An artificial neuron implemented on an actual quantum processor}. npj Quantum Information 5(1),
1–8 (2019)


\bibitem{useche2022quantum}
Useche, D.H., Giraldo-Carvajal, A., Zuluaga-Bucheli, H.M., Jaramillo-Villegas, J.A., Gonz{\'a}lez, F.A.: \href{https://link.springer.com/article/10.1007/s11128-021-03363-y}{Quantum measurement classification with qudits}. Quantum Information Processing 21(1), 1–12 (2022)


\bibitem{herr2021anomaly}
Herr, D., Obert, B., Rosenkranz, M.: \href{https://iopscience.iop.org/article/10.1088/2058-9565/ac0d4d/pdf?casa_token=hYkJSJ9deGsAAAAA:9FCkzhxqXPIzxNSBpVutSK1XLKzolaZOJ0tbA1r98GsZmTyMUx_4qbycaNbW1YF-X5ocbJdpi5SWbg}{Anomaly detection with variational quantum generative adversarial networks}. Quantum Science and Technology 6(4), 045004 (2021)


\bibitem{moro2023anomaly}
 Moro, L., Prati, E.: \href{https://www.nature.com/articles/s42005-023-01390-y.pdf}{Anomaly detection speed-up by quantum restricted Boltzmann machines}. Communications Physics 6(1), 269 (2023)


\bibitem{harrow2009quantum}
 Harrow, A.W., Hassidim, A., Lloyd, S.: \href{https://journals.aps.org/prl/pdf/10.1103/PhysRevLett.103.150502}{Quantum algorithm for linear systems of equations}. Physical review letters 103(15), 150502 (2009)

\bibitem{albarran2018measurement}
Albarr{\'a}n-Arriagada, F., Retamal, J.C., Solano, E., Lamata, L.: \href{https://journals.aps.org/pra/pdf/10.1103/PhysRevA.98.042315?casa_token=a72YsnVBWLcAAAAA%3AFWsOcjQx84mq9BNJB-1ZgygfVt6YDVueCnyIssAqB8AegQDGwDIjJCZ40sAU9cD4mbjWJjgXa2MX8A}{Measurement-based adaptation protocol with quantum reinforcement learning}. Physical Review A 98(4), 042315 (2018)


\bibitem{payares2021quantum}
Payares, E., Martinez-Santos, J.: \href{https://repositorio.utb.edu.co/bitstream/handle/20.500.12585/10426/116990B.pdf?sequence=1}{Quantum machine learning for intrusion detection of distributed denial of service attacks: a comparative overview}. Quantum Computing, Communication, and Simulation 11699, 35–43 (2021)


\bibitem{wang2022integrating}
Wang, H., Wang, W., Liu, Y., Alidaee, B.: \href{https://ieeexplore.ieee.org/stamp/stamp.jsp?arnumber=9829748&tag=1}{Integrating Machine Learning Algorithms With Quantum Annealing Solvers for Online Fraud Detection}. IEEE Access 10, 75908–75917 (2022)


\bibitem{liu2018quantum}
 Liu, N., Rebentrost, P.: \href{https://journals.aps.org/pra/pdf/10.1103/PhysRevA.97.042315}{Quantum machine learning for quantum anomaly detection}. Physical Review A 97(4), 042315 (2018)


\bibitem{suryotrisongko2022evaluating}
 Suryotrisongko, H., Musashi, Y.: \href{https://www.sciencedirect.com/science/article/pii/S1877050921023590}{Evaluating hybrid quantum-classical deep learning for cybersecurity botnet DGA detection}. Procedia Computer Science
197, 223–229 (2022)


\bibitem{ngairangbam2022anomaly}
Ngairangbam, V.S., Spannowsky, M., Takeuchi, M.: \href{https://journals.aps.org/prd/pdf/10.1103/PhysRevD.105.095004}{Anomaly detection in high-energy physics using a quantum autoencoder}. Physical Review D 105(9), 095004 (2022)


\bibitem{wozniak2023quantum}
Wo{\'z}niak, K.A., Belis, V., Puljak, E., Barkoutsos, P., Dissertori, G., Grossi, M., Pierini, M., Reiter, F., Tavernelli, I., Vallecorsa, S.: \href{https://arxiv.org/pdf/2301.10780}{Quantum anomaly detection in the latent space of proton collision events at the LHC}. arXiv preprint arXiv:2301.10780 (2023)


\bibitem{abedi2012support}
Abedi, M., Norouzi, G.-H., Bahroudi, A.: \href{https://www.sciencedirect.com/science/article/pii/S0098300411004389?casa_token=jDbkAzW9XqAAAAAA:FKtTvPH3MYL8yHNoQ0DggTE22VhgEuBKVmFb-X0TJEY7BuV9zpVUVBZNoqAV3H2127IYgiS4Ew}{Support vector machine for multi-classification of mineral prospectivity areas}. Computers \& Geosciences 46, 272–283 (2012)


\bibitem{wu2021application}
Wu, S.L., Sun, S., Guan, W., Zhou, C., Chan, J., Cheng, C.L., Pham, T., Qian, Y., Wang, A.Z., Zhang, R., et al.: \href{https://journals.aps.org/prresearch/pdf/10.1103/PhysRevResearch.3.033221}{Application of quantum machine learning using the quantum kernel algorithm on high energy physics analysis at the LHC}. Physical Review Research 3(3), 033221 (2021)


\bibitem{schuhmacher2023unravelling}
 Schuhmacher, J., Boggia, L., Belis, V., Puljak, E., Grossi, M., Pierini, M., Vallecorsa, S., Tacchino, F., Barkoutsos, P., Tavernelli, I.: \href{https://iopscience.iop.org/article/10.1088/2632-2153/ad07f7}{Unravelling physics beyond the standard model with classical and quantum anomaly detection}. Machine Learning: Science and Technology 4(4), 045031 (2023)


\bibitem{davy2002detection}
Davy, M., Godsill, S.: \href{https://ieeexplore.ieee.org/stamp/stamp.jsp?arnumber=5744044}{ Detection of abrupt spectral changes using support vector machines an application to audio signal segmentation}. In: 2002 IEEE International Conference on Acoustics, Speech, and Signal Processing, vol. 2, p. 1313 (2002). IEEE


\bibitem{chai2024quantum}
Chai, Z., Liu, Y., Wang, M., Guo, Y., Shi, F., Li, Z., Wang, Y., Du, J.: \href{https://onlinelibrary.wiley.com/doi/epdf/10.1002/qute.202300385?saml_referrer=}{ Quantum Anomaly Detection with a Spin Processor in Diamond}. Advanced Quantum Technologies, 2300385 (2024)


\bibitem{sarker2021deep}
Sarker, I.H.: \href{https://link.springer.com/article/10.1007/s42979-021-00535-6}{ Deep cybersecurity: a comprehensive overview from neural network and deep learning perspective}. SN Computer Science 2(3), 1–16 (2021)


\bibitem{li2021comprehensive}
Li, Y., Liu, Q.: \href{https://www.sciencedirect.com/science/article/pii/S2352484721007289}{ A comprehensive review study of cyber-attacks and cyber security; Emerging trends and recent developments}. Energy Reports 7, 8176–8186 (2021)


\bibitem{ravinder2023review}
Ravinder, M., Kulkarni, V.: \href{https://ieeexplore.ieee.org/document/10060871?denied=}{ A Review on Cyber Security and Anomaly Detection Perspectives of Smart Grid}. In: 2023 5th International Conference on Smart Systems and Inventive Technology (ICSSIT), pp. 692–697 (2023). IEEE


\bibitem{wang2022efficient}
Wang, F., Chai, G., Li, Q., Wang, C.: \href{https://www.mdpi.com/2073-8994/14/2/296}{An efficient deep unsupervised domain adaptation for unknown malware detection}. Symmetry 14(2), 296 (2022)


\bibitem{ayodeji2020new}
Ayodeji, A., Liu, Y.-k., Chao, N., Yang, L.-q.: \href{https://www.sciencedirect.com/science/article/pii/S1738573320300590}{ A new perspective towards the development of robust data-driven intrusion detection for industrial control systems}. Nuclear engineering and technology 52(12), 2687–2698 (2020)


\bibitem{gomez2023susan}
G{\'o}mez, {\'A}.L.P., Maim{\'o}, L.F., Celdr{\'a}n, A.H., Clemente, F.J.G.: \href{https://www.sciencedirect.com/science/article/pii/S2210537922001731}{ SUSAN: A Deep Learning based anomaly detection framework for sustainable industry}. Sustainable Computing: Informatics and Systems, 100842 (2023)

\bibitem{tufan2021anomaly}
 Tufan, E., Tezcan, C., Acart{\"u}rk, C.: \href{https://ieeexplore.ieee.org/stamp/stamp.jsp?arnumber=9387304}{ Anomaly-based intrusion detection by machine learning: A case study on probing attacks to an institutional network}. IEEE Access 9, 50078–50092 (2021)


\bibitem{wirkuttis2017artificial}
Wirkuttis, N., Klein, H.: \href{https://intercore.net/wp-content/uploads/2021/06/Artificial-Intelligence-in-Cybersecurity-2021.pdf}{Artificial intelligence in cybersecurity}. Cyber, Intelligence, and Security 1(1), 103–119 (2017)


\bibitem{truong2020artificial}
 Truong, Thanh Cong and Zelinka, Ivan and Plucar, Jan and {\v{C}}and{\'\i}k, Marek and {\v{S}}ulc, Vladim{\'\i}r: \href{https://www.researchgate.net/profile/Srinivasan-Rajendran/publication/342638144_Book_Series_springerfeb_2020/links/5efde532458515505084b739/Book-Series-springerfeb-2020.pdf\#page=359}{Artificial intelligence and cybersecurity: Past, presence, and future}. In: Artificial Intelligence and Evolutionary Computations in Engineering Systems, pp. 351–363 (2020).
Springer


\bibitem{yuan2014droid}
 Yuan, Z., Lu, Y., Wang, Z., Xue, Y.: \href{https://dl.acm.org/doi/pdf/10.1145/2619239.2631434}{Droid-sec: deep learning in android malware detection}. In: Proceedings of the 2014 ACM Conference on SIGCOMM, pp. 371–372 (2014)


\bibitem{yuxin2019malware}
 Yuxin, D., Siyi, Z.: \href{https://link.springer.com/article/10.1007/s00521-017-3077-6}{Malware detection based on deep learning algorithm}. Neural Computing and Applications 31(2), 461–472 (2019)


\bibitem{vinayakumar2019robust}
 Vinayakumar, R., Alazab, M., Soman, K., Poornachandran, P., Venkatraman,
S.: \href{https://ieeexplore.ieee.org/stamp/stamp.jsp?arnumber=8681127}{ Robust intelligent malware detection using deep learning}. IEEE Access 7, 46717–46738 (2019)


\bibitem{bitter2010application}
Bitter, C., Elizondo, D.A., Watson, T.: \href{https://ieeexplore.ieee.org/stamp/stamp.jsp?arnumber=5596532}{Application of artificial neural networks and related techniques to intrusion detection}. In: The 2010 International Joint Conference on Neural Networks (IJCNN), pp. 1–8 (2010). IEEE


\bibitem{kou2004survey}
Kou, Y., Lu, C.-T., Sirwongwattana, S., Huang, Y.-P.: \href{https://ieeexplore.ieee.org/stamp/stamp.jsp?arnumber=1297040}{Survey of fraud detection techniques}. In: IEEE International Conference on Networking, Sensing and Control, 2004, vol. 2, pp. 749–754 (2004). IEEE


\bibitem{ahmad2021network}
Ahmad, Z., Shahid Khan, A., Wai Shiang, C., Abdullah, J., Ahmad, F.: \href{https://onlinelibrary.wiley.com/doi/pdfdirect/10.1002/ett.4150}{Network intrusion detection system: A systematic study of machine learning and deep learning approaches}. Transactions on Emerging Telecommunications
Technologies 32(1), 4150 (2021)


\bibitem{sewak2021deep}
Sewak, M., Sahay, S.K., Rathore, H.: \href{https://arxiv.org/pdf/2206.02733.pdf}{Deep Reinforcement Learning for Cybersecurity Threat Detection and Protection: A Review}. In: International Conference On Secure Knowledge Management In Artificial Intelligence Era, pp. 51–72 (2021). Springer


\bibitem{sjarif2019endpoint}
Sjarif, N.N.A., Chuprat, S., Mahrin, M.N., Ahmad, N.A., Ariffin, A., Senan, F.M., Zamani, N.A., Saupi, A.: \href{https://ieeexplore.ieee.org/stamp/stamp.jsp?arnumber=8939836}{Endpoint Detection and Response: Why Use Machine Learning?}. In: 2019 International Conference on Information and Communication Technology Convergence (ICTC), pp. 283–288 (2019). IEEE


\bibitem{silver2014deterministic}
 Silver, D., Lever, G., Heess, N., Degris, T., Wierstra, D., Riedmiller, M.: \href{http://proceedings.mlr.press/v32/silver14.pdf}{Deterministic policy gradient algorithms}. In: International Conference on Machine Learning, pp. 387–395 (2014). Pmlr


\bibitem{schulman2015trust}
Schulman, J., Levine, S., Abbeel, P., Jordan, M., Moritz, P.: \href{http://proceedings.mlr.press/v37/schulman15.pdf}{Trust region policy optimization}. In: International Conference on Machine Learning, pp. 1889–1897 (2015). PMLR


\bibitem{schulman2017proximal}
 Schulman, J., Wolski, F., Dhariwal, P., Radford, A., Klimov, O.: \href{https://arxiv.org/pdf/1707.06347.pdf}{Proximal policy optimization algorithms}. arXiv preprint arXiv:1707.06347 (2017)


\bibitem{schulman2015high}
Schulman, J., Moritz, P., Levine, S., Jordan, M., Abbeel, P.: \href{https://arxiv.org/pdf/1506.02438.pdf}{High-dimensional continuous control using generalized advantage estimation}. arXiv preprint arXiv:1506.02438 (2015)


\bibitem{campbell2000linear}
Campbell, C., Bennett, K.: \href{https://proceedings.neurips.cc/paper/2000/file/0e087ec55dcbe7b2d7992d6b69b519fb-Paper.pdf}{ A linear programming approach to novelty detection}. Advances in neural information processing systems 13 (2000)


\bibitem{petsche1995neural}
Petsche, T., Marcantonio, A., Darken, C., Hanson, S., Kuhn, G., Santoso, N.: \href{https://proceedings.neurips.cc/paper/1995/file/062ddb6c727310e76b6200b7c71f63b5-Paper.pdf}{A neural network autoassociator for induction motor failure prediction}. Advances in neural information processing systems 8 (1995)


\bibitem{fujimaki2005approach}
Fujimaki, R., Yairi, T., Machida, K.: \href{https://dl.acm.org/doi/pdf/10.1145/1081870.1081917}{An approach to spacecraft anomaly detection problem using kernel feature space}. In: Proceedings of the Eleventh ACM SIGKDD International Conference on Knowledge Discovery in Data Mining, pp. 401–410 (2005)


\bibitem{brotherton2001anomaly}
Brotherton, T., Johnson, T.: \href{https://ieeexplore.ieee.org/stamp/stamp.jsp?arnumber=931329}{Anomaly detection for advanced military aircraft using neural networks}. In: 2001 IEEE Aerospace Conference Proceedings (Cat. No. 01TH8542), vol. 6, pp. 3113–3123 (2001). IEEE


\bibitem{manevitz2001one}
Manevitz, L.M., Yousef, M.: \href{https://www.jmlr.org/papers/volume2/manevitz01a/manevitz01a.pdf?ref=https://codemonkey.link}{One-class SVMs for document classification}. Journal of machine Learning research 2(Dec), 139–154 (2001)


\bibitem{augusteijn2002neural}
Augusteijn, M., Folkert, B.: \href{https://www.tandfonline.com/doi/pdf/10.1080/01431160110055804}{Neural network classification and novelty detection}. International Journal of Remote Sensing 23(14), 2891–2902 (2002)


\bibitem{singh2004approach}
Singh, S., Markou, M.: \href{https://ieeexplore.ieee.org/stamp/stamp.jsp?arnumber=1269665}{An approach to novelty detection applied to the classification of image regions}. IEEE Transactions on Knowledge and Data Engineering 16(4), 396–407 (2004)


\bibitem{srivastava2006enabling}
Srivastava, A.N.: \href{https://ieeexplore.ieee.org/stamp/stamp.jsp?arnumber=1656136}{Enabling the discovery of recurring anomalies in aerospace problem reports using high-dimensional clustering techniques}. In: 2006 IEEE Aerospace Conference, p. 17 (2006). IEEE


\bibitem{srivastava2005discovering}
Srivastava, A.N., Zane-Ulman, B.: \href{https://ieeexplore.ieee.org/stamp/stamp.jsp?arnumber=1559692}{Discovering recurring anomalies in text reports regarding complex space systems}. In: 2005 IEEE Aerospace  conference, pp. 3853–3862 (2005). IEEE


\bibitem{gowtham2014comprehensive}
Gowtham, R., Krishnamurthi, I.: \href{https://www.sciencedirect.com/science/article/pii/S0167404813001442}{A comprehensive and efficacious architecture for detecting phishing webpages}. Computers \& Security 40, 23–37 (2014)


\bibitem{yu2017network}
Yu, Y., Long, J., Cai, Z.: \href{https://www.hindawi.com/journals/scn/2017/4184196/}{Network intrusion detection through stacking dilated convolutional autoencoders}. Security and Communication Networks 2017 (2017)


\bibitem{deutsch1985quantum}
Deutsch, D.: \href{https://www.jstor.org/stable/pdf/2397601.pdf}{Quantum theory, the Church–Turing principle and the universal quantum computer}. Proceedings of the Royal Society of London. A. Mathematical and Physical Sciences 400(1818), 97–117 (1985)


\bibitem{montanaro2016quantum}
Montanaro, A.: \href{https://www.nature.com/articles/npjqi201523}{Quantum algorithms: an overview}. npj Quantum Information 2(1), 1–8 (2016)


\bibitem{jager2023universal}
J{\"a}ger, J., Krems, R.V.: \href{https://www.nature.com/articles/s41467-023-36144-5}{Universal expressiveness of variational quantum classifiers and quantum kernels for support vector machines}. Nature Communications 14(1), 576 (2023)


\bibitem{ronnow2014defining}
Rønnow, T.F., Wang, Z., Job, J., Boixo, S., Isakov, S.V., Wecker, D., Martinis, J.M., Lidar, D.A., Troyer, M.: \href{https://www.science.org/doi/full/10.1126/science.1252319}{Defining and detecting quantum speedup}. science 345(6195), 420–424 (2014)


\bibitem{schuld2019quantum}
Schuld, M., Killoran, N.: \href{https://journals.aps.org/prl/pdf/10.1103/PhysRevLett.122.040504}{Quantum machine learning in feature Hilbert spaces}. Physical review letters 122(4), 040504 (2019)


\bibitem{havlivcek2019supervised}
Havl{\'\i}{\v{c}}ek, Vojt{\v{e}}ch and C{\'o}rcoles, Antonio D and Temme, Kristan and Harrow, Aram W and Kandala, Abhinav and Chow, Jerry M and Gambetta, Jay M: \href{https://www.nature.com/articles/s41586-019-0980-2}{Supervised learning with quantum-enhanced feature
spaces}. Nature 567(7747), 209–212 (2019)


\bibitem{agliardi2022optimal}
Agliardi, G., Prati, E.: Optimal tuning of quantum generative adversarial networks for multivariate distribution loading. Quantum Reports 4(1), 75–105 (2022)


\bibitem{duan2020survey}
 Duan, B., Yuan, J., Yu, C.-H., Huang, J., Hsieh, C.-Y.: \href{https://reader.elsevier.com/reader/sd/pii/S037596012030462X?token=18508C72051830F98FBBF9B6AAC31D092E572FB8449F668D070A81C432EA53F02C06C071693D0B02F721C945E1E14702&originRegion=eu-west-1&originCreation=20230512145510}{A survey on HHL algorithm: From theory to application in quantum machine learning}. Physics Letters A 384(24), 126595 (2020)


\bibitem{giovannetti2008quantum}
 Giovannetti, V., Lloyd, S., Maccone, L.: \href{https://journals.aps.org/prl/pdf/10.1103/PhysRevLett.100.160501}{Quantum random access memory}. Physical review letters 100(16), 160501 (2008)


\bibitem{jaques2023qram}
 Jaques, S., Rattew, A.G.: \href{https://arxiv.org/pdf/2305.10310.pdf}{QRAM: A Survey and Critique}. arXiv preprint
arXiv:2305.10310 (2023)


\bibitem{perez2020data}
 P{\'e}rez-Salinas, A., Cervera-Lierta, A., Gil-Fuster, E., Latorre, J.I.: \href{https://quantum-journal.org/papers/q-2020-02-06-226/?utm_source=researcher_app&utm_medium=referral&utm_campaign=RESR_MRKT_Researcher_inbound}{Data reuploading for a universal quantum classifier}. Quantum 4, 226 (2020)


\bibitem{konar2021qutrit}
 Konar, D., Bhattacharyya, S., Panigrahi, B.K., Behrman, E.C.: \href{https://arxiv.org/pdf/2009.06767.pdf}{Qutrit-inspired fully self-supervised shallow quantum learning network for brain tumor segmentation}. IEEE Transactions on Neural Networks and Learning Systems (2021)


\bibitem{srivastava2016modelling}
Srivastava, D.P., Sahni, V., Satsangi, P.S.: \href{https://arxiv.org/pdf/1505.00774.pdf}{Modelling microtubules in the brain as n-qudit quantum Hopfield network and beyond}. International Journal of General Systems 45(1), 41–54 (2016)


\bibitem{wang2020qudits}
Wang, Y., Hu, Z., Sanders, B.C., Kais, S.: \href{https://www.frontiersin.org/articles/10.3389/fphy.2020.589504/full}{Qudits and high-dimensional quantum computing}. Frontiers in Physics 8, 589504 (2020)


\bibitem{bravyi2022hybrid}
Bravyi, S., Kliesch, A., Koenig, R., Tang, E.: \href{https://quantum-journal.org/papers/q-2022-03-30-678/pdf/}{Hybrid quantum-classical algorithms for approximate graph coloring}. Quantum 6, 678 (2022)


\bibitem{danos2007measurement}
Danos, V., Kashefi, E., Panangaden, P.: \href{https://dl.acm.org/doi/pdf/10.1145/1219092.1219096?casa_token=q9tOFI8E0DcAAAAA:4yZrucaJImwu76wcg7TTuJ_6o_M5Chlva5exjUQAZgKfrrhRCnn4i3lordL-QPeobI9U-Rdrl9M}{The measurement calculus}. Journal of the ACM (JACM) 54(2), 8 (2007)


\bibitem{morita2007convergence}
 Morita, S., Nishimori, H.: \href{https://journals.jps.jp/doi/pdf/10.1143/JPSJ.76.064002}{Convergence of quantum annealing with real-time Schr{\"o}dinger dynamics}. Journal of the Physical Society of Japan 76(6), 064002 (2007)


\bibitem{arute2019quantum}
Arute, F., Arya, K., Babbush, R., Bacon, D., Bardin, J.C., Barends, R., Biswas, R., Boixo, S., Brandao, F.G., Buell, D.A., et al.: \href{https://arxiv.org/pdf/1910.11333.pdf}{Quantum supremacy using a programmable superconducting processor}. Nature 574(7779), 505–510 (2019)


\bibitem{zhong2021phase}
Zhong, H.-S., Deng, Y.-H., Qin, J., Wang, H., Chen, M.-C., Peng, L.-C.,
Luo, Y.-H., Wu, D., Gong, S.-Q., Su, H., et al.: \href{https://journals.aps.org/prl/pdf/10.1103/PhysRevLett.127.180502}{Phase-programmable gaussian boson sampling using stimulated squeezed light}. Physical review letters 127(18), 180502 (2021)


\bibitem{wu2021strong}
Wu, Y., Bao, W.-S., Cao, S., Chen, F., Chen, M.-C., Chen, X., Chung, T.-H., Deng, H., Du, Y., Fan, D., et al.: \href{https://journals.aps.org/prl/pdf/10.1103/PhysRevLett.127.180501}{Strong quantum computational advantage using a superconducting quantum processor}. Physical review letters 127(18), 180501 (2021)


\bibitem{madsen2022quantum}
Madsen, L.S., Laudenbach, F., Askarani, M.F., Rortais, F., Vincent, T., Bulmer, J.F., Miatto, F.M., Neuhaus, L., Helt, L.G., Collins, M.J., et al.: \href{https://www.nature.com/articles/s41586-022-04725-x}{Quantum computational advantage with a programmable photonic processor}. Nature 606(7912), 75–81 (2022)


\bibitem{humble2021quantum}
Humble, T.S., McCaskey, A., Lyakh, D.I., Gowrishankar, M., Frisch, A., Monz, T.: \href{https://ieeexplore.ieee.org/stamp/stamp.jsp?arnumber=9537178}{Quantum computers for high-performance computing}. IEEE Micro 41(5), 15–23 (2021)


\bibitem{bergholm2018pennylane}
Bergholm, V., Izaac, J., Schuld, M., Gogolin, C., Ahmed, S., Ajith, V., Alam, M.S., Alonso-Linaje, G., AkashNarayanan, B., Asadi, A., et al.: \href{https://arxiv.org/pdf/1811.04968.pdf}{Pennylane: Automatic differentiation of hybrid quantum-classical computations}. arXiv preprint arXiv:1811.04968 (2018)


\bibitem{peruzzo2014variational}
 Peruzzo, A., McClean, J., Shadbolt, P., Yung, M.-H., Zhou, X.-Q., Love, P.J., Aspuru-Guzik, A., O’brien, J.L.: \href{https://www.nature.com/articles/ncomms5213}{A variational eigenvalue solver on a photonic quantum processor}. Nature communications 5(1), 4213 (2014)


\bibitem{fahri2014quantum}
Farhi, E., Goldstone, J., Gutmann, S.:  \href{https://openreview.net/pdf?id=BJJsrmfCZ}{A quantum approximate optimization algorithm}. arXiv preprint arXiv:1411.4028 (2014)


\bibitem{paszke2017automatic}
Paszke, A., Gross, S., Chintala, S., Chanan, G., Yang, E., DeVito, Z., Lin, Z., Desmaison, A., Antiga, L., Lerer, A.:  \href{https://openreview.net/pdf?id=BJJsrmfCZ}{Automatic differentiation in pytorch} (2017)


\bibitem{abadi2016tensorflow}
Abadi, M., Agarwal, A., Barham, P., Brevdo, E., Chen, Z., Citro, C., Corrado, G.S., Davis, A., Dean, J., Devin, M., et al.: \href{https://arxiv.org/pdf/1603.04467.pdf}{Tensorflow: Large-scale machine learning on heterogeneous distributed systems}. arXiv preprint arXiv:1603.04467 (2016)


\bibitem{heurtel2023perceval}
 Heurtel, N., Fyrillas, A., Gliniasty, G., Le Bihan, R., Malherbe, S., Pailhas, M., Bertasi, E., Bourdoncle, B., Emeriau, P.-E., Mezher, R., et al.: \href{https://quantum-journal.org/papers/q-2023-02-21-931/}{Perceval: A software platform for discrete variable photonic quantum computing}. Quantum 7, 931 (2023)


\bibitem{vincent2022jet}
 Vincent, T., O’Riordan, L.J., Andrenkov, M., Brown, J., Killoran, N., Qi, H., Dhand, I.: \href{https://quantum-journal.org/papers/q-2022-05-09-709/pdf/}{Jet: Fast quantum circuit simulations with parallel task-based tensor-network contraction}. Quantum 6, 709 (2022)


\bibitem{hornik1991approximation}
 Hornik, K.: \href{http://www.vision.jhu.edu/teaching/learning/deeplearning18/assets/Hornik-91.pdf}{Approximation capabilities of multilayer feedforward networks}. Neural networks 4(2), 251–257 (1991)


\bibitem{kubler2021inductive}
 K{\"u}bler, J., Buchholz, S., Sch{\"o}lkopf, B.: \href{https://proceedings.neurips.cc/paper/2021/file/69adc1e107f7f7d035d7baf04342e1ca-Paper.pdf}{The inductive bias of quantum kernels}. Advances in Neural Information Processing Systems 34, 12661–12673 (2021)


\bibitem{jerbi2024shadows}
 Jerbi, S., Gyurik, C., Marshall, S.C., Molteni, R., Dunjko, V.: \href{https://www.nature.com/articles/s41467-024-49877-8}{Shadows of quantum machine learning}. Nature Communications 15(1), 5676 (2024)


\bibitem{suzuki2022natural}
 Suzuki, Y., Gao, Q., Pradel, K.C., Yasuoka, K., Yamamoto, N.: \href{https://www.nature.com/articles/s41598-022-05061-w}{Natural quantum reservoir computing for temporal information processing}. Scientific reports 12(1), 1353 (2022)


\bibitem{huang2020quantum}
Huang, Y., Lei, H., Li, X., Zhu, Q., Ren, W., Liu, X.: \href{https://cdn.techscience.cn/uploads/attached/file/20200722/20200722054435_68381.pdf}{Quantum generative model with variable-depth circuit}. CMC-COMPUTERS MATERIALS \& CONTINUA
65(1), 445–458 (2020)


\bibitem{chalumuri2021hybrid}
 Chalumuri, A., Kune, R., Manoj, B.: \href{https://link.springer.com/article/10.1007/s11128-021-03029-9}{A hybrid classical-quantum approach for multi-class classification}. Quantum Information Processing 20(3), 119 (2021)


\bibitem{li2019orthogonal}
 Li, S., Jia, K., Wen, Y., Liu, T., Tao, D.: \href{https://ieeexplore.ieee.org/stamp/stamp.jsp?arnumber=8877742}{Orthogonal deep neural networks}. IEEE transactions on pattern analysis and machine intelligence 43(4), 1352–1368 (2019)


\bibitem{li2020quantum}
Li, Y., Zhou, R.-G., Xu, R., Luo, J., Hu, W.: \href{https://iopscience.iop.org/article/10.1088/2058-9565/ab9f93}{A quantum deep convolutional neural network for image recognition}. Quantum Science and Technology 5(4), 044003 (2020)


\bibitem{srikumar2021clustering}
Srikumar, M., Hill, C.D., Hollenberg, L.C.: \href{https://iopscience.iop.org/article/10.1088/2058-9565/ac3c53/pdf}{Clustering and enhanced classification using a hybrid quantum autoencoder}. Quantum Science and Technology 7(1), 015020 (2021)


\bibitem{chen2020quantum}
 Chen, B.-Q., Niu, X.-F.: \href{https://link.springer.com/article/10.1007/s10773-020-04470-9}{Quantum neural network with improved quantum learning algorithm}. International Journal of Theoretical Physics 59, 1978–1991 (2020)


\bibitem{maronese2021continuous}
Maronese, M., Prati, E.:\href{https://www.worldscientific.com/doi/abs/10.1142/S0219749921400025?journalCode=ijqi}{A continuous rosenblatt quantum perceptron}. International Journal of Quantum Information 19(04), 2140002 (2021)


\bibitem{mcclean2018barren}
McClean, J.R., Boixo, S., Smelyanskiy, V.N., Babbush, R., Neven, H.: \href{https://www.nature.com/articles/s41467-018-07090-4}{Barren plateaus in quantum neural network training landscapes}. Nature communications 9(1), 1–6 (2018)


\bibitem{grant2019initialization}
Grant, E., Wossnig, L., Ostaszewski, M., Benedetti, M.: \href{https://quantum-journal.org/papers/q-2019-12-09-214/}{An initialization strategy for addressing barren plateaus in parametrized quantum circuits}. Quantum 3, 214 (2019)


\bibitem{pesah2021absence}
 Pesah, A., Cerezo, M., Wang, S., Volkoff, T., Sornborger, A.T., Coles, P.J.: \href{https://journals.aps.org/prx/pdf/10.1103/PhysRevX.11.041011}{Absence of barren plateaus in quantum convolutional neural networks}. Physical Review X 11(4), 041011 (2021)


\bibitem{cao2017quantum}
 Cao, Y., Guerreschi, G.G., Aspuru-Guzik, A.: \href{https://arxiv.org/pdf/1711.11240.pdf}{Quantum neuron: an elementary building block for machine learning on quantum computers}. arXiv preprint arXiv:1711.11240 (2017)


\bibitem{hu2018towards}
Hu, W.: \href{https://www.scirp.org/journal/paperinformation.aspx?paperid=83091}{Towards a real quantum neuron}. Natural Science 10(3), 99–109 (2018)


\bibitem{da2016weightless}
 Silva, A.J., Oliveira, W.R., Ludermir, T.B.: \href{https://www.sciencedirect.com/science/article/pii/S0925231215020433}{Weightless neural network parameters and architecture selection in a quantum computer}. eurocomputing 183, 13–22 (2016)


\bibitem{matsui2009qubit}
Matsui, N., Nishimura, H., Isokawa, T.: \href{https://www.igi-global.com/chapter/qubit-neural-network/6774}{Qubit neural network: Its performance and applications}. In: Complex-Valued Neural Networks: Utilizing High-Dimensional Parameters, pp. 325–351. IGI Global (2009)


\bibitem{da2016quantum}
 Silva, A.J., Ludermir, T.B., Oliveira, W.R.: \href{https://www.sciencedirect.com/science/article/pii/S0893608016000034}{Quantum perceptron over a field and neural network architecture selection in a quantum computer}. Neural Networks 76, 55–64 (2016)


\bibitem{ventura2000quantum}
Ventura, D., Martinez, T.: \href{https://www.sciencedirect.com/science/article/pii/S0020025599001012}{Quantum associative memory}. Information Sciences 124(1-4), 273–296 (2000)


\bibitem{da2017neural}
 Silva, A.J., Oliveira, R.L.F.: \href{https://ieeexplore.ieee.org/stamp/stamp.jsp?arnumber=8247047}{Neural networks architecture evaluation in a quantum computer}. In: 2017 Brazilian Conference on Intelligent Systems (BRACIS), pp. 163–168 (2017). IEEE


\bibitem{schuld2014quest}
 Schuld, M., Sinayskiy, I., Petruccione, F.: \href{https://link.springer.com/article/10.1007/s11128-014-0809-8}{The quest for a quantum neural network}. Quantum Information Processing 13(11), 2567–2586 (2014)


\bibitem{shao2018quantum}
 Shao, C.: \href{https://arxiv.org/pdf/1808.10561.pdf}{A quantum model for multilayer perceptron}. arXiv preprint arXiv:1808.10561 (2018)


\bibitem{kamruzzaman2019quantum}
 Kamruzzaman, A., Alhwaiti, Y., Leider, A., Tappert, C.C.: \href{http://csis.pace.edu/~ctappert/srd2018/2018PDF/a3.pdf}{Quantum deep learning neural networks}. In: Future of Information and Communication Conference, pp. 299–311 (2019). Springer


\bibitem{tacchino2020quantum}
 Tacchino, F., Barkoutsos, P., Macchiavello, C., Tavernelli, I., Gerace, D., Bajoni, D.: \href{https://arxiv.org/pdf/1912.12486.pdf}{Quantum implementation of an artificial feed-forward neural network}. Quantum Science and Technology 5(4), 044010 (2020)


\bibitem{molteni2023optimization}
 Molteni, R., Destri, C., Prati, E.: \href{https://www.sciencedirect.com/science/article/pii/S0375960123000932}{Optimization of the memory reset rate of a quantum echo-state network for time sequential tasks}. Physics Letters A 465, 128713 (2023)


\bibitem{pritt2017satellite}
Pritt, M., Chern, G.: \href{https://ieeexplore.ieee.org/stamp/stamp.jsp?arnumber=8457969\&casa_token=_1FKtgFjZ30AAAAA:3RPlPiGv4xJK1XuTf55HURetF2EEfcwoYpr4Gfi8G72bB4-6-RrYx9Z8M0gN5JUmCpiReXgU&tag=1}{Satellite image classification with deep learning}. In: 2017 IEEE Applied Imagery Pattern Recognition Workshop (AIPR), pp. 1–7 (2017). IEEE


\bibitem{tacchino2020variational}
 Tacchino, F., Barkoutsos, P.K., Macchiavello, C., Gerace, D., Tavernelli, I., Bajoni, D.: \href{https://ieeexplore.ieee.org/stamp/stamp.jsp?tp=&arnumber=9364892}{Variational learning for quantum artificial neural networks}. In: 2020 IEEE International Conference on Quantum Computing and Engineering (QCE), pp. 130–136 (2020). IEEE


\bibitem{rebentrost2014quantum}
Rebentrost, P., Mohseni, M., Lloyd, S.: \href{https://journals.aps.org/prl/pdf/10.1103/PhysRevLett.113.130503}{Quantum support vector machine for big data classification}. Physical review letters 113(13), 130503 (2014)


\bibitem{delilbasic2021quantum}
 Delilbasic, A., Cavallaro, G., Willsch, M., Melgani, F., Riedel, M., Michielsen, K.: \href{https://ieeexplore.ieee.org/stamp/stamp.jsp?arnumber=9554802}{Quantum support vector machine algorithms for remote sensing data classification}. In: 2021 IEEE International Geoscience and Remote Sensing Symposium IGARSS, pp. 2608–2611 (2021). IEEE


\bibitem{cervantes2020comprehensive}
Cervantes, J., Garcia-Lamont, F., Rodr{\'i}guez-Mazahua, L., Lopez, A.: \href{https://www.sciencedirect.com/science/article/pii/S0925231220307153}{A comprehensive survey on support vector machine classification: Applications, challenges and trends}. Neurocomputing 408, 189–215 (2020)


\bibitem{zhang2004wavelet}
Zhang, L., Zhou, W., Jiao, L.: \href{https://ieeexplore.ieee.org/stamp/stamp.jsp?arnumber=1262479}{Wavelet support vector machine}. IEEE Transactions on Systems, Man, and Cybernetics, Part B (Cybernetics) 34(1), 34–39 (2004)


\bibitem{ding2021quantum}
 Ding, C., Bao, T.-Y., Huang, H.-L.: \href{https://ieeexplore.ieee.org/stamp/stamp.jsp?arnumber=9451546}{Quantum-inspired support vector machine}. IEEE Transactions on Neural Networks and Learning Systems (2021)


\bibitem{li2015experimental}
Li, Z., Liu, X., Xu, N., Du, J.: \href{https://journals.aps.org/prl/pdf/10.1103/PhysRevLett.114.140504}{Experimental realization of a quantum support vector machine}. Physical review letters 114(14), 140504 (2015)


\bibitem{lazzarin2022multi}
 Lazzarin, M., Galli, D.E., Prati, E.: \href{https://www.sciencedirect.com/science/article/pii/S0375960122001384}{Multi-class quantum classifiers with tensor network circuits for quantum phase recognition}. Physics Letters A 434, 128056 (2022)


\bibitem{jeon2024anomaly}
 Jeon, H.-J., Lang, S., Vogel, C., Behrens, R.: \href{https://ieeexplore.ieee.org/stamp/stamp.jsp?arnumber=10589753}{Anomaly Detection from Image Classification}. In: 2024 9th International Conference on Control and Robotics Engineering (ICCRE), pp. 377–381 (2024). IEEE


\bibitem{wei2018anomaly}
 Wei, Q., Ren, Y., Hou, R., Shi, B., Lo, J.Y., Carin, L.: \href{https://www.bibliosearch.polimi.it/discovery/openurl?institution=39PMI_INST&vid=39PMI_INST:VU1&volume=10575&date=2018&aulast=Wei&issn=0143-0343&spage=375&id=doi:10.1117\%2F12.2293408&auinit=Q&title=SPI&atitle=Anomaly\%20detection\%20for\%20medical\%20images\%20based\%20on\%20a\%20one-class\%20classification&sid=google}{Anomaly detection for medical images based on a one-class classification}. In: Medical Imaging 2018: Computer-Aided Diagnosis, vol. 10575, pp. 375–380 (2018). SPIE


\bibitem{liu2024deep}
 Liu, J., Xie, G., Wang, J., Li, S., Wang, C., Zheng, F., Jin, Y.: \href{https://link.springer.com/content/pdf/10.1007/s11633-023-1459-z.pdf}{Deep industrial image anomaly detection: A survey}. Machine Intelligence Research 21(1), 104–135 (2024)


\bibitem{maronese2022quantum}
Maronese, M., Destri, C., Prati, E.: \href{https://link.springer.com/article/10.1007/s11128-022-03466-0}{Quantum activation functions for quantum neural networks}. Quantum Information Processing 21(4), 1–24 (2022)


\bibitem{hinton2002training}
Hinton, G.E.: \href{https://cseweb.ucsd.edu//~gary/guru/tr00-004.pdf}{Training products of experts by minimizing contrastive divergence}. Neural computation 14(8), 1771–1800 (2002)


\bibitem{tieleman2008training}
Tieleman, T.: \href{https://dl.acm.org/doi/pdf/10.1145/1390156.1390290}{Training restricted Boltzmann machines using approximations to the likelihood gradient}. In: Proceedings of the 25th International Conference on Machine Learning, pp. 1064–1071 (2008)

\bibitem{ning2018lcd}
 Ning, L., Pittman, R., Shen, X.: \href{}{LCD: A fast contrastive divergence based algorithm for restricted Boltzmann machine}. Neural Networks 108, 399–410
(2018)


\bibitem{maronese2022quco}
 Maronese, M., Moro, L., Rocutto, L., Prati, E.: \href{https://arxiv.org/pdf/2112.00187}{Quantum compiling}, 39–74 (2022)


\bibitem{rocutto2021quantum}
Rocutto, L., Destri, C., Prati, E.: \href{https://onlinelibrary.wiley.com/doi/pdf/10.1002/qute.202000133?casa_token=yeKhwXLYQmYAAAAA\%3AK45T4Pf5sx4p6A2jT2cj1VXz6bjESnWPhPGF5lgwVj7RNN2mIMNuPmsPQrQqt7SHE3mgVCCzqvZFZQ}{Quantum semantic learning by reverse annealing of an adiabatic quantum computer}. Advanced Quantum Technologies 4(2), 2000133 (2021)


\bibitem{rocutto2021complete}
 Rocutto, L., Prati, E.: \href{https://www.worldscientific.com/doi/pdf/10.1142/S0219749921410033?casa_token=G3QDh2mma40AAAAA\%3Aio5vZ66fwWXD47bfh-tYyg2_fXOulwBeIdh-2YOWqnFfgY_dWpjheRhkl4khdESQlmEyaRrz4HLKeg}{A complete restricted Boltzmann machine on an adiabatic quantum computer}. International Journal of Quantum Information 19(04), 2141003 (2021)


\bibitem{rocutto2023fast}
 Rocutto, L., No{\'e}, D., Moro, L., Prati, E.: \href{https://ieeexplore.ieee.org/stamp/stamp.jsp?arnumber=10313707}{Fast training of fully-connected Boltzmann Machines on an Adiabatic Quantum Computer}. In: 2023 IEEE International Conference on Quantum Computing and Engineering (QCE), vol. 1, pp. 630–635 (2023). IEEE


\bibitem{gonzalez2021classification}
 Gonz{\'a}lez, F.A., Vargas-Calder{\'o}n, V., Vinck-Posada, H.: \href{https://www.researchgate.net/profile/Vladimir-Vargas-Calderon/publication/350889050_Classification_with_Quantum_Measurements/links/607ed709881fa114b4151bb3/Classification-with-Quantum-Measurements.pdf}{Classification with quantum measurements}. Journal of the Physical Society of Japan 90(4), 044002 (2021)


\bibitem{sutherland2015error}
 Sutherland, D.J., Schneider, J.: \href{https://auai.org/uai2015/proceedings/supp/168_supp.pdf}{On the error of random Fourier features}. arXiv preprint arXiv:1506.02785 (2015)


\bibitem{barrue2023quantum}
Barru{\'e}, G., Quertier, T.: \href{https://arxiv.org/pdf/2305.09674.pdf}{Quantum Machine Learning for Malware Classification}. arXiv preprint arXiv:2305.09674 (2023)


\bibitem{kariya2021investigation}
Kariya, A., Behera, B.K.: \href{https://arxiv.org/pdf/2112.06912.pdf}{Investigation of Quantum Support Vector Machine for Classification in NISQ era}. arXiv preprint arXiv:2112.06912 (2021)


\bibitem{boser1992training}
Boser, B.E., Guyon, I.M., Vapnik, V.N.: \href{https://dl.acm.org/doi/pdf/10.1145/130385.130401}{A training algorithm for optimal margin classifiers}. In: Proceedings of the Fifth Annual Workshop on Computational Learning Theory, pp. 144–152 (1992)


\bibitem{suykens1999least}
Suykens, J.A., Vandewalle, J.: \href{https://link.springer.com/content/pdf/10.1023/A:1018628609742.pdf}{Least squares support vector machine classifiers}. Neural processing letters 9(3), 293–300 (1999)


\bibitem{cao2020connecting}
 Cao, H., Guo, X., Laurière, M.: \href{https://arxiv.org/pdf/2002.04112.pdf}{Connecting GANs, MFGs, and OT}. arXiv preprint arXiv:2002.04112 (2020)


\bibitem{gulrajani2017improved}
 Gulrajani, I., Ahmed, F., Arjovsky, M., Dumoulin, V., Courville, A.C.: \href{https://proceedings.neurips.cc/paper/2017/file/892c3b1c6dccd52936e27cbd0ff683d6-Paper.pdf}{Improved training of wasserstein gans}. Advances in neural information processing systems 30 (2017)


\bibitem{aaronson2011computational}
Aaronson, S., Arkhipov, A.: \href{https://dl.acm.org/doi/pdf/10.1145/1993636.1993682}{The computational complexity of linear optics}. In: Proceedings of the Forty-third Annual ACM Symposium on Theory of Computing, pp. 333–342 (2011)

\bibitem{bremner2016average}
Bremner, M.J., Montanaro, A., Shepherd, D.J.: \href{https://journals.aps.org/prl/pdf/10.1103/PhysRevLett.117.080501?casa_token=aENfjh6RA7oAAAAA\%3AnDkDewKQJ5uaAjMSiyXECaOf74yErPppD3taKE9yoQ09W4u0DlliU5ekEv-1_J-hQWRmGw7zHT9bjQ}{Average-case complexity versus approximate simulation of commuting quantum computations}. Physical review letters 117(8), 080501 (2016)


\bibitem{sweke2021quantum}
Sweke, R., Seifert, J.-P., Hangleiter, D., Eisert, J.: \href{https://quantum-journal.org/papers/q-2021-03-23-417/pdf/}{On the quantum versus classical learnability of discrete distributions}. Quantum 5, 417 (2021)


\bibitem{aaronson2015read}
Aaronson, S.: \href{https://www.nature.com/articles/nphys3272}{Read the fine print}. Nature Physics 11(4), 291–293 (2015)


\bibitem{alvi2022quantum}
Alvi, S., Bauer, C., Nachman, B.: \href{https://link.springer.com/content/pdf/10.1007/JHEP02(2023)220.pdf}{Quantum Anomaly Detection for Collider Physics}. arXiv preprint arXiv:2206.08391 (2022)


\bibitem{liang2019quantum}
 Liang, J.-M., Shen, S.-Q., Li, M., Li, L.: \href{https://journals.aps.org/pra/abstract/10.1103/PhysRevA.99.052310}{Quantum anomaly detection with density estimation and multivariate Gaussian distribution}. Physical Review A 99(5), 052310 (2019)


\bibitem{guo2022quantum}
 Guo, M., Liu, H., Li, Y., Li, W., Gao, F., Qin, S., Wen, Q.: \href{https://www.sciencedirect.com/science/article/pii/S0378437122005957}{Quantum algorithms for anomaly detection using amplitude estimation}. Physica A: Statistical Mechanics and its Applications 604, 127936 (2022)


\bibitem{kyriienko2022unsupervised}
 Kyriienko, O., Magnusson, E.B.: \href{https://arxiv.org/pdf/2208.01203.pdf}{Unsupervised quantum machine learning for fraud detection}. arXiv preprint arXiv:2208.01203 (2022)


\bibitem{omran2007overview}
Omran, M.G., Engelbrecht, A.P., Salman, A.: \href{https://www.ire.pw.edu.pl/~arturp/Dydaktyka/PPO/pomoce/clustering.pdf}{An overview of clustering methods}. Intelligent Data Analysis 11(6), 583–605 (2007)


\bibitem{li2022robust}
Li, W., Zhang, X.-Y., Bao, H., Wang, Q., Li, Z.: \href{https://www.sciencedirect.com/science/article/pii/S1389128622004029?casa_token=mNQ4goBT02sAAAAA:wBtrbIJV9_MLtzGjLXieZYaBhXUO0j0nYECoYPSm17UUkCbi85cHCRmFHxpBoZ83EBWMIG55-g}{Robust network traffic identification with graph matching}. Computer Networks 218, 109368 (2022)


\bibitem{lagraa2023review}
 Lagraa, S., Hus{\'a}k, M., Seba, H., Vuppala, S., State, R., Ouedraogo, M.: \href{https://link.springer.com/article/10.1007/s10207-023-00742-7}{A review on graph-based approaches for network security monitoring and botnet detection}. International Journal of Information Security, 1–22 (2023)

\bibitem{lagraa2017botgm}
 Lagraa, S., Fran¸cois, J., Lahmadi, A., Miner, M., Hammerschmidt, C., State, R.: \href{https://ieeexplore.ieee.org/stamp/stamp.jsp?arnumber=8241990}{BotGM: Unsupervised graph mining to detect botnets in traffic flows}. In: 2017 1st Cyber Security in Networking Conference (CSNet), pp. 1–8 (2017). IEEE


\bibitem{festa2002randomized}
 Festa, P., Pardalos, P.M., Resende, M.G., Ribeiro, C.C.: \href{https://citeseerx.ist.psu.edu/viewdoc/download?doi=10.1.1.19.9810&rep=rep1&type=pdf}{Randomized heuristics for the MAX-CUT problem}. Optimization methods and software 17(6), 1033–1058 (2002)


\bibitem{burer2002rank}
Burer, S., Monteiro, R.D., Zhang, Y.: \href{https://epubs.siam.org/doi/pdf/10.1137/s1052623400382467}{Rank-two relaxation heuristics for max-cut and other binary quadratic programs}. SIAM Journal on Optimization 12(2), 503–521 (2002)


\bibitem{zhou2020quantum}
 Zhou, L., Wang, S.-T., Choi, S., Pichler, H., Lukin, M.D.: \href{https://journals.aps.org/prx/pdf/10.1103/PhysRevX.10.021067}{Quantum approximate optimization algorithm: Performance, mechanism, and implementation on near-term devices}. Physical Review X 10(2), 021067 (2020)

\bibitem{ding2001min}
Ding, C.H., He, X., Zha, H., Gu, M., Simon, H.D.: \href{https://ieeexplore.ieee.org/stamp/stamp.jsp?arnumber=989507}{A min-max cut algorithm for graph partitioning and data clustering}. In: Proceedings 2001 IEEE International Conference on Data Mining, pp. 107–114 (2001). IEEE


\bibitem{beaulieu2021max}
 Beaulieu, D., Pham, A.: \href{https://arxiv.org/pdf/2108.13464.pdf}{Max-cut clustering utilizing warm-start qaoa and ibm runtime}. arXiv preprint arXiv:2108.13464 (2021)


\bibitem{proietti2022native}
 Proietti, M., Cerocchi, F., Dispenza, M.: \href{https://journals.aps.org/pra/pdf/10.1103/PhysRevA.106.022437}{Native measurement-based quantum approximate optimization algorithm applied to the Max K-Cut problem}. Physical Review A 106(2), 022437 (2022)


\bibitem{corli2023max}
 Corli, S., Dragoni, D., Proietti, M., Dispenza, M., Cavazzoni, C., Prati, E.: \href{https://ieeexplore.ieee.org/document/10313696?denied=}{A Max K-Cut implementation for QAOA in the measurement based quantum computing formalism}. In: 2023 IEEE International Conference on Quantum Computing and Engineering (QCE), vol. 2, pp. 284–285 (2023). IEEE


\bibitem{an2022quantum}
An, D., Lin, L.: \href{https://dl.acm.org/doi/full/10.1145/3498331\#sec-8}{Quantum linear system solver based on time-optimal adiabatic quantum computing and quantum approximate optimization algorithm}. ACM Transactions on Quantum Computing 3(2), 1–28 (2022)


\bibitem{sun2018adiabatic}
 Sun, Y., Zhang, J.-Y., Byrd, M.S., Wu, L.-A.: \href{https://arxiv.org/pdf/1805.11568}{Adiabatic quantum simulation using trotterization}. arXiv preprint arXiv:1805.11568 (2018)


\bibitem{streif2019comparison}
 Streif, M., Leib, M.: \href{https://arxiv.org/pdf/1901.01903.pdf}{Comparison of QAOA with quantum and simulated annealing}. arXiv preprint arXiv:1901.01903 (2019)


\bibitem{weggemans2022solving}
Weggemans, J.R., Urech, A., Rausch, A., Spreeuw, R., Boucherie, R., Schreck, F., Schoutens, K., Min{\'a}{\v{r}}, Speelman, F.: \href{https://quantum-journal.org/papers/q-2022-04-13-687/}{Solving correlation clustering with QAOA and a Rydberg qudit system: a full-stack approach}. Quantum 6, 687 (2022)

\bibitem{choi2019tutorial}
Choi, J., Kim, J.: \href{https://ieeexplore.ieee.org/stamp/stamp.jsp?arnumber=8939749}{A tutorial on quantum approximate optimization algorithm (QAOA): Fundamentals and applications}. In: 2019 International Conference on Information and Communication Technology Convergence (ICTC), pp. 13 8–142 (2019). IEEE


\bibitem{lotshaw2021empirical}
 Lotshaw, P.C., Humble, T.S., Herrman, R., Ostrowski, J., Siopsis, G.: \href{https://link.springer.com/article/10.1007/s11128-021-03342-3}{Empirical performance bounds for quantum approximate optimization}. Quantum Information Processing 20, 1–32 (2021)


\bibitem{lee2021parameters}
 Lee, X., Saito, Y., Cai, D., Asai, N.: \href{https://ieeexplore.ieee.org/stamp/stamp.jsp?arnumber=9605323}{Parameters fixing strategy for quantum approximate optimization algorithm}. In: 2021 IEEE International Conference on Quantum Computing and Engineering (QCE), pp. 10–16 (2021). IEEE

\bibitem{wurtz2021fixed}
Wurtz, J., Lykov, D.: \href{https://journals.aps.org/pra/pdf/10.1103/PhysRevA.104.052419}{Fixed-angle conjectures for the quantum approximate optimization algorithm on regular MaxCut graphs}. Physical Review A 104(5),
052419 (2021)


\bibitem{pan2022automatic}
 Pan, Y., Tong, Y., Yang, Y.: \href{https://journals.aps.org/pra/pdf/10.1103/PhysRevA.105.032433}{Automatic depth optimization for a quantum approximate optimization algorithm}. Physical Review A 105(3), 032433 (2022)


\bibitem{wang2018quantum}
Wang, Z., Hadfield, S., Jiang, Z., Rieffel, E.G.: \href{https://journals.aps.org/pra/pdf/10.1103/PhysRevA.97.022304}{Quantum approximate optimization algorithm for MaxCut: A fermionic view}. Physical Review A 97(2), 022304 (2018)


\bibitem{sack2021quantum}
Sack, S.H., Serbyn, M.: \href{https://quantum-journal.org/papers/q-2021-07-01-491/pdf/}{Quantum annealing initialization of the quantum approximate optimization algorithm}. quantum 5, 491 (2021)


\bibitem{barak2015beating}
 Barak, B., Moitra, A., O’Donnell, R., Raghavendra, P., Regev, O., Steurer, D., Trevisan, L., Vijayaraghavan, A., Witmer, D., Wright, J.: \href{https://arxiv.org/pdf/1505.03424.pdf}{Beating the random assignment on constraint satisfaction problems of bounded degree}. arXiv preprint arXiv:1505.03424 (2015)


\bibitem{cardenas2018multiqubit}
C{\'a}rdenas-L{\'o}pez, F.A., Lamata, L., Retamal, J.C., Solano, E.: \href{https://journals.plos.org/plosone/article?id=10.1371/journal.pone.0200455}{Multiqubit and multilevel quantum reinforcement learning with quantum technologies}. PloS one 13(7), 0200455 (2018)


\bibitem{mishra2021quantum}
 Mishra, N., Kapil, M., Rakesh, H., Anand, A., Mishra, N., Warke, A., Sarkar, S., Dutta, S., Gupta, S., Prasad Dash, A., et al.: \href{https://www.researchgate.net/profile/Bikash-Behera/publication/335836247_Quantum_Machine_Learning_A_Review_and_Current_Status/links/5d8268fc299bf1996f7749d1/Quantum-Machine-Learning-A-Review-and-Current-Status.pdf}{Quantum machine learning: A review and current status}. Data Management, Analytics and Innovation, 101–145 (2021)


\bibitem{martin2022quantum}
Mart{\'i}n-Guerrero, J.D., Lamata, L.: \href{https://www.sciencedirect.com/science/article/pii/S0925231221011000}{Quantum machine learning: A tutorial}. Neurocomputing 470, 457–461 (2022)


\bibitem{chen2022variational}
 Chen, S.Y.-C., Huang, C.-M., Hsing, C.-W., Goan, H.-S., Kao, Y.-J.: \href{https://iopscience.iop.org/article/10.1088/2632-2153/ac4559/meta}{Variational quantum reinforcement learning via evolutionary optimization}. Machine Learning: Science and Technology 3(1), 015025 (2022)


\bibitem{acuto2022variational}
Acuto, A., Barillà, P., Bozzolo, L., Conterno, M., Pavese, M., Policicchio, A.: \href{https://arxiv.org/pdf/2212.11681.pdf}{Variational Quantum Soft Actor-Critic for Robotic Arm Control}. arXiv preprint arXiv:2212.11681 (2022)


\bibitem{dalla2022quantum}
Dalla Pozza, N., Buffoni, L., Martina, S., Caruso, F.: \href{https://link.springer.com/article/10.1007/s42484-022-00068-y}{Quantum reinforcement learning: the maze problem}. Quantum Machine Intelligence 4(1), 1–10 (2022)


\bibitem{cherrat2022quantum}
Cherrat, E.A., Kerenidis, I., Prakash, A.: \href{https://arxiv.org/pdf/2203.01889.pdf}{Quantum Reinforcement Learning via Policy Iteration}. arXiv preprint arXiv:2203.01889 (2022)


\bibitem{dunjko2016quantum}
Dunjko, V., Taylor, J.M., Briegel, H.J.: \href{https://journals.aps.org/prl/pdf/10.1103/PhysRevLett.117.130501}{Quantum-enhanced machine learning}. Physical review letters 117(13), 130501 (2016)

\bibitem{dong2008quantum}
Dong, D., Chen, C., Li, H., Tarn, T.-J.: \href{https://ieeexplore.ieee.org/stamp/stamp.jsp?arnumber=4579244&casa_token=eO_vEg1vqO4AAAAA:0xBSS5wWfIn9NilcsBM-_AZkQKo09DgccnVJ1nl8h6rdYrQE15ZrO-z8kZoGoQgtN91ZtmKz}{Quantum reinforcement learning}. IEEE Transactions on Systems, Man, and Cybernetics, Part B (Cybernetics) 38(5), 1207–1220 (2008)


\bibitem{yu2019reconstruction}
 Yu, S., Albarr{\'a}n-Arriagada, F., Retamal, J.C., Wang, Y.-T., Liu, W., Ke, Z.-J., Meng, Y., Li, Z.-P., Tang, J.-S., Solano, E., et al.: \href{https://onlinelibrary.wiley.com/doi/pdfdirect/10.1002/qute.201800074}{Reconstruction of a photonic qubit state with reinforcement learning}. Advanced Quantum Technologies 2(7-8), 1800074 (2019)


\bibitem{albarran2020reinforcement}
 Albarr{\'a}n-Arriagada, F., Retamal, J.C., Solano, E., Lamata, L.: \href{https://iopscience.iop.org/article/10.1088/2632-2153/ab43b4/pdf}{Reinforcement learning for semi-autonomous approximate quantum eigensolver}. Machine Learning: Science and Technology 1(1), 015002 (2020)


\bibitem{tavallaee2009detailed}
 Tavallaee, M., Bagheri, E., Lu, W., Ghorbani, A.A.: \href{https://ieeexplore.ieee.org/document/5356528}{A detailed analysis of the KDD CUP 99 data set}. In: 2009 IEEE Symposium on Computational Intelligence for Security and Defense Applications, pp. 1–6 (2009). Ieee


\bibitem{sharafaldin2018toward}
 Sharafaldin, I., Lashkari, A.H., Ghorbani, A.A., et al.: \href{https://www.scitepress.org/Papers/2018/66398/66398.pdf}{Toward generating a new intrusion detection dataset and intrusion traffic characterization}. ICISSp 1, 108–116 (2018)

\bibitem{dal2015calibrating}
 Dal Pozzolo, A., Caelen, O., Johnson, R.A., Bontempi, G.: \href{https://dalpozz.github.io/static/pdf/SSCI_calib_final_noCC.pdf}{Calibrating probability with undersampling for unbalanced classification}. In: 2015 IEEE Symposium Series on Computational Intelligence, pp. 159–166 (2015). IEEE


\bibitem{farhi2014quantum}
Farhi, Edward, Jeffrey Goldstone, and Sam Gutmann: \href{https://arxiv.org/pdf/1411.4028.pdf}{A quantum approximate optimization algorithm}. arXiv preprint arXiv:1411.4028 (2014).









\end{thebibliography}
\end{document}